\documentclass[twocolumn,showpacs,preprintnumbers,superscriptaddress,floatfix,amsmath,prb]{revtex4}
\usepackage{graphicx,hyperref}
\usepackage{graphicx,amsfonts,amssymb,amsmath,xspace}
\usepackage{bm}
\usepackage[FIGTOPCAP]{subfigure}
\usepackage{color}

\def\nbC{{\mathchoice {\setbox0=\hbox{$\displaystyle\rm C$}%
\hbox{\hbox to0pt{\kern0.4\wd0\vrule height0.9\ht0\hss}\box0}}
{\setbox0=\hbox{$\textstyle\rm C$}\hbox{\hbox
to0pt{\kern0.4\wd0\vrule height0.9\ht0\hss}\box0}}
{\setbox0=\hbox{$\scriptstyle\rm C$}\hbox{\hbox
to0pt{\kern0.4\wd0\vrule height0.9\ht0\hss}\box0}}
{\setbox0=\hbox{$\scriptscriptstyle\rm C$}\hbox{\hbox
to0pt{\kern0.4\wd0\vrule height0.9\ht0\hss}\box0}}}}
\def\nbQ{{\mathchoice {\setbox0=\hbox{$\displaystyle\rm
Q$}\hbox{\raise
0.15\ht0\hbox to0pt{\kern0.4\wd0\vrule height0.8\ht0\hss}\box0}}
{\setbox0=\hbox{$\textstyle\rm Q$}\hbox{\raise
0.15\ht0\hbox to0pt{\kern0.4\wd0\vrule height0.8\ht0\hss}\box0}}
{\setbox0=\hbox{$\scriptstyle\rm Q$}\hbox{\raise
0.15\ht0\hbox to0pt{\kern0.4\wd0\vrule height0.7\ht0\hss}\box0}}
{\setbox0=\hbox{$\scriptscriptstyle\rm Q$}\hbox{\raise
0.15\ht0\hbox to0pt{\kern0.4\wd0\vrule height0.7\ht0\hss}\box0}}}}
\def\nbT{{\mathchoice {\setbox0=\hbox{$\displaystyle\rm
T$}\hbox{\hbox to0pt{\kern0.3\wd0\vrule height0.9\ht0\hss}\box0}}
{\setbox0=\hbox{$\textstyle\rm T$}\hbox{\hbox
to0pt{\kern0.3\wd0\vrule height0.9\ht0\hss}\box0}}
{\setbox0=\hbox{$\scriptstyle\rm T$}\hbox{\hbox
to0pt{\kern0.3\wd0\vrule height0.9\ht0\hss}\box0}}
{\setbox0=\hbox{$\scriptscriptstyle\rm T$}\hbox{\hbox
to0pt{\kern0.3\wd0\vrule height0.9\ht0\hss}\box0}}}}
\def\nbS{{\mathchoice
{\setbox0=\hbox{$\displaystyle     \rm S$}\hbox{\raise0.5\ht0%
\hbox to0pt{\kern0.35\wd0\vrule height0.45\ht0\hss}\hbox
to0pt{\kern0.55\wd0\vrule height0.5\ht0\hss}\box0}}
{\setbox0=\hbox{$\textstyle        \rm S$}\hbox{\raise0.5\ht0%
\hbox to0pt{\kern0.35\wd0\vrule height0.45\ht0\hss}\hbox
to0pt{\kern0.55\wd0\vrule height0.5\ht0\hss}\box0}}
{\setbox0=\hbox{$\scriptstyle      \rm S$}\hbox{\raise0.5\ht0%
\hboxto0pt{\kern0.35\wd0\vrule height0.45\ht0\hss}\raise0.05\ht0%
\hbox to0pt{\kern0.5\wd0\vrule height0.45\ht0\hss}\box0}}
{\setbox0=\hbox{$\scriptscriptstyle\rm S$}\hbox{\raise0.5\ht0%
\hboxto0pt{\kern0.4\wd0\vrule height0.45\ht0\hss}\raise0.05\ht0%
\hbox to0pt{\kern0.55\wd0\vrule height0.45\ht0\hss}\box0}}}}
\def\nbZ{{\mathchoice {\hbox{$\sf\textstyle Z\kern-0.4em Z$}}
{\hbox{$\sf\textstyle Z\kern-0.4em Z$}}
{\hbox{$\sf\scriptstyle Z\kern-0.3em Z$}}
{\hbox{$\sf\scriptscriptstyle Z\kern-0.2em Z$}}}}

\begin{document}

\title{The random anisotropy model revisited}

\author{Dominique Mouhanna} \email{mouhanna@lptmc.jussieu.fr}
\affiliation{LPTMC, CNRS-UMR 7600, Universit\'e Pierre et Marie Curie,
bo\^ite 121, 4 Pl. Jussieu, 75252 Paris c\'edex 05, France}
\date{\today}

\author{Gilles Tarjus} \email{tarjus@lptmc.jussieu.fr}
\affiliation{LPTMC, CNRS-UMR 7600, Universit\'e Pierre et Marie Curie,
bo\^ite 121, 4 Pl. Jussieu, 75252 Paris c\'edex 05, France}

\begin{abstract}
We revisit the thermodynamic behavior of the random-anisotropy O($N$) model by investigating its large-$N$ limit. We focus on the system at zero temperature where the mean-field-like artifacts of the large-$N$ limit are less severe. We analyze the connection between the description in terms of self-consistent Schwinger-Dyson equations and the functional renormalization group. We provide a unified description of the phase diagram and critical behavior of the model and clarify the nature of the possible ``glassy'' phases. Finally we discuss the implications of our findings for the finite-$N$ and finite-temperature systems.
\end{abstract}

\pacs{11.10.Hi, 75.40.Cx}

\maketitle

\section{Introduction}

The random anisotropy O($N$) model (RAO($N$)M) was introduced in the 70's to describe the magnetic properties of amorphous alloys with on-site random uniaxial anisotropy.\cite{harris73,harris_review} Such physical realizations involve the Heisenberg ($N=3$) and the XY ($N=2$) versions. The quenched disorder associated with the random axes is relevant and therefore modifies the properties of the pure O($N$) counterpart. However, as for many disordered systems, the nature of the long-distance physics, including the phase diagram and the critical behavior, is still disputed.

Issues that have been settled concern the absence of long-range ferromagnetic order in dimension $d<4$,\cite{pelcovits78,aharony80,goldschmidt83} at least for an isotropic distribution of the random axes,\cite{dudka_review} and the possibility instead of a quasi-long range order (QLRO) for weak enough disorder strength.\cite{feldman,itakura03,tissier_2loop} On the other hand, an issue that has not been settled is the existence of a spin-glass phase for strong enough disorder when $d>4$ and for any disorder strength when $d<4$. Such a phase has been predicted through a variety of theoretical approaches, and sometimes associated with a so-called spontaneous replica-symmetry breaking as in the Sherrington-Kirkpatrick mean-field model of Ising spin glasses,\cite{pelcovits78,boyanovsky83,goldschmidt83,goldschmidt84,khurana84} but this result has been challenged.\cite{fisher85,fisher_pathologies} Furthermore, experiments and simulations have been inconclusive.\cite{itakura03,dudka_review,toldin06}

We revisit the model by considering its large-$N$ limit. Even this limit has led to controversies, which we clarify and resolve in this work. We apply and compare two formalisms: on the one hand, the $2$-particle irreducible (2-PI) formalism, and the associated Schwinger-Dyson-like self-consistent  equations for the pair correlation functions, which allows one to make contact with the putative replica-symmetry breaking; on the other hand, the $1$-particle irreducible (1-PI)  functional renormalization group (FRG), which has proven a powerful tool to study the emergence of nonanalytic renormalized disorder cumulants near zero-temperature fixed points and to unveil the associated physics.\cite{fisher85,fisher86b,nattermann92,narayan92,FRGledoussal-giamarchi,FRGledoussal,ledoussal-wiese_review,feldman,tarjus04,tissier06,tissier_2loop,tissier11,balog-tarjus} Both formalisms are {\it a priori} exact in the large $N$ limit but, as shown in the case of a manifold pinned in a random environment,\cite{doussal_largeN,MDW_largeN} studying the 1-PI FRG flow provides a systematic way to find solutions in the regions of parameter space where the self-consistent Schwinger-Dyson-like equations apparently cease to have stable solutions. Quite importantly, it is also generalizable to finite $N$ cases. 

We focus on the model at zero temperature. The main reason is that one expects that the possible pathologies of the large-$N$ limit in random-anisotropy models\cite{fisher_pathologies} will then be minimized. For instance, one knows from the study of the random-manifold model\cite{doussal_largeN} that the $N \to \infty$ limit is anomalous as far as temperature is concerned: temperature is dangerously irrelevant and leads to a thermal boundary layer in finite $N$, with a deep connection with the phenomenological description in terms of droplet excitations,\cite{fisher_review, FRGledoussal-giamarchi,balents-doussal}  whereas it does not when $N \to \infty$. Restricting oneself to zero temperature therefore avoids some aspects of the large-$N$ limit  (but not all, as will be further discussed) that may not be generic to all values of $N$.

By introducing  in the theory an infrared regulator that suppresses integration of modes with momentum below some cutoff and studying the resulting flow of the observables when decreasing this cutoff, we are able to make the connection between the two formalisms, the 2-PI approach with the associated Schwinger-Dyson equations and the 1-PI FRG, and present a unified description of the RAO($N$)M in the large-$N$ limit at zero temperature. We find in particular the presence of ``glassy" paramagnetic and, above $d=4$, ``glassy" ferromagnetic phases, but no {\it bona fide} spin-glass phase with a spontaneous emergence of a nonzero Edwards-Anderson order parameter. There is no QLRO phase when $d<4$, as anticipated from the finite-$N$ FRG result that predicts the disappearance of such a phase above a critical value $N_c\simeq 9.4412\cdots$.\cite{feldman,tissier_2loop} Finally, we discuss the consequences of our findings for the finite-$N$ model at finite temperature.

\section{Model and naive phase diagram} 
\label{section_model}

The RAO($N$)M is described at a field-theoretical level by the following action: 
\begin{equation}
\begin{aligned}
\label{eq_RAO(N)M_disorder}
S[\boldsymbol{\chi}]= \frac 1T\int_x & \bigg \{\frac 12 (\partial \boldsymbol{\chi}(x))^2+\frac{m^2}{2} \boldsymbol{\chi}(x)^2+ \frac{w}{4! N}(\boldsymbol{\chi}(x)^2)^2\\&  - \mathcal V(x;\boldsymbol{\chi}(x))\bigg \}\,,
\end{aligned}
\end{equation}
where $\int_x \equiv \int d^d x$, $\boldsymbol{\chi}$ is an $N$-component vector, and we have introduced a bare temperature $T$  for bookkeeping. Quenched disorder appears in the form of a random potential $\mathcal V$ that describes a (generalized) random anisotropy and is invariant in the inversion $\boldsymbol{\chi}(x)\to - \boldsymbol{\chi}(x)$ (this symmetry distinguishes the model from its random-field counterpart). The original model considers a random uniaxial anisotropy,\cite{harris73,pelcovits78,aharony80,dudka_review}
\begin{equation}
\mathcal V(x;\boldsymbol{\chi}(x))=D\, \bigg [\sum_{\mu=1}^N \bigg(\hat n^\mu(x)-\frac 1{\sqrt N}\bigg)\chi^\mu(x) \bigg ]^2\,,
\end{equation}
where $\boldsymbol{\hat n}$ a random $N$-dimensional unit vector uncorrelated in space and sampled from a given distribution, but a tensorial model has been studied as well,\cite{goldschmidt83,goldschmidt84,khurana84} with 
\begin{equation}
\mathcal V(x;\boldsymbol{\chi}(x))=\sum_{\mu,\nu=1}^N \tau^{\mu\nu}(x) \chi^\mu(x)\chi^\nu(x)\,,
\end{equation}
where the $\tau^{\mu\nu}$'s are uncorrelated in space and chosen from a Gaussian distribution with zero mean and variance $\overline{\tau^{\mu\nu}(x)\tau^{\mu'\nu'}(x')}=\delta^{(d)}(x-x')(1/2)(\delta_{\mu\mu'}\delta_{\nu\nu'}+\delta_{\mu\nu'}\delta_{\nu\mu'})(\Delta_2/N)$.

It has been argued that the two choices of disorder are not equivalent.\cite{fisher_pathologies} Provided that one restricts oneself to an isotropic distribution of the random axes,\cite{dudka_review} which we will do, the symmetries are the same, but when considering copies or replicas of the systems and averaging over disorder (see below), the random-axis disorder provides an infinite sum of cumulants of all orders whereas the other one is truncated at the second cumulant. We have found that the latter case indeed leads to nongeneric behavior when $N\to \infty$ (as in some sense already alluded to by studies\cite{khurana84,goldschmidt84} of the next-to-leading order in $1/N$). We therefore consider a more general random potential taken from a Gaussian distribution with zero mean, $\overline{\mathcal V}=0$, and variance 
\begin{equation}
\overline{\mathcal V(x;\boldsymbol{\chi}(x))\mathcal V(x';\boldsymbol{\chi'}(x'))}=N \delta^{(d)}(x-x')R\Big(\frac{\boldsymbol{\chi}(x)\mathbf{.} \boldsymbol{\chi'}(x)}{N}\Big)\,,
\end{equation}
where $R(u)$ is a regular function of $u^2$. In the simplest tensorial model, $R(u)=\Delta_2 u^2$. In general, for finite $N$, this simple disorder variance should anyhow be renormalized to a full function, so it seems more generic to directly consider such a function at the bare level. 

\begin{figure}
\hspace{-0.2cm}\includegraphics[width=8.5cm]{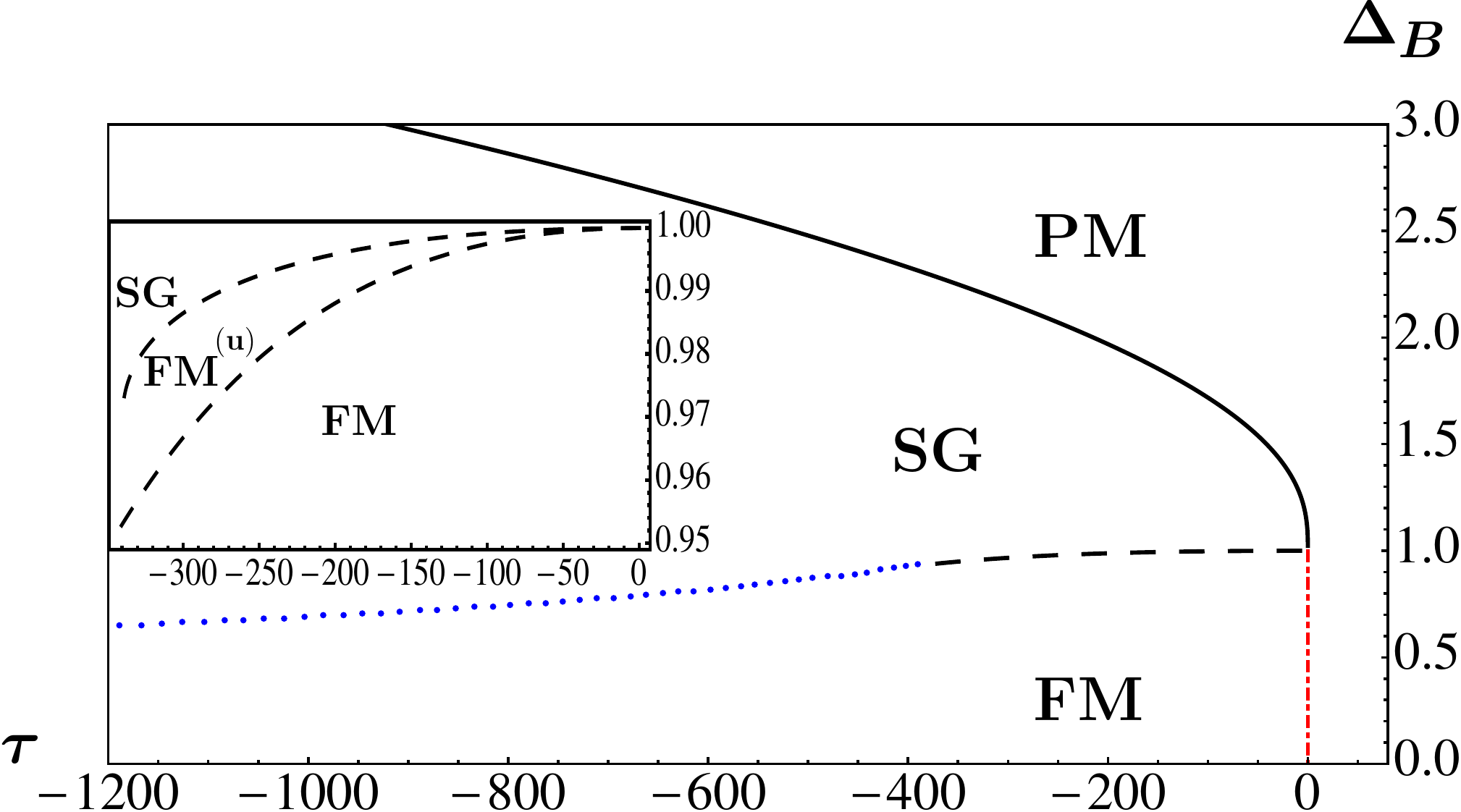}
\caption{(Color on line) Zero-temperature phase diagram of the RAO($N$)M model when $N\to\infty$, as obtained from the ``naive'' solution in $d>4$, illustrated here for $d=5$ and a specific choice of bare disorder variance given in section \ref{section_naive}.  In the main figure, PM, FM,  and SG refer to paramagnetic, ferromagnetic  and  spin-glass phases, respectively. The vertical axis correspond to a measure of the bare disorder strength and the horizontal axis to a measure of the bare mass. There is no applied source (magnetic field). The (red) dashed-dotted vertical line corresponds to the PM-FM transition. The (black) dashed lines are where the ``naive'' solution becomes unstable. (Note that the whole SG phase is unstable.) The (blue) dotted line is the limit of existence of meaningful solutions in the FM region. More details are given in section \ref{section_naive}. 
The inset is a magnification of the SG-FM region near $\Delta_B=1$ and at small negative $\tau$: There is a line separating a stable FM phase from an unstable FM phase, noted FM$^{\scriptsize\hbox{(u)}}$, and another line separating the latter  from the SG phase. }
\label{fig_naive_diagram_a}
\end{figure}

\begin{figure}[h]
\hspace{-0.3cm}
\includegraphics[width=8.5cm,height=5cm]{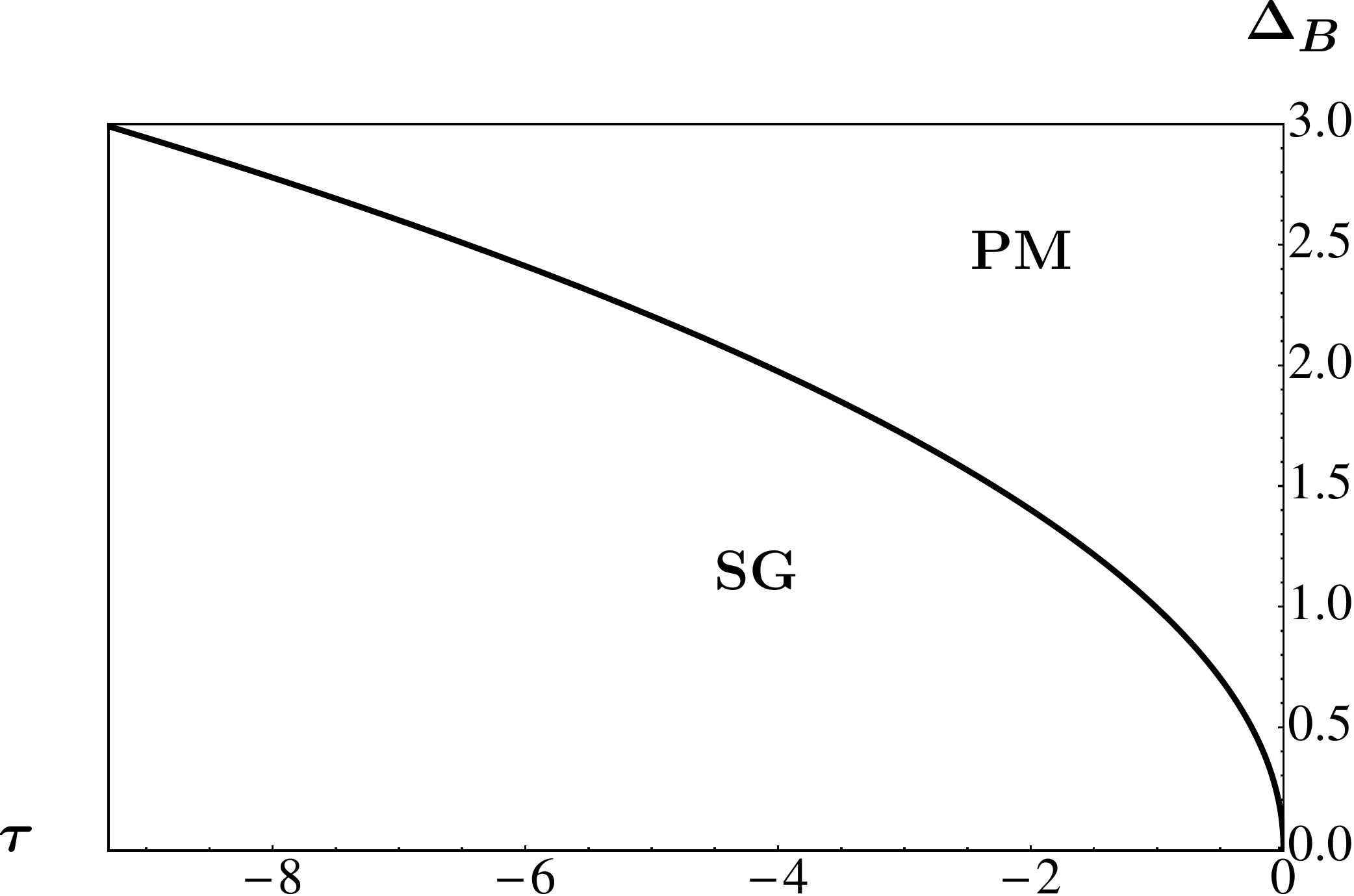}
\caption{(Color on line) Zero-temperature phase diagram of the RAO($N$)M model when $N\to\infty$, as obtained from the ``naive'' solution for $2<d<4$, illustrated here for $d=3$. The whole SG phase is unstable. Details are given in section \ref{section_naive}. Note that due to the definition of the renormalized mass, the scale for the horizontal axis is very different from that of Fig. \ref{fig_naive_diagram_a} for $d=5$.}
\label{fig_naive_diagram_b}
\end{figure}

To sum up the problem, it is convenient to first sketch the putative phase diagram of the RAO($N\to \infty$) model, as obtained from the ``naive'' solution, without bothering with its stability. As already stressed, we consider the model at zero temperature, $T=0$. However, there are still two control parameters (in addition to the coupling constants), the bare mass $m^2$ and the bare disorder strength, $\propto R''(0)$.  We illustrate in Fig.~\ref{fig_naive_diagram_a} the zero-temperature phase diagram in the mass/disorder-strength plane in the absence of applied magnetic field (or source, in field-theoretical language). The critical PM-FM line is described by classical, mean-field exponents, but the FM-SG one is characterized by those of the pure model in 2 dimensions less: this is the $d\to (d-2)$ dimensional-reduction property. The details of the calculation are given below in section~\ref{section_naive}. For $d>4$ (Fig.~\ref{fig_naive_diagram_a}) we find three phases, paramagnetic (PM) for a large bare mass, ferromagnetic (FM) for a small or negative mass and small enough disorder, and spin glass (SG) for a small or negative mass and large enough disorder. There is a whole region of the phase diagram where the  ``naive'' solution is actually unstable, and this includes all of the SG phase and part of the FM one. For $2<d<4$ (Fig.~\ref{fig_naive_diagram_b}), ferromagnetism disappears and only the PM and SG phases survive, with the latter again being obtained from an unstable solution. These phase diagrams are similar to those obtained by several authors before, except for the instability of the SG and part of the FM phase that was not found with the simple tensorial model.\cite{pelcovits78,goldschmidt83}


\section{2-PI formalism for the large-$N$ limit of the random anisotropy O($N$) model}
\label{section_formalism}

As mentioned above, in some limiting cases, closed self-consistent equations can be derived for one type or another of correlation functions. This is for instance the case for the large-$N$ limit of the random-manifold model, in which a manifold of internal dimension $d$ is placed  in a disordered environment embedded in a space of dimension $N$.\cite{mezard-parisi91}  In this system, the relation between closed self-consistent equations and the 1-PI FRG has been studied in great detail by Le Doussal, Wiese, and M\"uller.\cite{doussal_largeN,MDW_largeN} Here we consider the related but somewhat more involved case of the RAO($N$)M.

\subsection{Replicated action and 2-PI formalism}

The first step is to introduce copies or replicas of the system with the same disorder. After averaging over the disorder one obtains a ``replicated'' action
\begin{equation}
\begin{aligned}
\label{eq_RAO($N$)M_replica}
&S_{rep}[\{\boldsymbol{\chi}_a\}]=\frac 1T \int_x\bigg \{\sum_a \bigg [\frac 12 (\partial \boldsymbol{\chi}_a(x))^2+\frac{m^2}{2} \boldsymbol{\chi}_a(x)^2 \\&+ 
\frac{w}{4! N}(\boldsymbol{\chi}_a(x)^2)^2 \bigg ] 
- \frac{N}{2T} \sum_{a,b} R\Big(\frac{\boldsymbol{\chi}_a(x)\mathbf{.}\boldsymbol{\chi}_b(x)}{N}\Big)\bigg \},
\end{aligned}
\end{equation}
where $a,b$ label the replicas. 

The 2-PI effective action can be obtained by first introducing two kinds of sources, $\mathbf J_a(x)$ that linearly couples to $\boldsymbol{\chi}_a(x)$ and 
$\mathbf K_{ab}(x,x')$ that quadratically couples to $(1/2)\boldsymbol{\chi}_a(x)\boldsymbol{\chi}_b(x')$, and then performing a double Legendre transform from $\boldsymbol{\chi}_a$ to the local order parameter field $\boldsymbol{\phi}_a$ and from $\mathbf K_{ab}(x,x')$ to the connected correlation function (considered as a dynamical field) $\mathbf{G}_{ab}(x,x')$. This leads to\cite{luttinger-ward,cornwall74} 
\begin{equation}
\begin{aligned}
\label{eq_RAO($N$)M_Gamma2}
\Gamma_{2PI}&[\{\boldsymbol{\phi}_a\},\{\mathbf{G}_{ab}\}]=S_{rep}[\{\boldsymbol{\phi}_a\} + \frac 12 \mathrm{Tr} \log \mathbf G^{-1}\\& 
+ \frac 12 \mathrm{Tr} \mathbf G\, \mathbf G_0^{-1}[\{\boldsymbol{\phi}_a\}] + \Gamma_2[\{\boldsymbol{\phi}_a\},\{\mathbf{G}_{ab}\}]
\end{aligned}
\end{equation}
where $\mathbf G_0$ is the classical (bare) propagator obtained from the second functional derivative of the bare action 
and $\Gamma_2$ is the sum of all 2-PI diagrams. The stationary condition 
on $\Gamma_{2PI}$ provides the following ``equation of motions'':\cite{luttinger-ward,cornwall74}
\begin{equation}
\begin{aligned}
\label{eq_RAO($N$)M_stationary1}
\frac{\delta \Gamma_{2PI}}{\delta \phi_{a}^{\mu}(x)}-J_{a}^{\mu}(x)=0
\end{aligned}
\end{equation}
and
\begin{equation}
\begin{aligned}
\label{eq_RAO($N$)M_stationary2}
&\frac{\delta \Gamma_{2PI}}{\delta G_{ab}^{\mu\nu}(x,x')}=0 \,,
\end{aligned}
\end{equation}
the latter equation leading to
\begin{equation}
 \left (\mathbf{G}^{-1}\right )_{ab}^{\mu\nu}=\left (\mathbf{G}_0^{-1}\right )_{ab}^{\mu\nu} + \Sigma_{ab}^{\mu\nu}\,,
 \end{equation}
where we have introduced the self-energy functions, 
$\Sigma_{ab}^{\mu\nu}(x,x')=2 \delta \Gamma_2/\delta G_{ab}^{\mu\nu}(x,x')$, which are 1-PI correlation functions. 
We have kept nonzero sources $\mathbf J_a$ that {\it explicitly} break the replica permutational symmetry 
in order to perform expansions in increasing number of free replica sum and access the (renormalized) disorder-averaged cumulants characterizing the system.\cite{doussal_largeN,tarjus04,mouhanna-tarjus} On the other hand, the sources $\mathbf K_{ab}$ are set to zero to obtain Eq.~(\ref{eq_RAO($N$)M_stationary2}).

In the $N\rightarrow \infty$ limit, an explicit expression can be derived for $\Gamma_{2PI}$. A short-cut is provided by using the Gaussian 
variational method, which is known to be exact in the large-$N$ limit.\cite{mezard-parisi91} The variational action, which is a functional of the replica fields $\boldsymbol \phi_a$ and correlation functions $\mathbf  G_{ab}$, gives the expression of $\Gamma_{2PI}$ in the $N\to \infty$ limit. Note that we consider the large-$N$ limit where all replica fields and their differences are of order $\sqrt N$ (this is called the ``thermodynamic regime'' in Ref.~[\onlinecite{MDW_largeN}]). Details of the calculation as well as the resulting expressions are given in Appendix~\ref{appendixA}. In Appendix~\ref{appendixB} we also show how the same result can be derived from the more standard diagrammatic expansion and Feynman graphical representation.

\subsection{Schwinger-Dyson equations}

 From the expression of $\Gamma_{2PI}$, one then obtains the stationary conditions, Eqs.~(\ref{eq_RAO($N$)M_stationary1},\ref{eq_RAO($N$)M_stationary2}). To go further, one splits the correlation functions according to $\mathbf G_{ab}[\{\boldsymbol \phi_c\}]=\widehat{\mathbf G}_a[\{\boldsymbol \phi_c\}] \delta_{ab} + \widetilde{\mathbf G}_{ab}[\{\boldsymbol \phi_c\}]$  (and similarly for $\mathbf G_0$, $\mathbf \Sigma$). Taking advantage of the fact that the symmetry between replicas is explicitly broken by the sources, each component can then be expanded in free replica sums as done in previous work.\cite{tarjus04,tissier11,mouhanna-tarjus} The details are given in Appendix~\ref{appendixA}. Simplifications resulting from the large-$N$ limit are that, first, only fully transverse functions are needed (the longitudinal direction is  as usual defined parallel to the order parameter $\boldsymbol \phi_a$) and, second, that all elements of the self-energy are purely local in space. For uniform replica fields $\boldsymbol \phi_a$, one finds in momentum space 
 \begin{equation}
\begin{aligned}
\label{eq_RAO(Ninfty)M_SDhat}
&\widehat G_T^{[0]}(q^2;\rho_1)=\displaystyle{T\over  \displaystyle{q^2 + m^2 + {w\over 6}\rho_1 +T \widehat \Sigma_T^{[0]}(\rho_1)} }
\end{aligned}
\end{equation}
\begin{equation}
\begin{aligned}
\label{eq_RAO(Ninfty)M_SDtilde}
\widetilde G_{TT}^{[0]}(q^2;\rho_1,&\rho_2,z)= -\frac 1{T^2} \widehat G_T^{[0]}(q^2;\rho_1)\,\widehat G_T^{[0]}(q^2;\rho_2)\\&
\times  \left (-R'(\sqrt{\rho_1\rho_2}z)+ \widetilde \Sigma_{TT}^{[0]}(\rho_1,\rho_2,z) \right )
\end{aligned}
\end{equation}
where we have parametrized the replica fields by $\rho_a=(\boldsymbol \phi_a)^2/N$ (note that there is a factor of 2 of difference with the definition often used for $\rho$: see, {\it e.g.}, Ref.~[\onlinecite{berges02}]) and $z$, which is the cosine of the angle between the two replica fields $\boldsymbol \phi_1$ and $\boldsymbol \phi_2$. The self-energies are known functionals of the transverse correlation functions, $\widehat \Sigma_T^{[0]}(\rho_1)\equiv \widehat \Sigma_T^{[0]}[\rho_1;\widehat G_T]$ and $\widetilde \Sigma_{TT}^{[0]}(\rho_1,\rho_2,z)\equiv  \widetilde \Sigma_{TT}^{[0]}[\rho_1,\rho_2,z;\widehat G_T^{[0]},\widetilde G_{TT}^{[0]}]$, which are given in Appendix~\ref{appendixA}.

To provide explicit expressions, we consider from now on $\rho_1=\rho_2=\rho$, but we 
keep $z$ general so that the important physics associated with the potential nonanalytic behavior of the correlation functions 
in the limit of equal replica sources (or replica fields) is still described: the two replica fields are equal when $z=1$ but are otherwise distinct. It is convenient to introduce a renormalized mass term $y(\rho)=m^2 + (w/6) \rho + T \widehat \Sigma_T(\rho)$, so that 
\begin{equation}
\widehat G_T^{[0]}(q^2;\rho)=\frac T{q^2+y(\rho)}\,, 
\end{equation}
and a renormalized disorder function $\Delta(\rho, z )= R'(\rho z) - \widetilde \Sigma_{TT}(\rho,\rho,z)$, so that
\begin{equation}
\widetilde G_{TT}^{[0]}(q^2;\rho,\rho,z)= \bigg(\frac{\widehat G_T^{[0]}(q^2;\rho)}{T}\bigg)^2 \Delta(\rho, z )\,. 
\end{equation}
Their expressions are given in the form of two coupled self-consistent equations:
\begin{equation}
\begin{aligned}
\label{eq_RAO(Ninfty)M_self-energy_hat}
y(\rho)=   m^2 &+\frac{w}{6}\Big(\rho + \Delta(\rho,1)I_{2} [y(\rho)] \Big) \\
&+\frac{1}{T} \Big [R'\Big(\rho+\Delta(\rho,1) I_2[y(\rho)] \Big) \\&
- R'\Big(\rho+T I_1[y(\rho)]+ \Delta(\rho,1) I_2[y(\rho)] \Big)\Big]
\end{aligned}
\end{equation}
and
\begin{equation}
\begin{aligned}
\label{eq_RAO(Ninfty)M_self-energy_tilde}
&\Delta \left (\rho, z \right )=R'\Big(\rho z+\Delta \left (\rho, z \right )I_2[y(\rho)] \Big)
\end{aligned}
\end{equation}
where
\begin{equation}
\begin{aligned}
\label{eq_I_n(tau)}
I_p[y(\rho)]=T^{-p} \int_q \widehat G_T^{[0]}(q^2;\rho)^p=\int_q \left (\frac{1}{q^2+y(\rho)}\right )^p
\end{aligned}
\end{equation}
with $\int_q \equiv  \int  d^d q/(2\pi)^d$ and $p$ a nonzero integer. An ultra-violet (UV) cutoff $\Lambda$, associated with the inverse of the underlying microscopic length scale, {\it e.g.}, a lattice spacing,  is implicitly considered in the above integrals to ensure the convergence at large momentum $q$. On the other hand, the convergence in the infrared (IR) when $y=0$ depends on the spatial dimension. We will refer to self-consistent equations for correlation functions,  such as Eqs.~(\ref{eq_RAO(Ninfty)M_self-energy_hat},\ref{eq_RAO(Ninfty)M_self-energy_tilde}), as Schwinger-Dyson (SD) equations.

In the limit $T \to 0$, Eq.~(\ref{eq_RAO(Ninfty)M_self-energy_hat}) can be further simplified to
\begin{equation}
\begin{aligned}
\label{eq_RAO(Ninfty)M_self-energy_hat_T0}
y(\rho)=  m^2& +\frac{w}{6}\Big(\rho + \Delta(\rho,1)I_{2} [y(\rho)] \Big)\\
& -I_1[y(\rho)]R''\Big(\rho+\Delta(\rho,1) I_2[y(\rho)]\Big )\, .
\end{aligned}
\end{equation}
The renormalized (transverse) mass $y(\rho)$ is the derivative with respect to $\rho$ of the effective 1-replica potential (or mean Gibbs free energy in the language of magnetic systems) $\mathcal U(\boldsymbol \phi)\equiv N U(\rho=\boldsymbol \phi^2/N)$, {\it i.e.}, $y(\rho)=2U'(\rho)$.

\section{The ``naive'' solution and its instability}
\label{section_naive}

The above set of SD equations is already closed when considering Eq.~(\ref{eq_RAO(Ninfty)M_self-energy_tilde}) for the renormalized disorder function $\Delta(\rho,z)$ in $z=1$ only. This corresponds to the limit in which the two replicas appearing in this renormalized disorder variance are equal (since we have already taken $\rho_1=\rho_2=\rho$, having $z=1$ corresponds to $\boldsymbol{\phi}_1=\boldsymbol{\phi}_2$). It is then the same as what is obtained in the conventional replica framework (where replica symmetry is not explicitly broken by considering distinct sources) when replica symmetry is assumed. 

For the sake of concreteness, we consider a model with a bare variance of the disorder $R(u)$ whose derivative $R'(u)$ is obtained from the inverse function $u=R'^{-1}(Y(u))$ with $R'^{-1}(Y)=\lambda Y-\mu Y^3$. When the disorder strength goes to zero, $Y\to 0$ and $u\to 0$, this form goes back to the simple random tensorial anisotropy case, $R(u)=(1/\lambda)(u^2/2)$, with corrections due to higher-order anisotropies, $(\mu/\lambda^4)(u^4/4)+{\rm O}(u^6)$,  which become increasingly important as $u$ increases. The form of the disorder function makes physical sense at small disorder provided $\lambda, \, \mu>0$ (which correspond to positive variances of the random anisotropies), and the requirement that $u(Y)$ is single-valued imposes the restriction that $\partial_Y R'^{-1}(Y)\geq 0$.

With the above choice of disorder variance, the SD equations, Eqs.~(\ref{eq_RAO(Ninfty)M_self-energy_hat},\ref{eq_RAO(Ninfty)M_self-energy_tilde}),  restricted to $z=1$ read
\begin{equation}
\label{eq_y_0}
y(\rho)= m^2+\frac{w}{6}\Big(\rho + \Delta(\rho,1)I_{2} [y(\rho)] \Big)- \frac{I_{1} [y(\rho) ]}{\lambda-3\mu \Delta (\rho,1)^2}
\end{equation}
\begin{equation}
\begin{aligned}
\label{eq_Delta_0_z1}
\hspace{-1.5cm}\rho =\Big(\lambda-I_{2}[y(\rho)]\Big) \Delta (\rho, 1)-\mu \Delta(\rho, 1 )^3\,,
\end{aligned}
\end{equation}
where we have used $R''(u)=1/[\lambda-3\mu R'(u)^2]$ together with Eq.~(\ref{eq_RAO(Ninfty)M_self-energy_tilde}) to obtain the first equation.

There are {\it a priori} three possible phases in the $N\to \infty$ limit of the RAO($N$)M: PM, FM, and SG (see also section \ref{section_model}).  A phase with quasi-long-range order is not observed in this limit, as will be verified below (see also Refs. [\onlinecite{feldman,tissier_2loop}]).  The  PM phase corresponds to a situation where the only minimum of the effective potential $\mathcal U(\boldsymbol \phi)$ is in $\boldsymbol \phi=\mathbf 0$ (and $y(\rho)=U'(\rho)>0$ $\forall \rho$, except at the critical point where the mass vanishes in $\rho=0$). The order parameter of the FM phase is the spontaneous magnetization, {\it i.e.}, a nonzero value of the field at the minimum of the effective potential: $y(\rho_m)=U'(\rho_m)=0$ with $\rho_m>0$. Finally, the order parameter for a SG phase is the Edwards-Anderson-like parameter: $Q=\int_q \widetilde G_{TT}^{[0]}(q^2;\rho,\rho,z=1)=I_2[y(\rho)]\Delta(\rho,1)$ evaluated in the absence of sources, when $\rho=0$ (and $y(\rho=0)>0$).

\subsection{The ``naive'' phase diagram in $d>4$}

We first consider the case where $d>4$. Then, both $I_1[0]$ and $I_2[0]$ are finite without the need to introduce an IR cutoff (but this is not the case for $I_3[0]$ unless $d>6$). For a large bare mass $m^2$ the system is always in the PM phase. By decreasing $m^2$ for low enough disorder strength, one eventually reaches a critical line separating the PM from the FM phase. This corresponds to $\rho_m=0$, $y(0)=0$, and $Q=0$ (which implies $\Delta(0,1)=0$). The critical line is given from Eq.~(\ref{eq_y_0}) by $0=m^2-I_1[0]/\lambda$. The FM phase exists when $y(\rho_m)=0$ with $\rho_m>0$. Even when $\rho_m \to 0^+$, Eq.~(\ref{eq_Delta_0_z1}) tells us that this can only take place when $\lambda>I_2[0]$. We define for convenience a modified bare mass and a modified bare disorder strength as \begin{equation}
\tau=m^2-\frac{I_1[0]}{\lambda}                                                                                                                                                                                                                                                                                                                                                                                                                                                                                                                                                                                                                                                                                                                                                                                                                                                                                                                                                                                                                                                                                                                                                                                                                                                                                                                                                                                                                                                                                                                                                                                                                                                                  \end{equation}
and 
\begin{equation}
\Delta_B=\frac{I_2[0]}{\lambda}\,.                                                                                                                                                                                                                                                                                                                                                                                                                                                                                                                                                                                                                                                                                                                                                                                                                                                                                                                                                                                                                                                                                                                                                                                                                                                                                                                                                                                                                                                                                                                                                                                                                                                                   \end{equation}
The PM phase thus exists for $\tau\geq 0$ and the critical PM-FM line is located in $\tau=0$ for $\Delta_B \leq 1$ (see Fig. \ref{fig_naive_diagram_a}). It is easily found that the critical exponents characterizing the transition are the classical ones, $\eta=0$, $\nu=1/2$, $\beta=1/2$, etc., which corresponds to the mean-field behavior of a pure $O($N$)$ ferromagnet above its upper critical dimension $d=4$.

The SG phase corresponds to $\rho=0$, $y(0)>0$ and $Q>0$, {\it i.e.}, $\Delta(0,1)>0$. From Eq.~(\ref{eq_Delta_0_z1}), this requires $\Delta(0,1)^2=(\lambda-I_2[y(0)])/\mu$ with $y(0)$ given by Eq.~(\ref{eq_y_0}). A transition between the SG and the FM phase takes place when $y(0)=0$ and $\Delta(0,1)>0$: then, $\Delta(0,1)=\sqrt{(\lambda-I_2[0])/\mu}$ and 
\begin{equation}
\begin{aligned}
-\tau&={w\over 6}\Big({I_2[0]^3\over \mu}\Big)^{1/2}\sqrt{{1-\Delta_B\over\Delta_B}} -3 {I_1[0]\over I_2[0]}{\Delta_B(1-\Delta_B)\over 3\Delta_B-2}
\end{aligned}
\end{equation}
 with $1\geq \Delta_B>2/3$: see the inset in Fig. \ref{fig_naive_diagram_a}.

 Interestingly, when $d<6$, due to the divergence of $I_3[0]$, the critical exponents along this line are not the classical ones and still depend on the dimension $d$. They can be computed from the analysis of the SD equations in the vicinity of the critical line and are found to be those of the pure O($N$) model in the $N\to \infty$ limit in dimension $d-2$: $\eta=0$, $\nu=1/(d-4)$, $\beta=1/2$, etc. This corresponds to the so-called dimensional-reduction property which is predicted for the RAO($N$)M from supersymmetry arguments,\cite{fisher85} but was shown to be wrong from FRG results.\cite{fisher85,feldman,tissier06,tissier_2loop} We will come back to this point in more detail below. Note that for $d>6$ one recovers the mean-field exponents along the SG-FM line: $d=6$ is thus the upper critical dimension. In the following we will mostly consider the case $d<6$, which is more interesting.

\subsection{Replicon operator and stability}

So far the analysis has not taken into account the stability of the SD equations nor the SG susceptibility. In a previous work\cite{mouhanna-tarjus} we have shown that the linear stability operator of the SD equations obtained from the 2-PI formalism is related to the ``replicon'' operator\cite{SGbook,almeida78} which signals the instability of the replica-symmetric solution. In the present case where the self-energies are purely local, it is convenient to consider the 2-PI effective action $\Gamma_{2PI}$ as a functional of the fields $\{\boldsymbol \phi_a\}$ and the self-energies $\{\Sigma_{ab}\}$ and study the stability of the stationary solutions by looking at the second derivatives of $\Gamma_{2PI}$ with respect to the self-energies (see also [\onlinecite{mezard-parisi91,doussal_largeN}]), $(1/NT^2)\sum_{\mu,\nu}2(\delta^2 \Gamma_{2PI}/\delta \Sigma_{ab}^{\mu\mu}\delta \Sigma_{cd}^{\nu\nu})$ with $a<b$ and $c<d$, evaluated when all replica fields are equal: see also Appendix~\ref{appendixA}. The replicon eigenvalue is then simply given by
\begin{equation}
\begin{aligned}
\label{eq_replicon_z=1}
\widetilde \Lambda_{rep} \left (\rho, 1 \right )&= I_{2}[y(\rho)]\Big [1-I_{2}[y(\rho)] R''\Big(\rho+\Delta(\rho, 1)I_{2}[y(\rho)]\Big) \Big ]\\&
=\frac{I_{2}[y(\rho)]}{\lambda-3\mu \Delta (\rho,1)^2}\Big [\lambda-I_{2}[y(\rho)]-3\mu \Delta(\rho, 1)^2 \Big ]
\end{aligned}
\end{equation}
where the second line is valid for the present choice of bare disorder variance. Note that $\lambda-3\mu \Delta (\rho,1)^2$ must be strictly positive (see above) and $I_{2}[y(\rho)]$ is strictly positive as well, so that without loss of generality we can study the second term only, {\it i.e.},
\begin{equation}
\begin{aligned}
\label{eq_replicon_z=1simple}
\Lambda_{rep} \left (\rho, 1 \right )=\lambda-I_{2}[y(\rho)]-3\mu \Delta(\rho, 1)^2\,,
\end{aligned}
\end{equation}
which for simplicity we will keep referring to as the replicon eigenvalue.

The replica-symmetric solution is valid when $\Lambda_{rep} \left (\rho, 1 \right )\geq 0$ and its instability to a putative spontaneous breaking of replica symmetry (the ``Almeida-Thouless'' line\cite{almeida78,SGbook}) corresponds to $\Lambda_{rep} \left (\rho, 1 \right )= 0$. On the other hand, one also has from deriving Eq.~(\ref{eq_RAO(Ninfty)M_self-energy_tilde}) with respect to $z$
\begin{equation}
\begin{aligned}
\label{eq_stability}
\Lambda_{rep} (\rho, 1 )\partial_z \Delta(\rho,z)\vert_{z=1}=\rho\,,
\end{aligned}
\end{equation}
so that when the replicon goes to zero, $\partial_z \Delta(\rho,z)\to \infty$ in the limit $z\to 1$, which signals the appearance of a nonanalytic behavior in $\sqrt{1-z}$ in $\Delta_k(\rho,z)$.\cite{mouhanna-tarjus}

The replicon eigenvalue is positive for a large bare mass $\tau$ and is still positive at the PM-FM critical line. There indeed,  $\Lambda_{rep} (0, 1 )=\lambda-I_{2}[0]$, which is positive when $\Delta_B \leq 1$. When $\rho=0$ the replicon becomes zero when $\Delta(0,1)^2=(\lambda-I_2[y(0)])/3\mu $. It is then negative in the SG phase (where, see above, $\Delta(0,1)^2=(\lambda-I_2[y(0)])/\mu $) and is zero right at the PM-SG transition where  $\Delta(0,1)=0$ and $I_2[y(0)]=\lambda$. The SG susceptibility, which is given by the inverse of the replicon eigenvalue when $\rho=Q=0$, hence, when $\Delta(0,1)=0$, therefore diverges on this line, and the latter is defined by $\tau_{{\rm G}}=y(0)-(I_1(0)-I_1[y(0)])/I_2[y(0)]$ with $I_2[y(0)]=\lambda$. When approaching the transition line from the PM phase, the SG susceptibility diverges as $(\tau-\tau_{{\rm G}})^{-1/2}$, which implies a critical exponent $\gamma=1/2$.

We next determine the locus of the points where $\Lambda_{rep} (\rho_m, 1 )=0$ in the FM phase. This corresponds to $\Delta(\rho_m, 1)^2=(\lambda-I_{2}[0])/(3\mu)$, which from Eq.~(\ref{eq_Delta_0_z1}) gives
\begin{equation}
\hspace{-0.5cm}\Delta(\rho_m, 1)=\Big({I_2[0]\over 3\mu}\Big)^{1/2}\bigg({1\over \Delta_B}-1\bigg)^{1/2}
\end{equation}
and 
\begin{equation}
\rho_m=(2/3)\Big({I_2[0]^3\over 3\mu}\Big)^{1/2}\bigg({1\over \Delta_B}-1\bigg)^{3/2}\ .
\end{equation}

 The additional SD equation leads to another Almeida-Thouless instability line defined by
 \begin{equation}
-\tau_{{\rm G}}={w\over 18}\Big({I_2[0]^3\over 3\mu}\Big)^{1/2}(1-\Delta_B)^{1/2}{(2+\Delta_B)\over \Delta_B^{3/2}}-{I_1[0]\over I_2[0]}(1-\Delta_B)\ . 
\end{equation}

The phase diagram obtained from the above calculation is illustrated in Fig.~\ref{fig_naive_diagram_a} (with $\mu=16$, $w=90$, and $\Lambda=100$).  Note that due to the field-theoretical nature of the formulation, there is only a restricted interval of the bare parameters that leads to meaningful solutions.  At zero temperature, a too large negative masse or a too large disorder strength may lead to no or spurious solutions. This of course would not be the case for a more realistic lattice model but does not alter the qualitative nature of the phase diagram nor the critical behavior. As an illustration of this problem, the blue dotted line in the FM region of the diagram marks the limit of existence of acceptable solutions; the replicon eigenvalue is positive along this line and to the right of it.
 
We stress that the ``naive'' solution, either replica-symmetric in the conventional replica method or analytic in $z$ in the present formalism with explicit breaking of the replica symmetry, is not valid in the region at negative $\tau$ delimited by the two Almeida-Thouless instability lines (the associated ``naive'' replicon eigenvalue is negative there). In particular, one cannot conclude on the existence of a SG phase, and the SG-FM critical line with the dimensional-reduction property has no validity.\cite{footnote_firstorder}

\subsection{The ``naive'' phase diagram in $2<d<4$}

Before delving more into the problem of solving the model when one reaches the lines where the replicon vanishes, we consider the case where $2<d<4$ (only the PM exists when $d<2$). Then, $I_1[0]$ is finite but $I_2[0]$ diverges.  The latter divergence prevents the occurrence of a FM phase and one can check as well that there is no pseudo-FM phase with quasi-long range order. We are left with only two possible phases, PM and SG, in which the renormalized mass stays nonzero. The line separating the PM and SG phases again corresponds to one where the replicon vanishes (and the SG susceptibility diverges). It is again given by $\tau_{{\rm G}}=y(0)-(I_1(0)-I_1[y(0)])/I_2[y(0)]$, with $I_2[y(0)]=\lambda$ for all values of the bare disorder, and the SG susceptibility diverges when approaching the transition line with an exponent $\gamma=1/2$. Note that the definition $\Delta_B=I_2[0]/\lambda$ is no longer appropriate and that we have used instead $\Delta_B=J_d/\lambda$, with $J_d=\lim_{y\to 0}(y^{(4-d)/2} I_2[y])$, in the phase diagram displayed in Fig.~\ref{fig_naive_diagram_b}.

Finally we note that if we set $\mu=0$ in the above results, whether for $2<d<4$ or for $d>4$, the SG phase corresponds to $\lambda-I_2[y(0)]=0$ and the replicon eigenvalue is then equal to zero in the whole phase, instead of becoming negative when $\mu>0$. This corresponds to the simple tensorial model studied in Refs.~[\onlinecite{pelcovits78,boyanovsky83,goldschmidt83}] where it was argued that the SG is stable (or marginal) in the $N\to \infty$ limit.

\section{Cutoff-dependent SD equations and $1$-PI FRG flow}
\label{section_1PI}

Guided by the treatment of the large-$N$ random manifold model,\cite{doussal_largeN} it is tempting to try to go from the SD 
equations to the FRG ones. To do this, one must first add  to the 1-PI 2-point correlation functions (proper vertices) an infrared regulator $R_k(q^2)$ that plays the role of an additional mass term suppressing the small-momentum fluctuations below some running IR scale $k$:
\begin{equation}
\begin{aligned}
\label{eq_RAO(Ninfty)M_SD_k}
T \widehat G_{T,k}^{[0]}(q^2;\rho)^{-1}&=\\
 q^2 &+ R_k(q^2) + m^2 + \frac{\lambda}{3}\rho +\widehat \Sigma_T^{[0]}[\rho;\widehat G_{T,k}^{[0]}(\rho) ]\end{aligned}
\end{equation}
with $R_k(q^2=0)\sim k^2$ and $R_k(q^2 > k^2)\simeq 0$. Note that the functional form of the self-energy $\widehat \Sigma_{T}^{[0]}$ 
does not explicitly depend on $k$: the dependence only comes through the transverse correlation function (or ``propagator'') $\widehat G_{T,k}^{[0]}$. The same property applies to the self-energy $\widetilde \Sigma_{TT}^{[0]}$. Eqs. (\ref{eq_RAO(Ninfty)M_self-energy_hat}) and (\ref{eq_RAO(Ninfty)M_self-energy_tilde}) are thus still valid with $y(\rho)$ replaced by $y_k(\rho)$, $\Delta \left (\rho, z \right )$ replaced by 
$\Delta_k \left (\rho, z \right )$, and $I_p[y(\rho)]$ by 
\begin{equation}
\begin{aligned}
\label{eq_running_Ip}
I_{p,k}[y_k(\rho)]=\int_q \left (\frac 1{q^2+R_k(q^2)+y_k(\rho)}\right )^{p}\,. 
\end{aligned}
\end{equation}
In the limit $T\to 0$ this gives
\begin{equation}
\begin{aligned}
\label{eq_running_y}
y_k(\rho)=  m^2& +{w\over 6}\Big(\rho + \Delta_k(\rho,1)I_{2,k} [y_k(\rho)] \Big) \\ 
& - I_{1,k}[y_k(\rho)]R''\Big(\rho+\Delta_k(\rho,1) I_{2,k}[y_k(\rho)] \Big)
\end{aligned}
\end{equation}
and
\begin{equation}
\label{eq_running_Delta}
\Delta_k \left (\rho, z \right )=R'\Big(\rho z+\Delta_k \left (\rho, z \right )I_{2,k}[y_k(\rho)]\Big)\ .
\end{equation}
Note that the above equations contain {\it both} renormalized 1-PI quantities, $y_k(\rho)$ and $\Delta_k (\rho, z)$, and bare quantities, $m^2$, $w$, and $R(u)$.

Associated with these equations is a cutoff-dependent replicon eigenvalue,
\begin{equation}
\label{eq_running_replicon}
\Lambda_{rep,k}(\rho,1)=\frac 1{R''\Big(\rho +\Delta_k (\rho, 1)I_{2,k}[y_k(\rho)] \Big)}-I_{2,k}[y_k(\rho)] \,.
\end{equation}
This eigenvalue controls the stability of the replica-symmetric solution of the $k$-dependent SD equations with respect to spontaneous replica-symmetry breaking (RSB). It also controls the stability of Eq. (\ref{eq_running_Delta}) when considering its $z$ dependence. Indeed, and similarly to what discussed above [see Eq. (\ref{eq_stability})], deriving the equation with respect to $z$ and evaluating it in $z=1$  leads to
\begin{equation}
\label{eq_running_replicon_nonanalyticity}
\Lambda_{rep,k} (\rho, 1)\, \partial_z \Delta_k(\rho, z )\vert_{z=1}=\rho \,.
\end{equation}
The right-hand side is finite so that the vanishing of the replicon implies the divergence of $\partial_z \Delta_k(\rho, z )$ in $z=1$  and the emergence of a nonanalytic $z$-dependence of $\Delta_k \left (\rho, z \right )$ (see also above). 

Within the present formalism, the replicon can be generalized to any value of $z$, with:
\begin{equation}
\Lambda_{rep,k}(\rho, z)=\frac 1{R''\Big(\rho z+\Delta_k (\rho, z)I_{2,k}[y_k(\rho)] \Big)} -I_{2,k}[y_k(\rho)]
\end{equation}
However, with the physical requirement that  $R''(u)\geq 0$ and $R'''(u)\geq 0$ (and of course $R'(u)\geq 0$), one finds that $\Lambda_{rep,k} (\rho, z)\geq \Lambda_{rep,k} (\rho, z=1)$, at least so long as the replicon in $z=1$ is positive. For given control parameters and a given $\rho$, an instability therefore first appears in $z=1$.

RG flow equations for the 1-PI functions $y_k(\rho)$ and $\Delta_k(\rho,z)$ can be obtained by taking derivatives of the above $k$-dependent SD equations with respect to the IR cutoff $k$ and by eliminating all bare quantities in the expressions to retain only renormalized 1-PI ones. It turns out that this can be done by using the information provided by the partial derivatives of  the equations with respect to $\rho$ and $z$. Focusing first on Eq.~(\ref{eq_running_Delta}) for $\Delta_k(\rho,z)$ one arrives at 
\begin{equation}
\partial_k  \Delta_k(\rho,z)= {1\over \rho} \partial_z \Delta_k(\rho,z)\, \Delta_k(\rho,z)\,  \partial_k I_{2,k}[y_k(\rho)]
\label{eqderivdelta_z}
\end{equation}
and
\begin{equation}
\partial_k \Delta_k(\rho,1)={\partial_{\rho}\Delta_k(\rho,1)\, \Delta_k(\rho,1)\,  \partial_k I_{2,k}[y_k(\rho)]\over 1+\Delta_k(\rho,1) \partial_{\rho}I_{2,k}[y_k(\rho)]}\,,
\label{eqderivdelta_1}
\end{equation}
which are indeed compatible as the partial derivatives of $I_{2,k}$ and $\Delta_k$ are related through
\begin{equation}
\begin{aligned}
\partial_{\rho} I_{2,k}[y_k(\rho)]&=\frac{\rho\, \partial_{\rho}  \Delta_k(\rho,z)-z\, \partial_z  \Delta_k(\rho,z)}{\Delta_k(\rho,z)\,  \partial_z  \Delta_k(\rho,z)} \,.
\label{eq_I2partial}
\end{aligned}
\end{equation}
To go further, one also needs to take into account Eq.~(\ref{eq_running_y}) for $y_k(\rho)$. Details are given in Appendix \ref{appendixC}. The final flow equations are obtained after introducing the operator $\widehat \partial_k$ acting only on the regulator function $R_k(q^2)$, such that $\widehat \partial_k I_{p,k}[y_k(\rho)]\equiv -p \int_q [q^2+R_k(q^2)+y_k(\rho)]^{-(p+1)}\partial_k R_k(q^2)$. They read (recall that we work in the limit $T=0$)
\begin{equation}
\begin{aligned}
\label{eq_RG_y_T0}
\partial_k y_k(\rho)=&-\widehat \partial_k I_{1,k}[y_k(\rho)]\, \partial_{\rho}\Delta_k \left (\rho, 1 \right )\\
&+ \widehat \partial_k I_{2,k}[y_k(\rho)]\, \Delta_k \left (\rho, 1 \right )
\partial_{\rho} y_k(\rho)
\end{aligned}
\end{equation}
and
\begin{equation}
\begin{aligned}
&\partial_k   \Delta_k(\rho,z) =\frac 1\rho  \widehat\partial_k I_{2,k}[y_k(\rho)] 
\Big(\left[\Delta_k(\rho,z)-z\,  \Delta_k(\rho,1)\right]\times \\&
\partial_z  \Delta_k(\rho,z)  + \,   \Delta_k(\rho,1)\, \rho\,  \partial_{\rho} \Delta_k(\rho,z) \Big) 
 - \frac 1 \rho   \widehat\partial_k I_{1k}[y_k(\rho)] \times \\&
\frac{\partial_{\rho}  \Delta_k(\rho,1)}{\partial_{\rho}y_k(\rho)} \Big[\rho\, \partial_{\rho}\,  \Delta_k(\rho,z) -z\, \partial_z  \Delta_k(\rho,z)\Big]\,.
\label{eq_RG_Delta_T0}
\end{aligned}
\end{equation}
Note that the above flow equations are still in a ``dimensionful'' form. To obtain proper RG equations, where transformations associated with both coarse-graining and rescaling are performed, one must also introduce scaling dimensions to define ``dimensionless'' quantities as well as a dimensionless RG scale, $t=\log(k/\Lambda)$ with $\Lambda$ the UV cutoff. This is easily done but for most of the present study it is more convenient to stay with the dimensionful flow equations, which will nonetheless be referred to as FRG equations.

As in the simpler case of the large-$N$ random manifold model,\cite{doussal_largeN} the present derivation  of the 1-PI FRG flow equations from the cutoff-dependent SD equations relies on the structure of the self-energy functions. The latter indeed depend on the arguments $\rho$, $z$, $\widehat G_{T,k}^{[0]}$, $\widetilde G_{TT,k}^{[0]}$  through two combinations only: $\rho z+\Delta_k \left (\rho, z \right )I_{2,k}[y_k(\rho)]$ and $\rho+T I_{1,k}[y_k(\rho)]+\Delta_k \left (\rho, 1 \right )I_{2,k}[y_k(\rho)]$. This is a property of the $N \rightarrow \infty$ limit and is not true in general. (For instance, this is no longer found at the next-to-leading order of the $1/N$ expansion.) The more general relation between SD equations and 1-PI FRG flow will be analyzed elsewhere.\cite{mouhanna-tarjus_new}

One can check that in the vicinity of the lower critical dimension for ferromagnetism,  $d=4$, the above equations coincide with the perturbative 1-PI FRG equations derived from the nonlinear sigma model.\cite{fisher85,feldman,tissier06,tissier_2loop} Indeed in $d=4+\epsilon$, the minimum of the effective potential, which corresponds to $y_k=0$,  grows as $\rho_{mk} \sim 1/\epsilon$ at and around the fixed point that controls the critical point. Asymptotically close to this fixed point,  when $k\to 0$, $\widehat \partial_t I_{1,k}[0]\sim - k^{2+\epsilon}$ and $\widehat \partial_t I_{2,k}[0]\sim - k^\epsilon$, and one can introduce dimensionless quantities through the following scaling: $\rho_{mk}\sim k^{\epsilon+\bar\eta}$, $\Delta_k\sim k^{\bar\eta}$, $y_k(\rho)\sim k^2$, where we have used that the anomalous dimension $\eta$ is zero in the limit $N \to \infty$. Furthermore, it can be  self-consistently checked that $\partial_{\rho}\,  \Delta_k(\rho,1)\vert_{\rho=\rho_{mk}} ={\rm O}(\epsilon)$ while ${\partial_{\rho}y_k(\rho)}\vert_{\rho=\rho_{mk}}={\rm O}(1)$. Eqs.~(\ref{eq_RG_y_T0},\ref{eq_RG_Delta_T0}) can finally be simplified to
\begin{equation}
\begin{aligned}
\label{eq_RG_rho_epsilon}
\partial_t \rho_{mk}=C_4 \,k^\epsilon \,\Delta_k \left (\rho_{mk}, 1 \right )
\end{aligned}
\end{equation}
and
\begin{equation}
\begin{aligned}
&\partial_t  \Delta_k(\rho_{mk},z) =\\&
-\frac {1}{\rho_{mk} }C_4 \,k^\epsilon\Big[[\Delta_k(\rho_{mk},z)- z\,  \Delta_k(\rho_{mk},1)\Big] 
\partial_z  \Delta_k(\rho_{mk},z)
\label{eq_RG_Delta_epsilon}
\end{aligned}
\end{equation}
where $C_4$ is a positive constant and the first equation follows from $\partial_t y_k(\rho_{mk})=0=\partial_t y_k(\rho)\vert_{\rho=\rho_{mk}}+\partial_\rho y_k(\rho)\vert_{\rho=\rho_{mk}} \partial_t \rho_{mk}$. After introducing $\widehat \Delta_{k}(z)=k^\epsilon \Delta_k(\rho_{mk},z)/\rho_{mk}$, which is of order $\epsilon$ at the fixed point, one arrives at the following flow equation: 
\begin{equation}
\begin{aligned}
\partial_t   \widehat \Delta_k(z) = \, \epsilon \widehat \Delta_k(z)
& -C_4 \Big (\big[\widehat \Delta_k(z) -z\widehat \Delta_k(1) \big] \widehat \Delta_k'(z) \\
&  + \widehat \Delta_k(z) \widehat \Delta_k(1) \Big )\,.
\label{eq_RG_hatDelta_epsilon}
\end{aligned}
\end{equation}
From Eq.~(\ref{eq_RG_rho_epsilon}) combined with the definition of $\widehat \Delta_k$ and the scaling of $\rho_{mk}$, one also immediately derives that $\epsilon+\bar\eta=C_4 \widehat \Delta_*(z=1)$, where $\widehat \Delta_*(z)$ is the fixed-point function. These equations are identical to the $N\to \infty$ limit of the 1-PI FRG equations derived at one loop in Refs.~[\onlinecite{fisher85,feldman,tissier06}]. The above equations are amenable to explicit solutions (see also [\onlinecite{tissier_2loop}]).

To conclude this section, what has been gained by turning the self-consistent equations (at the scale $k$) into differential flow equations is that the linear stability (replicon) operator has been partly eliminated through the manipulations and that a solution can now be found by following the evolution with decreasing $k$. We will now show the interest of such a change of perspective.

\section{Partial solution of the cutoff dependent SD equations and vanishing of the replicon}
\label{section_solution_runningSD}

We first consider the solution of the $k$-dependent SD equations. So long as the associated replicon eigenvalue, $\Lambda_{rep,k}(\rho,1)$, is positive, the solution of these equations coincides with that of the 1-PI FRG flow. It is therefore interesting to first locate the lines where the replicon vanishes. 

For the sake of concreteness we study the model already introduced with $R'^{-1}(Y)=\lambda Y-\mu Y^3$, but the results are generic. The $k$-dependent SD equations at $T=0$ then read
\begin{equation}
\begin{aligned}
\label{eq_y_k}
y_k(\rho)= m^2+\frac{w}{6}\Big(\rho + \Delta_{k} (\rho)I_{2,k}\left [y_k(\rho)\right ]\Big) - \frac{I_{1,k}\left [y_k(\rho)\right ]}{\lambda-3\mu \Delta_{k} (\rho)^2}
\end{aligned}
\end{equation}
\begin{equation}
\begin{aligned}
\label{eq_Delta_k}
\rho z=\Big(\lambda-I_{2,k}[y_k(\rho)]\Big) \Delta_k \left (\rho, z \right )-\mu \Delta_k \left (\rho, z \right )^3\,,
\end{aligned}
\end{equation}
which can be closed by looking only at $z=1$. In addition, the $k$- and $\rho$-dependent replicon eigenvalue is given by
\begin{equation}
\begin{aligned}
\label{eq_replicon_k_z}
\Lambda_{rep,k}(\rho,1)= \lambda- I_{2,k}\left [y_k(\rho)\right ]-3\mu \Delta_k(\rho,1)^2\,.
\end{aligned}
\end{equation}
For numerical applications, we choose for IR regulator the ``optimized'' one introduced in Ref.~[\onlinecite{litim}]: $R_k(q)=(k^2-q^2)\Theta(k^2-q^2)$. With this choice one has
\begin{equation}
\begin{aligned}
\label{eq_running_Ip_litim}
I_{p,k}[y]=v_d\bigg[\frac 2d \frac{k^d}{(k^2+y)^p}+\int_{k^2}^{\Lambda^2} dx \, \frac {x^{\frac d2 -1}}{(x+y)^p} \bigg]\,,
\end{aligned}
\end{equation}
where $v_d^{-1}=2^{d+1}\pi^{d/2}\Gamma(d/2)$.

\begin{figure}[h]
\hspace{-0.2cm}
\includegraphics[width=8.5cm,height=5.5cm]{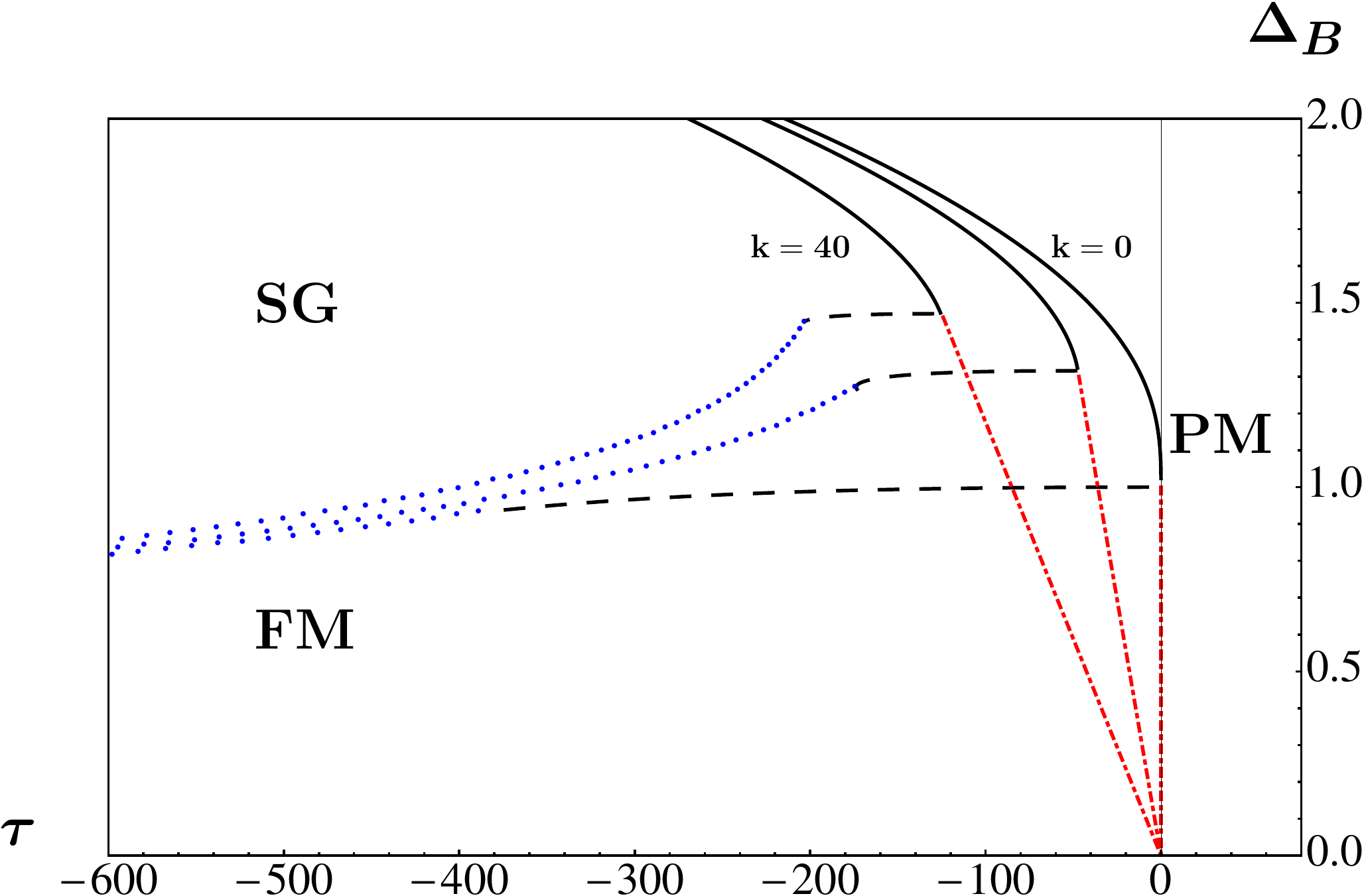}
\caption{(Color on line) Variation with $k$ of the zero-temperature phase diagram of the IR-regularized large-$N$ RAO($N$)M in $d=5$. It is illustrated from top-right to bottom left for $k=0,  30, 40$ (recall that $\Lambda=100$). Note that the shift of the transition lines to the lower right as $k$ decreases. The (black) full lines mark the PM-SG transition, the (black) dashed lines are where the $k$-dependent replicon vanishes in the FM phase, and the (red) dashed-dotted lines indicate the PM-FM transition. The (blue) dotted lines show the limit of existence of meaningful solutions (the $k$-dependent replicon is positive along and to the right of the line). The transition between the SG and the unstable FM phases shown in the inset of Fig. \ref{fig_naive_diagram_a} is not displayed here. The $k$-dependent phase diagram for $2<d<4$ is topologically similar to that shown in the figure, except that the FM phase vanishes when $k \to 0$. }
\label{fig_running_diagram}
\end{figure}

The phase diagram of the IR-regularized $k$-dependent model at zero temperature is similar to that obtained above for $k=0$. For $d>4$, it has the same topology, except that it is globally shifted to more negative values of $\tau$ and larger values of $\Delta_B$: the meeting point of the Almeida-Thouless instability lines is displaced from $(\tau=0,\Delta_B=1)$ to  $(\tau=-(I_{1,0}[0]-I_{1,k}[0])/\lambda \leq 0, \Delta_B=I_{2,0}[0]/I_{2,k}[0]\geq 1)$, and the PM-FM critical line is also moved from $\tau=0$ to $\tau=-(I_{1,0}[0]-I_{1,k}[0])/\lambda$. This is illustrated in Fig.~\ref{fig_running_diagram}. For $2<d<4$, an FM phase appears at nonzero $k$ because $I_{2,k}[0]$ is finite. The $k$-dependent phase diagram is similar to that for $d>4$, except that the tricritical point where the Almeida-Thouless lines also meet is at $(\tau=-(I_{1,0}[0]-I_{1,k}[0])/\lambda \leq 0, \Delta_B=J_d/I_{2,k}[0]\geq 0)$ and the FM phase vanishes when $k\to 0$.


The above property of the $k$-dependent phase diagram implies that a point which is located in the PM phase at $k=0$ is in this phase for all larger values of $k$ and, similarly, a point which at $k=0$ is in the FM phase where the replicon is positive is in this same phase (or in the PM phase) for all $k>0$. It is therefore interesting to consider the situation for the region of the $k=0$ phase diagram corresponding to the SG and FM phases inside the Almeida-Thouless instability lines for $d>4$ or to the SG phase for $2<d<4$, all of which are in fact ill-defined in the naive calculation where the solutions come with a negative value of the replicon. In this region, we also consider nonzero sources so that $\rho$ can take any value, and we study the locus of the points $k_*(\rho)$ where the replicon vanishes,
\begin{equation}
\label{eq_replicon_kstar}
\Lambda_{rep,k_*(\rho)}(\rho,1)=0\,.
\end{equation}
A large enough UV cutoff $\Lambda$ guarantees that the replicon is positive at this scale and one therefore finds a nontrivial line in the $(k,\rho)$ plane in the relevant region of the $k=0$ phase diagram. When $k=k_*(\rho)$ one derives from the $k$-dependent SD equations that
\begin{equation}
\begin{aligned}
\label{eq_solution_k_star}
&\Delta_{k*}(\rho,1)\equiv  \Delta_{k_*(\rho)}(\rho,1)=\left (\frac{\rho}{2\mu}\right )^{1/3}\,,\\&
I_{2,k*} [\rho]\equiv I_{2,k_*(\rho)} [y_{k_*(\rho)}(\rho)]=\lambda-\frac 32(2\mu)^{1/3}\rho^{2/3}\,,
\end{aligned}
\end{equation}
while $y_{k*}(\rho)\equiv y_{k_*(\rho)}(\rho)$ is given by the solution of Eq. (\ref{eq_y_k}).

\begin{figure}[htbp] 
\centering
\subfigure[\hspace{-7cm} ]{
\includegraphics[width=.9\linewidth,origin=tl]{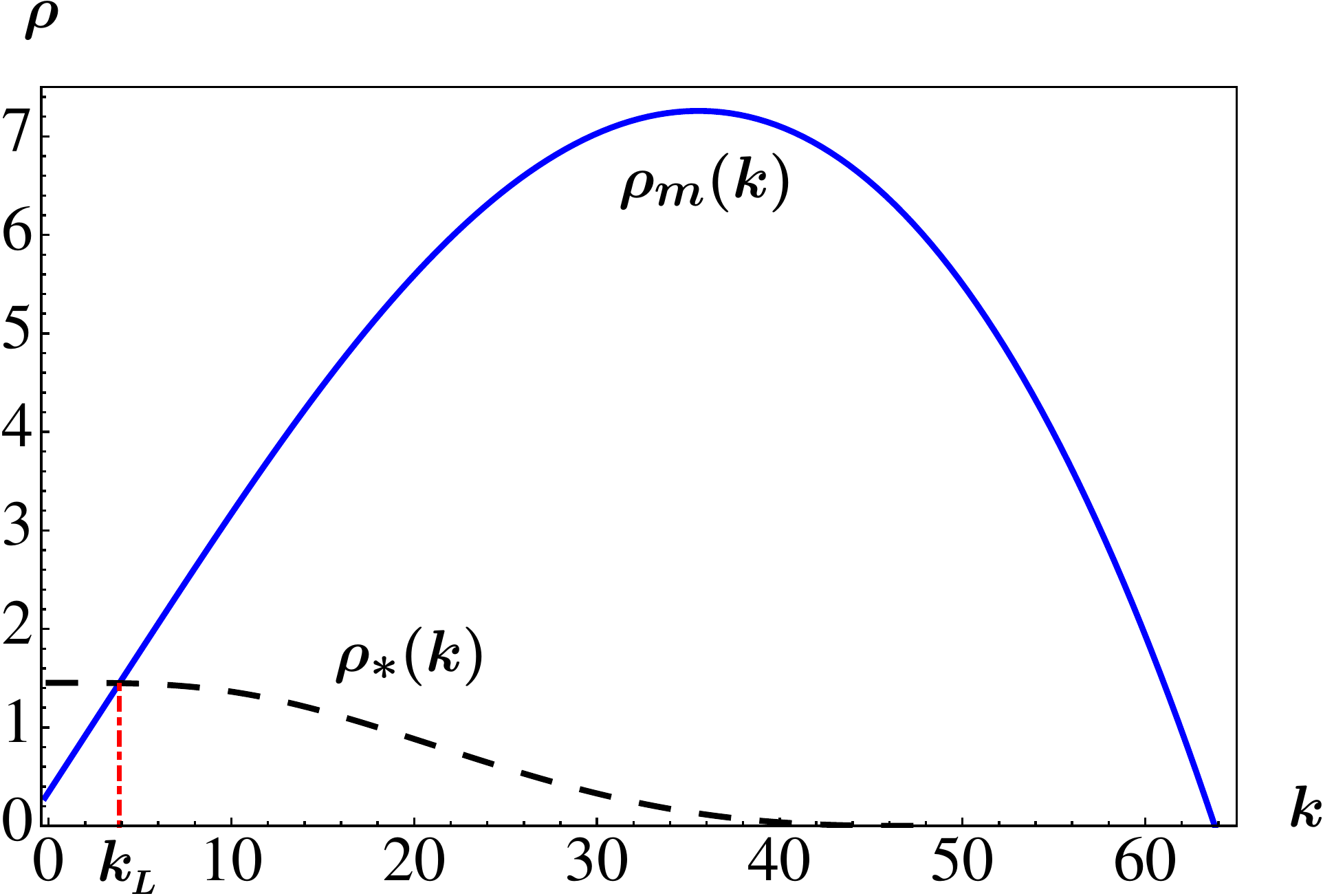}}\newline
\subfigure[\hspace{-7cm}]{
\hspace{-0.5cm}\includegraphics[width=.95\linewidth,origin=tl]{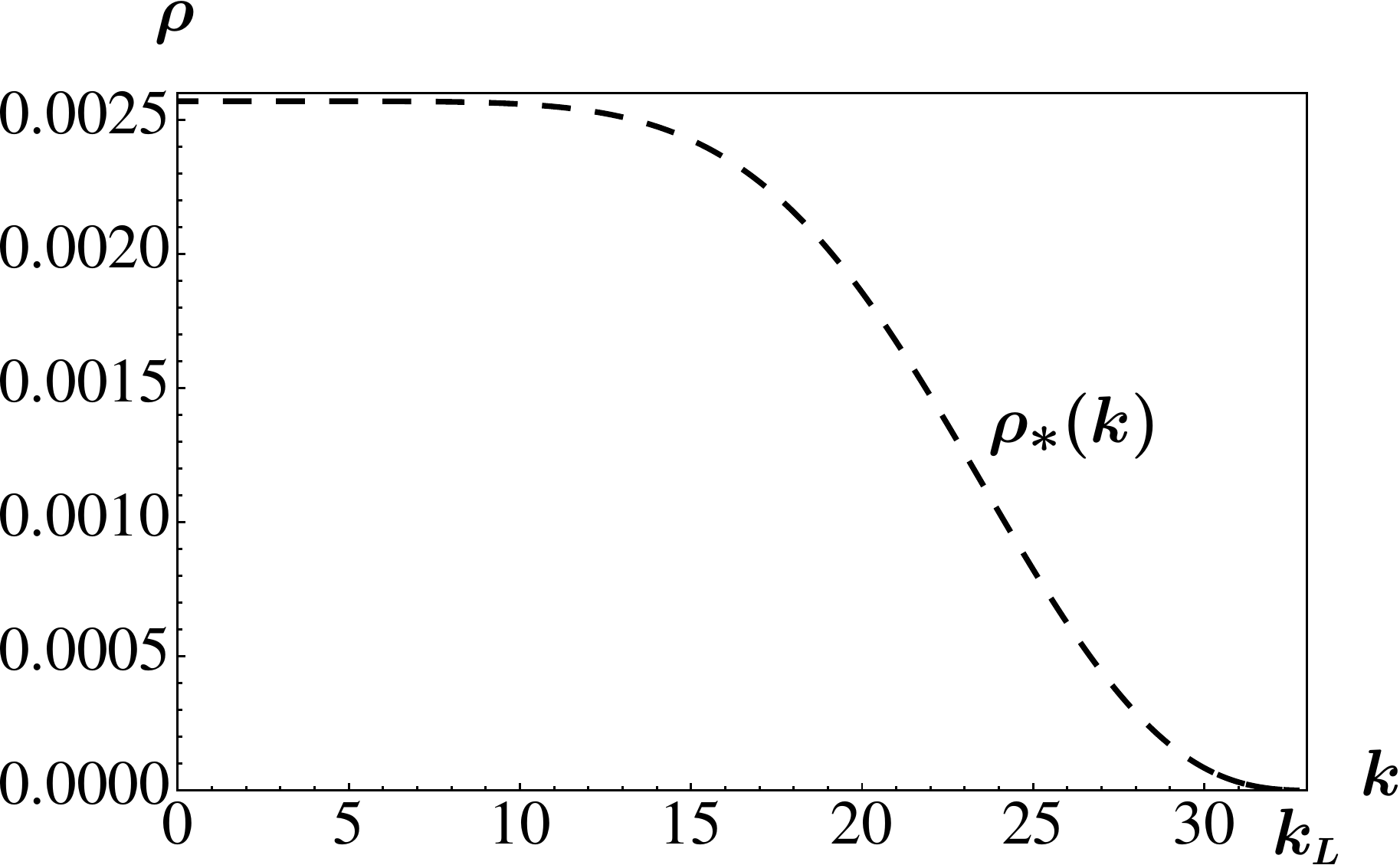}}
\caption{(Color on line) Locus of the points $\rho_*(k)$ where the replicon first vanishes in the IR-regulated model when $d>4$. Illustration for $d=5$ and for the model with bare disorder variance $R'^{-1}(Y)=\lambda Y -\mu Y^3$ with bare parameters $w=90$, $\mu=16$, and $\Lambda=100$. (a) Unstable FM region of the $k=0$ phase diagram (see Fig.~\ref{fig_naive_diagram_a}): here, $\Delta_B=0.971$ and $\tau=-336.25$. We also show the value of the nontrivial minimum of the potential $\rho_m(k)$. The two curves meet at the ``Larkin'' scale $k_L\simeq 3.874$ and below this scale the ``naive'' solution is unstable. Note that the renormalized mass $y_k(\rho)$ is positive for $\rho>\rho_m(k)$ and negative for $\rho<\rho_m(k)$. (b) Unstable SG region  of the $k=0$ phase diagram (see Fig.~\ref{fig_naive_diagram_a}): here, $\Delta_B=2$ and $\tau=-236.667$. The Larkin scale $k_L\simeq 32$  now corresponds to $\rho_*=0$  (we only plot $k\leq k_L$).}
\label{fig_rhostar(k)_d=5}
\end{figure}

\begin{figure}[h]
\includegraphics[width=.9\linewidth]{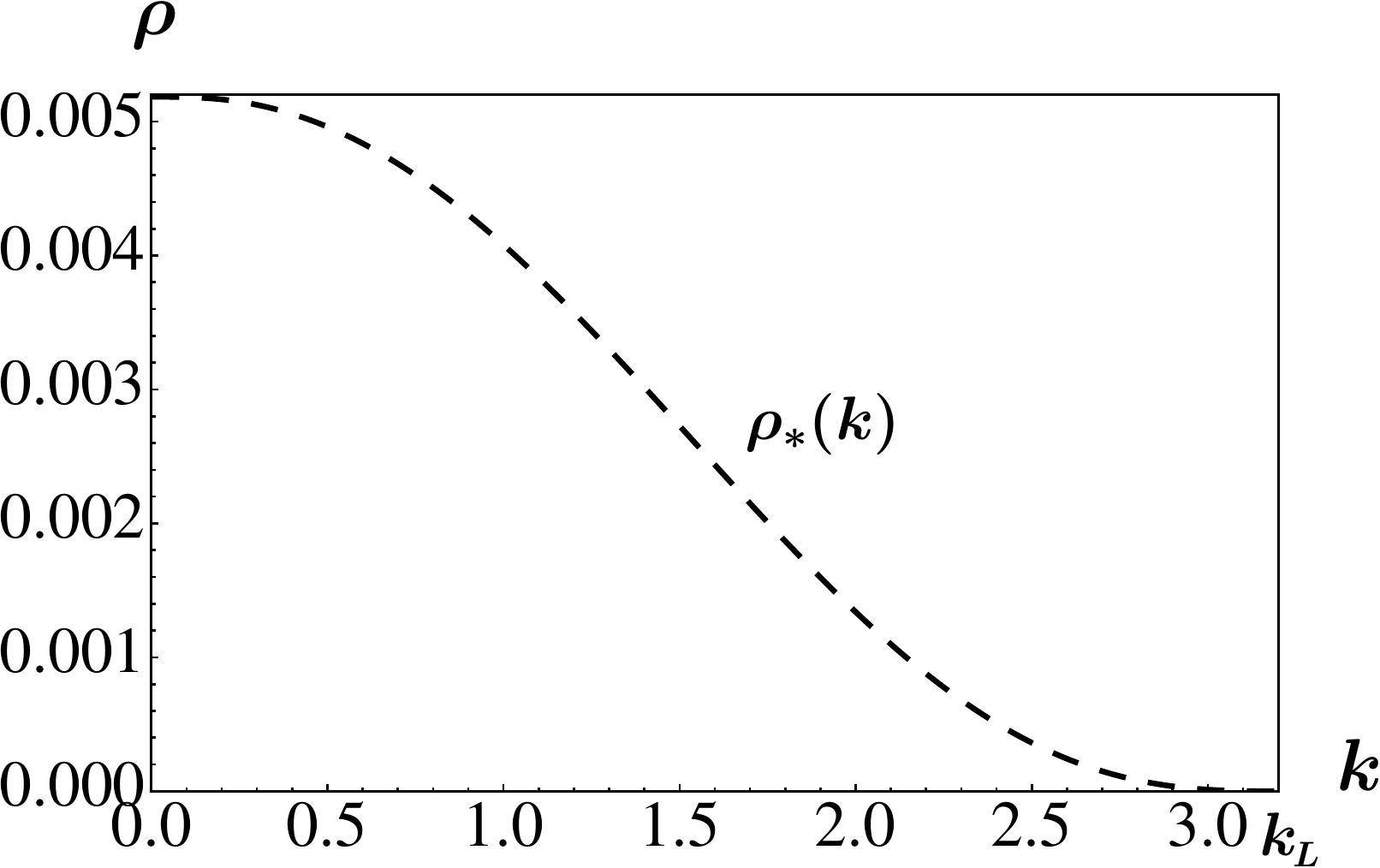}
\caption{(Color on line) Locus of the points $\rho_*(k)$ at which the replicon first vanishes in the IR-regulated model when $2<d<4$. Illustration for $d=3$ and for the unstable SG region  of the $k=0$ phase diagram (see Fig.~\ref{fig_naive_diagram_b}):  here, $\Delta_B=0.078$ and $\tau=-8$. The Larkin scale $k_L\simeq 3$ corresponds to $\rho_*=0$ (we only plot $k\leq k_L$). For the present choice of parameters the system is in the unstable SG phase at $k=0$ but moves through the FM phase for $k \gtrsim 0.15$ and the replicon then vanishes for a negative value of the renormalized mass.}
\label{fig_rhostar(k)_d=3}
\end{figure}

The results are illustrated in Fig.~\ref{fig_rhostar(k)_d=5} for $d=5$ (which is representative of the whole interval $4<d<6$) and in Fig.~\ref{fig_rhostar(k)_d=3} for $d=3$ (representative of $2<d<4$).  For convenience we plot $\rho_*(k)$, which is equivalent to $k_*(\rho)$ as the function is single-valued (one can show that $\rho_*$, or equivalently $k_*$, is a decreasing function of its argument).

Let us discuss first the case where $d>4$. We consider two situations, one corresponding to the (naive) FM phase and one to the (naive) SG phase. In the FM case, one finds that for a large enough $\Lambda$ the system at this scale is in the PM phase. When decreasing $k$ one then first encounters a transition to a FM phase characterized by a nonzero minimum $\rho_m(k)\equiv \rho_{mk}$ for which $y_k(\rho_m(k))=0$. For $\rho>\rho_m(k)$, $y_k(\rho)>0$ but for $\rho<\rho_m(k)$, $y_k(\rho)<0$. We find that a portion of the line $\rho_*(k)$ is located in the latter region. Indeed, since the fluctuations in the IR-regularized system considered at a nonzero scale $k$ are suppressed, the associated ``effective average potential'' $U_k(\rho)$  is not convex.\cite{berges02} This allows the presence of a region where the mass $y_k(\rho)$ is negative (but $y_k(\rho)+R_k(0)$ is always positive). The convexity of $U_k(\rho)$ is only restored in the limit $k\to 0$, as shown in the context of the FRG.\cite{berges02} The curves  $\rho_*(k)$ and $\rho_m(k)$ for a generic choice of bare parameters are  displayed in Fig.~\ref{fig_rhostar(k)_d=5} (a) for $d=5$. The point at which the two curves cross defines a ``Larkin'' scale $k_L$. In the case of the SG phase, $\rho_*(k)$ decreases from a nonzero value at $k=0$ to zero at a ``Larkin'' scale $k_L$: see Fig.~\ref{fig_rhostar(k)_d=5} (b). 
Finally for $2<d<4$ only the SG phase needs to be considered at $k=0$ (but the system may go through an FM phase when $k>0$) and the results are similar to those obtained for the SG phase when $d>4$: see Fig.~\ref{fig_rhostar(k)_d=3} for $d=3$.

In the region of the $(k,\rho)$ plane for $\rho$ below the curve $\rho_*(k)$, or alternatively for $k$ less than $k_*(\rho)$, the cutoff-dependent SD equations in their present form no longer have a solution. We will now show that the problem can be cured by switching to the 1-PI FRG equations.


\section{Full solution of the $1$-PI FRG equations}
\label{section_solution_1PI}

As already stressed, when the $k$-dependent replicon $\Lambda_{rep,k}(\rho,1)$ is positive, there is a complete equivalence between the cutoff-dependent SD equations and the $1$-PI FRG equations. What happens, then, when for a given $\rho$ the running IR scale $k$ reaches $k_*(\rho)$? 
We show below from the analysis of the 1-PI FRG equations that the flows of the two dimensionful quantities $\Delta_k(\rho,z)$ and $I_{2,k}(\rho)\equiv I_{2,k}[y_k(\rho)]$ {\it exactly freeze} at the scale $k_*(\rho)$. To demonstrate this property, it is useful to start not from the final form of the equations given above in Eqs.~(\ref{eq_RG_y_T0}, \ref{eq_RG_Delta_T0}) but from intermediate 1-PI expressions, such as those in Eqs.~(\ref{eqderivdelta_z}-\ref{eq_I2partial}).

Since $\partial_z \Delta_k(\rho,z) \to \infty$ when $z\to 1$ and $k\to k_*(\rho)$ while both $\Delta_k(\rho,1)$ and $\partial_{\rho}  \Delta_k(\rho,1)$ are expected to stay finite, Eq.~(\ref{eq_I2partial}) considered when $z\to 1$ gives that $\partial_{\rho} I_{2,k}[y_k(\rho)] \to -1/\Delta_k(\rho,1)$ when $k=k_*(\rho)$. Inserting this result in Eq.~(\ref{eqderivdelta_z}), together with the fact that $\partial_{k}  \Delta_k(\rho,z)$ is finite (not diverging), immediately leads to the result that $\partial_k I_{2,k}[y_k(\rho)] \to 0$ when $k \to k_*(\rho)$. One can see that Eq.~(\ref{eqderivdelta_1}) has then {\it a priori} an ambiguous $(\frac{0}{0})$ expression in $k_*(\rho)$, which will be clarified below. We can now consider Eq.~(\ref{eqderivdelta_z}) for $z\neq 1$, so that $\partial_z \Delta_k(\rho,z)$ is possibly large but stays finite when $k \to k_*(\rho)$. As we just found that $\partial_k I_{2,k}[y_k(\rho)] \to 0$, it follows that $\partial_k \Delta_k(\rho,z)\to 0$ when $k \to k_*(\rho)$ for $z\neq 1$. This proves that the flows of $\partial_k I_{2,k}[y_k(\rho)]$ and $\partial_k  \Delta_k(\rho,z)$ for $z \neq 1$ exactly freeze when $k=k_*(\rho)$.

The consistency of the result also implies an additional property. To see this it is convenient to derive a flow equation directly for $I_{2,k}[y_k(\rho)]$ from that for $y_k(\rho)$. By using that $\partial_k I_{2,k}[y_k(\rho)]=\widehat \partial_k I_{2,k}[y_k(\rho)] - 2 I_{3,k}[y_k(\rho)]\partial_k y_k(\rho)$ and $\partial_{\rho} I_{2,k}[y_k(\rho)]=-2 I_{3,k}[y_k(\rho)]\partial_\rho y_k(\rho)$, one obtains from Eq.~(\ref{eq_RG_y_T0})
\begin{equation}
\begin{aligned}
\label{eq_RG_I2_T0}
\partial_k I_{2,k}(\rho)& = 2 I_{3,k}(\rho)\, \widehat \partial_k I_{1,k}(\rho)\,  \partial_{\rho} \Delta_k(\rho, 1 )  \\
& + \widehat \partial_k I_{2,k}(\rho) \Big[1+\Delta_k(\rho, 1)\partial_{\rho} I_{2,k}(\rho) \Big]\,,
\end{aligned}
\end{equation}
where $I_{p,k}(\rho)\equiv I_{p,k}[y_k(\rho)]$. When $k\to k_*(\rho)$, the left-hand side and the second term of the right-hand side are zero (see above), which then implies that  $\partial_{\rho} \Delta_k(\rho, 1 )=0$ at $k=k_*(\rho)^+$. We have also checked these results by solving numerically the flow equations  for generic initial conditions of the flows down to $k_*(\rho)$ with the bare disorder function and the IR regulator given above.

We stress that it is the flow of $I_{2,k}[y_k(\rho)]$ that stops at $k_*(\rho)$, not that of $y_k(\rho)$ itself. Indeed, $\widehat \partial_k I_{2,k}[y_k(\rho)]$ is not zero when $k=k_*(\rho)$. Note also that it is the evolution of the dimensionful quantities that freezes, not that of their dimensionless counterparts which continue to flow due to the (trivial) scaling or dimensional part of the beta functions.

As one can see from Eq.~(\ref{eqderivdelta_z}) there is an ambiguity when both $z \to 1$ and $k \to k_*(\rho)^+$ since $\partial_z \Delta_k(\rho,z)$ diverges while  $\partial_k I_{2,k}[y_k(\rho)]$ goes to zero. This corresponds to a nonuniform convergence of the function $\Delta_k(\rho,z)$ and the presence of a boundary layer in $(1-z)/[k-k_*(\rho)]^\alpha$ with $\alpha >0$ some exponent to be determined. On the other hand, the solution for  $I_{2,k}[y_k(\rho)]$ and $\Delta_k(\rho,1)$ exactly at $k=k_*(\rho)$ is given by Eq.~(\ref{eq_solution_k_star}) as the SD equations are equivalent here with the 1-PI FRG equations.

One easily obtains that the boundary-layer solution has an exponent $\alpha=2$, {\it i.e.}, 
\begin{equation}
\label{eq_solution_boundary_layer_inner}
\Delta_k(\rho,z)=\Delta_{k*}(\rho,1)\bigg[1+\delta k \,G\bigg(\frac{1-z}{\delta k^2}\bigg) + {\rm O}(\delta  k^2,1-z)\bigg]\,, 
\end{equation}
where we have introduced $\delta k=[k-k_*(\rho)]/[\rho k'_*(\rho)]$ with $k'_*(\rho)$ the first derivative of $k_*$ with respect to $\rho$. Note that we are interested in $k \to k_*(\rho)^+$, which corresponds to $\delta k \to 0^-$ as $k'_*(\rho)$ is negative (see section \ref{section_solution_runningSD}). To give explicit expressions, we focus again on the bare disorder variance such that $R'^{-1}(Y)=\lambda Y-\mu Y^3$, in which case,  $\Delta_{k*}(\rho,1)=[\rho/(2\mu)]^{1/3}$. The set formed by Eqs.~(\ref{eqderivdelta_1}-\ref{eq_I2partial}) can be solved in the limit $z\to 1$, $\delta k \to 0^-$ with $x=(1-z)/\delta k^2 ={\rm O}(1)$, which leads to 
\begin{equation}
\label{eq_solution_boundary_layer}
G(x)=- \sqrt{C^2+\frac 23\, x}
\end{equation}
and
\begin{equation}
\label{eq_solution_I_2k_BL}
I_{2,k} (\rho)=\lambda-\frac 32(2\mu)^{1/3}\rho^{2/3}\Big[1+C^2 \delta k^2 +{\rm O}(\delta k^3)\Big]\,,
\end{equation}
where  $C $ is a nonzero constant that may {\it a priori} depend on $\rho$ and can be determined by looking at the $\rho$-derivative of $\Delta_k(\rho,1)$. From the expression $\Delta_k(\rho,1)=[\rho/(2\mu)]^{1/3}[1 + \vert C \vert  \delta k + {\rm O}(\delta k^2)]$, which is obtained by taking $x=0$ in Eq.~(\ref{eq_solution_boundary_layer}) with $\delta k$ negative, and the requirement $\partial_\rho \Delta_k(\rho,1)=0$ for $\delta k=0^-$, one finds that $\vert C\vert=1/3$. (Note the absence of linear term in $\delta k$ in the expression of $I_{2,k} (\rho)$, contrary to that of  $\Delta_k(\rho,1)$.)

One can also obtain the ``outer'' solution of Eq.~(\ref{eqderivdelta_z}) for $\delta k \to 0$ and $(1-z)={\rm O}(1)$ in the form
\begin{equation}
\label{eq_solution_boundary_layer_outer}
\Delta_k(\rho,z)=\left (\frac{\rho}{2\mu}\right )^{1/3}\Big[1+g(z) + h(z) \delta k + {\rm O}(\delta k^2)\Big]
\end{equation}
with $g(1)=h(1)=0$. From Eqs.~(\ref{eqderivdelta_1})-(\ref{eq_I2partial}), one finds that $h(z)=0$ and $g(z)$ is solution of $g(z)^3+3g(z)^2=2(1-z)$ (as also directly obtained from the $k$-dependent SD equations). As 
\begin{equation}
g(z) = - \sqrt{\frac 23(1-z)} +{\rm O}(1-z) 
\end{equation}
when $z \to 1$, the ``outer'' solution matches the ``inner'' one when $x \to \infty$, as required.

Since the evolution of $I_{2,k} (\rho)$ and of $\Delta_k(\rho,z)$ for $z\neq 1$ stop at $k=k_*(\rho)$, the 1-PI FRG equations for $k<k_*(\rho)$ simply read
\begin{equation}
\begin{aligned}
\label{eq_RG_I2_cusp}
\partial_k I_{2,k}[y_k(\rho)]\vert_{\rho}=0 \, ,
\end{aligned}
\end{equation}
which implies 
\begin{equation}
\begin{aligned}
\label{eq_RG_y_cusp}
\partial_k y_k(\rho) = \frac{\widehat \partial_k I_{2,k}[y_k(\rho)]}{2I_{3,k}[y_k(\rho)])}
= \widehat \partial_t I_{2,k}[y_k(\rho)] \Delta_k (\rho, 1 )\partial_{\rho} y_k(\rho) \,,
\end{aligned}
\end{equation}and
\begin{equation}
\begin{aligned}
&\partial_k   \Delta_k(\rho,z) =0\,.
\label{eq_RG_Delta_cusp}
\end{aligned}
\end{equation}
The solutions for $I_{2,k}$ and $\Delta_k$ are given by continuity in $k_*(\rho)$. For the bare disorder variance introduced above, one finds
\begin{equation}
\label{eq_solution_I_2_cusp}
I_{2,k} (\rho)=\lambda-\frac 32(2\mu)^{1/3}\rho^{2/3}\,,
\end{equation}
\begin{equation}
\label{eq_solution_Delta_cusp}
\Delta_k(\rho,z)=\Delta_{k*}(\rho,z)=\left (\frac{\rho}{2\mu}\right )^{1/3}[1+g(z)]\,.
\end{equation}
The function $\Delta_k(\rho,z)$ is nonanalytic near $z=1$ with a ``cusp'', {\it i.e.}, a term proportional to $\sqrt{1-z}$ [see above the behavior of $g(z)$]. Such a cusp in the functional dependence of the renormalized 1-PI disorder correlations as the two replica arguments become equal ($z\to 1$) is generated by jump discontinuities, called avalanches or shocks, in the ground state of the system as an applied source $J$ is varied.\cite{MDW_largeN,tissier06,ledoussal-wiese_review,BBM96}

In the above solution, $\Delta_k(\rho,1)$ is continuous in $k_*(\rho)$. However, its flow is {\it discontinuous} in $k_*(\rho)$, with $\partial_k \Delta_k(\rho,1)=-[1/(3 \rho k'_*(\rho)] [\rho/(2\mu)]^{1/3} \neq 0$ for $k\to k_*(\rho)^+$ and $\partial_k \Delta_k(\rho,1)=0$ for $k\to k_*(\rho)^-$. The derivative with respect to $\rho$ of $\Delta_k(\rho,1)$ is {\it discontinuous} as well in $k_*(\rho)$, with $\partial_\rho \Delta_k(\rho,1)= 0$ for $k\to k_*(\rho)^+$ and $\partial_\rho \Delta_k(\rho,1)=(1/3)(2\mu\rho^2)^{-1/3}$ for $k\to k_*(\rho)^-$. These features are general and do not depend on the choice of the bare variance of the disorder.

\begin{figure}[h]
\includegraphics[width=\linewidth]{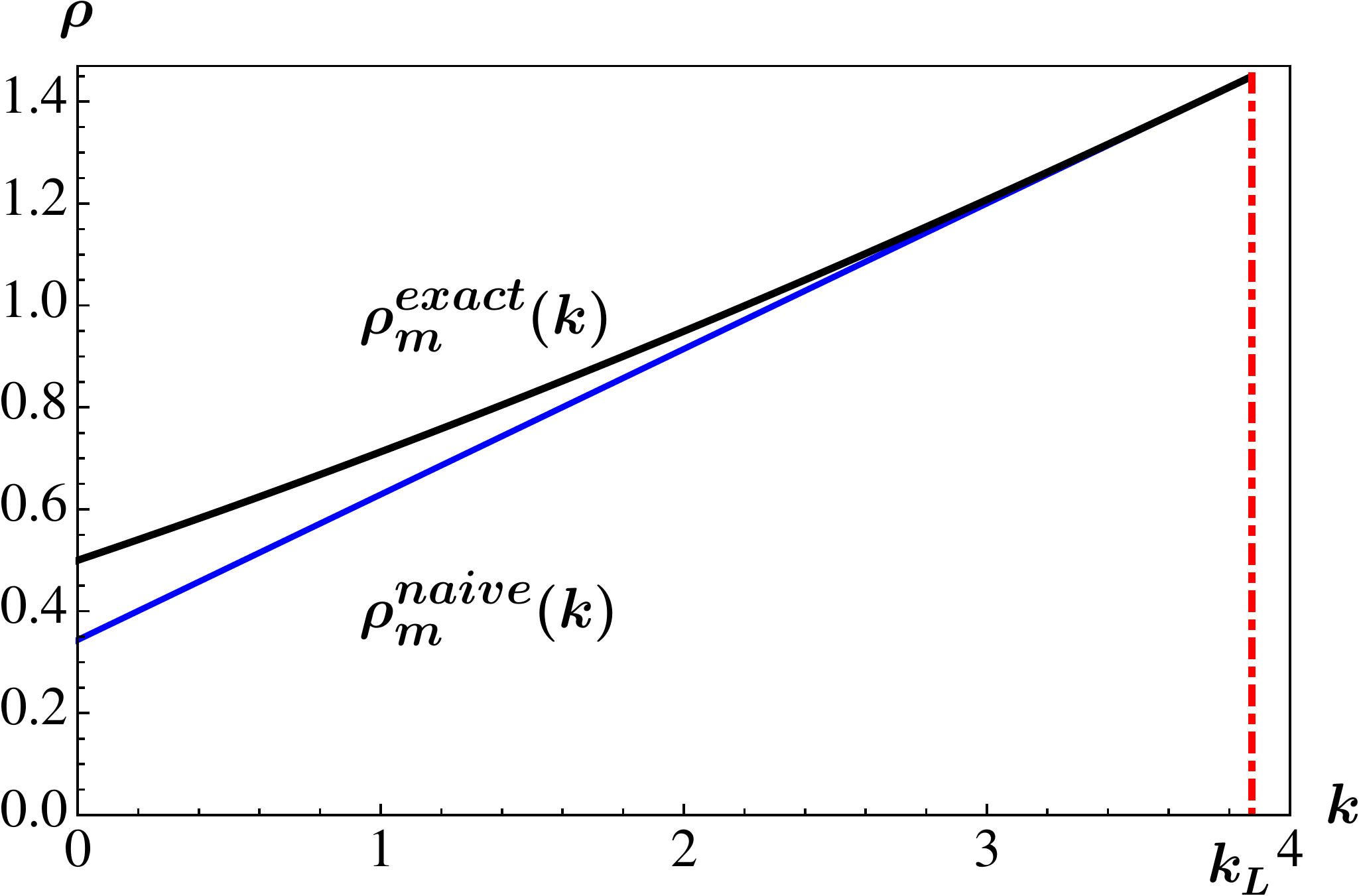}
\caption{(Color on line) Nontrivial minimum of the potential  $\rho_m(k)$ below the Larkin length $k_L$ in $d=5$: Comparison of the exact result $\rho_m^{exact}(k)$ describing a ``cuspy'' FM phase at the scale $k$ with the unstable naive solution $\rho_m^{naive}(k)$. The former is always larger than the latter and the difference decreases with increasing $k$ to vanish at $k_L$. The parameters are the same as in Fig. \ref{fig_rhostar(k)_d=5} and $k_L\simeq 3.874$.}
\label{fig_rhomin(k)_d=5}
\end{figure}

Finally, we note that the ``cuspy'' solution in Eqs.~(\ref{eq_solution_I_2_cusp},\ref{eq_solution_Delta_cusp}) is independent of the cutoff $k$ but that the boundary of the corresponding region, as characterized by $\rho_*(k)$, itself depends on the cutoff. Similarly, when a ``cuspy'' FM phase is present, which implies $d>4$, the spontaneous magnetization, or equivalently $\rho_m(k)$, is a running quantity, with
\begin{equation}
\label{eq_rho_mk}
\rho_{m}(k)=\frac 2{3\sqrt{3\mu}}\big (\lambda - I_{2,k}[0]\big )^{3/2}\,,
\end{equation}
where, with the choice of ``optimized'' IR regulator defined above [see Eq.~(\ref{eq_running_Ip_litim})],
\begin{equation}
\begin{aligned}
\label{eq_running_Ip_litim0}
I_{2,k}[0]=\frac{2 v_d}{(d-4)}\Lambda^{d-4}\Big [1- \left (\frac{4}{d}\right )\left ( \frac{k}{\Lambda} \right )^{d-4}\Big ]\,.
\end{aligned}
\end{equation}
The above expression for $\rho_{m}(k)$ is valid for $k\leq k_L$ [see section \ref{section_solution_runningSD}]. We plot this expression together with the value computed from the naive (analytic) solution in Fig.~\ref{fig_rhomin(k)_d=5} for $d=5$.


\section{ Reconciling the SD and FRG equations}

When deriving the above solution by following the flow of the 1-PI quantities with decreasing cutoff $k$, one finds that the renormalized mass $y_k(\rho)$ does not appear directly and one realizes that the corresponding $k$-dependent SD equations is no longer valid when $k<k_*(\rho)$. On the other hand, there is a condition of  ``marginality'' for the equation on $\Delta_k(\rho,z)$, which is associated with the existence of a nonanalyticity in $\sqrt{1-z}$ and is expressed by the fact that the (properly defined) replicon is equal to zero. This suggests to solve the set of two coupled equations,
\begin{equation}
\label{eq_running_SDDelta_cusp}
\Delta_k \left (\rho, z \right )=R'\Big(\rho z+\Delta_k \left (\rho, z \right )I_{2,k}[y_k(\rho)] \Big)\,,
\end{equation}
for all values of $z$ including $z=1$,
\begin{equation}
\begin{aligned}
\label{eq_running_SDreplicon_cusp}
\Lambda_{rep,k}(\rho,1)&= \frac 1{R''\Big(\rho +\Delta_k (\rho, 1)I_{2,k}[y_k(\rho)] \Big)}-I_{2,k}[y_k(\rho)] \\& =0 \,,
\end{aligned}
\end{equation}
and drop the SD equation for $y_k(\rho)$ whenever $k<k_*(\rho)$.

Eqs.~(\ref{eq_running_SDDelta_cusp},\ref{eq_running_SDreplicon_cusp}) admit a solution for $I_{2,k}(\rho)$ and $\Delta_k(\rho, z )$ that is independent of $k$ and coincides with that obtained from the flow equations. It is for instance given by Eqs.~(\ref{eq_solution_I_2_cusp}, \ref{eq_solution_Delta_cusp}) in the special case where $R'^{-1}(Y)=\lambda Y- \mu Y^3$. We also find that $\Delta_k(\rho, z )\equiv \Delta(\rho, z )$ behaves when $z\to 1$ as 
\begin{equation}
\begin{aligned}
\label{eq_Deltaz_cusp}
&\Delta(\rho, z )- \Delta(\rho, 1)=\\&-\sqrt{\left(\frac{2\rho}{I_2(\rho)^3 R'''\Big(\rho +\Delta(\rho, 1)I_{2}(\rho)\Big)}\right)(1-z)} + {\rm O}(1-z) \,,
\end{aligned}
\end{equation}
provided $R'''\Big(\rho +\Delta(\rho, 1)I_{2}(\rho)\Big)>0$ (it is equal to $3(2\mu)^{2/3} \rho^{1/3} I_{2}(\rho)^{-3}$ for the special illustrative case studied above).

We have already emphasized that the emergence of a nonanalytic behavior in the 1-PI functions characterizing the renormalized disorder, here $\Delta_k(\rho, z )$, coincides with the vanishing of the replicon eigenvalue that signals the instability of the replica-symmetric solution in the conventional replica method. In the $N \to \infty$ limit of the RAO($N$)M where we have at our disposal closed self-consistent equations provided by the SD equations one may look for solutions breaking {\it spontaneously} the replica symmetry in a setting where the sources are all equal so that replica symmetry is not {\it explicitly} broken. A thorough investigation of this kind has been carried out for the case of the random-manifold model in the large-$N$ limit.\cite{doussal_largeN,MDW_largeN} (Recall that we have taken here  the $N\to \infty$ limit first, with a scaling of the replica fields and of the differences between replica fields in $\sqrt N$, and next considered the limit $T\to 0$.) 

In the region where the (replica-symmetric) replicon becomes unstable, we find a solution displaying a thermal boundary layer as $T\to 0$. At infinitesimal but nonzero $T$, the solution displays a full (or continuous) replica-symmetry breaking of the Parisi-type,\cite{SGbook} but when $T=0$ the solution has an apparent replica-symmetric structure, albeit with a vanishing but nonnegative replicon. This marginal solution exactly coincides with the solution found above for $\Delta_k(\rho,z=1)$ and $I_{2,k}(\rho)$ (and is therefore independent of $k$). In the ultrametric picture, $\Delta_k(\rho,z=1)$ represents the 1-PI correlations between the most distant states;\cite{doussal_largeN,MDW_largeN} however, the correlations among any other types of states differ from the latter only by terms of the order O($\sqrt T$) and therefore vanish at $T=0$. The presence of the thermal boundary layer nonetheless generates a modification of the SD equation for $y_k(\rho)$, through an additional term to Eq.~(\ref{eq_y_k}) [or Eq.~(\ref{eq_running_y})], that does not vanish in the $T\to 0$ limit. The detailed calculations are provided in appendix \ref{appendixD}.

In the present case, as in the large-$N$ random manifold but not in the same form, one therefore finds an  equivalence between the $T=0$ nonanalytic solution of the 1-PI FRG equations and the full replica-symmetry breaking solution. Exactly at $T=0$, the former one, which describes the full functional $z$-dependence of the renormalized disorder correlations and its $\sqrt{1-z}$ behavior, is richer than the latter. It also directly leads to a connection to the physical underlying mechanism represented by the presence of avalanches or shocks in the ground state. At nonzero temperature, a full-blown Parisi-like replica-symmetry breaking solution develops as in the large-$N$ random manifold,\cite{mezard-parisi91} but we have already emphasized that the $N\to \infty$ limit is somehow anomalous at finite temperature due to the absence of thermal rounding of the cusp at criticality\cite{doussal_largeN,MDW_largeN} and is therefore not representative of the behavior at finite $N$.

\section{Exact phase diagram}
\label{section_exact phase diagram}

We are now in a position to establish the exact phase diagram of the model. It is obtained by taking the limit $k=0$ in the solutions derived above. The portions of the phase diagram for which the $k=0$ replicon eigenvalue is positive with the simplest ({\it i.e.}, ``analytic'' in the formalism with explicit replica-symmetry breaking or ``replica-symmetric'' in the conventional replica method) solution of the SD equations or, equivalently, of the 1-PI FRG equations are unchanged with respect to this solution: this concerns the whole PM phase and, for $d>4$, an FM phase separated from the PM one by a critical line ($\tau=0$, $\Delta_B<1$) with classical (mean-field) exponents.

More interesting are the regions for which the naively calculated replicon eigenvalue is negative. In the exact solution, we find that the properly calculated replicon is now always equal to zero, {\it i.e.}, the phases are {\it marginal} and some SG-like susceptibility defined as the inverse of the replicon diverges everywhere. As seen in section \ref{section_solution_1PI}, the renormalized disorder function, $\Delta(\rho,z)$, is then nonanalytic in its $z$-dependence, with a cusp in $\sqrt{1-z}$ as the two replica fields involved in the function, $\boldsymbol \phi_1, \boldsymbol \phi_2$, become equal (recall that $z=\cos(\boldsymbol \phi_1 \mathbf{.} \boldsymbol \phi_2/\rho)$ with $\rho=\boldsymbol \phi_1^2/N=\boldsymbol \phi_2^2/N$).

\subsection{Phase diagram with no applied sources}

\begin{figure}[h]
\hspace{-0.5cm}
\includegraphics[width=8.5cm,height=5.5cm]{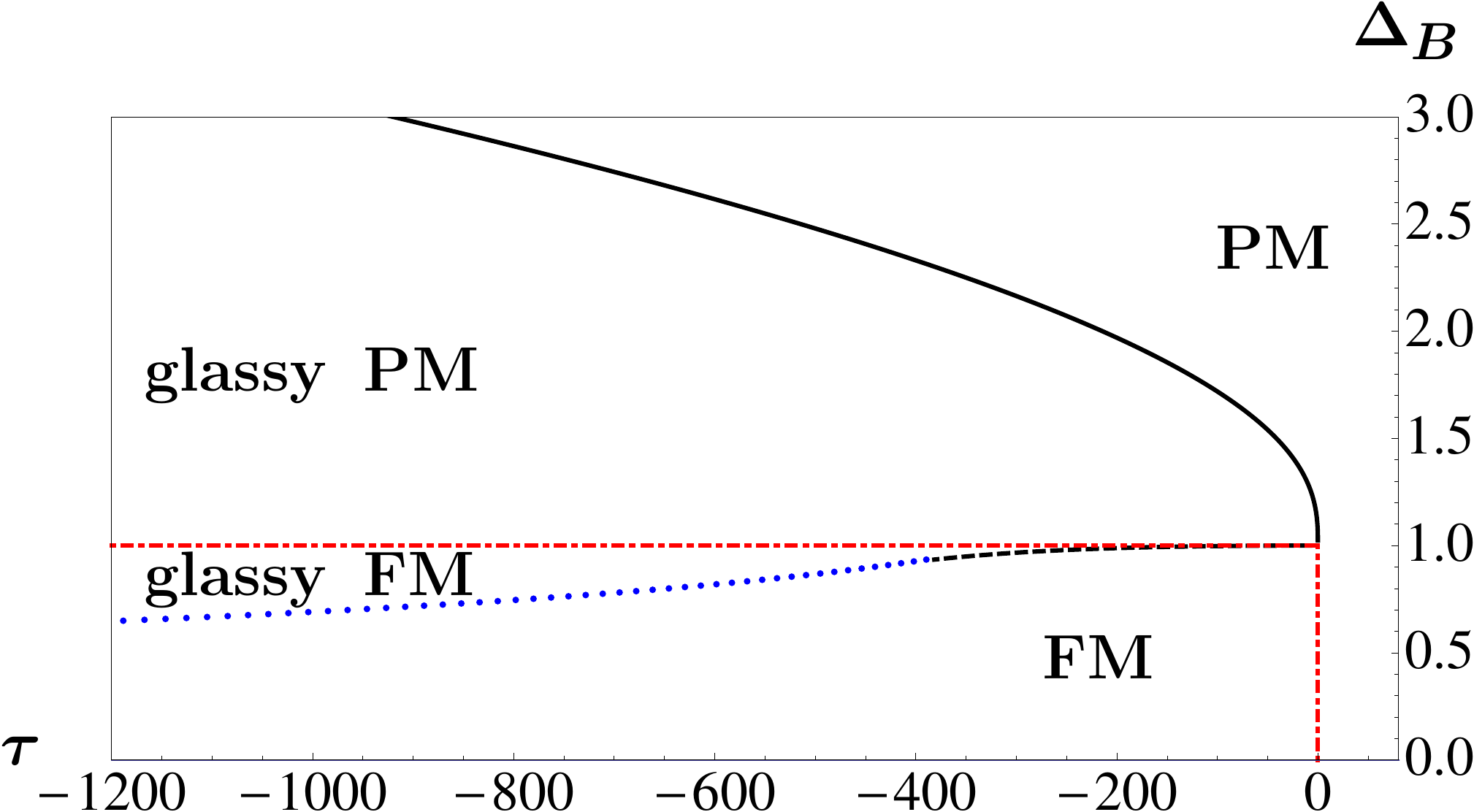}
\caption{(Color on line) Zero-temperature phase diagram of the RAO($N$)M model when $N\to\infty$ obtained from the exact solution in zero applied source  (magnetic field) for $d>4$, illustrated here for $d=5$. Same as Fig.~\ref{fig_naive_diagram_a}, except that one now has marginal ``glassy'' FM and  ``glassy'' PM phases separated by a critical line that does not coincide with the FM-SG line of the naive calculation.}
\label{fig_exactdiagram_a}
\end{figure}

Consider first the phase diagram in the absence of applied sources (no magnetic field), so that the spontaneous magnetization is either given by $\rho=\rho_m$ when ferromagnetism is present, or by $\rho=0$ otherwise. We focus on $d>4$ (the case $2<d<4$ is then simply obtained from the non-FM behavior) and, for the sake of concreteness, on the model with bare disorder variance given by $R'^{-1}(Y)=\lambda Y-\mu Y^3$. We have already noted (see Fig.~\ref{fig_rhomin(k)_d=5}) that the exact value for $\rho_m=\rho_m(k=0)$ is always larger than the naive one, which implies in particular that ferromagnetism in the former case  is more robust that in the latter. Accordingly, the critical line where ferromagnetism disappears in the ``cuspy'' region, which is characterized by $\rho_m=0$ and $y(\rho_m)=0$, is found from Eq.~(\ref{eq_rho_mk}) as $I_{2,0}[0]=\lambda$, {\it i.e.}, $\Delta_B=1$ ($\forall \tau\leq 0$). Along this line, one further has $\Delta(\rho_m,1)=0$, so that the SG Edwards-Anderson order parameter $Q$ is also equal to zero. The phase obtained above this critical line is also characterized by a vanishing spontaneous magnetization and, from Eq.~(\ref{eq_solution_Delta_cusp}),  by $\Delta(\rho=0,1)=0$ and $Q=0$. The SG phase found in the naive calculation therefore does not survive. It is replaced by a ``glassy'' PM phase in which both the magnetization {\it and} the SG Edwards-Anderson order parameter are equal to zero. The renormalized mass $y(\rho=0)$ is fixed by Eq.~(\ref{eq_solution_I_2_cusp}), {\it i.e.}, $I_{2,0}[y(0)]=\lambda=I_{2,0}[0]/\Delta_B$. We call this phase ``glassy'' because the susceptibility obtained as the inverse of the replicon is everywhere infinite (as stressed above, the phase is therefore marginal) and, as we will see, the behavior is somehow anomalous when one applies an infinitesimal magnetic field. The phase diagram is illustrated in Fig.~\ref{fig_exactdiagram_a} for $d=5$.


\begin{figure}[h]
\hspace{-0.5cm}
\includegraphics[width=8cm,height=5.5cm]{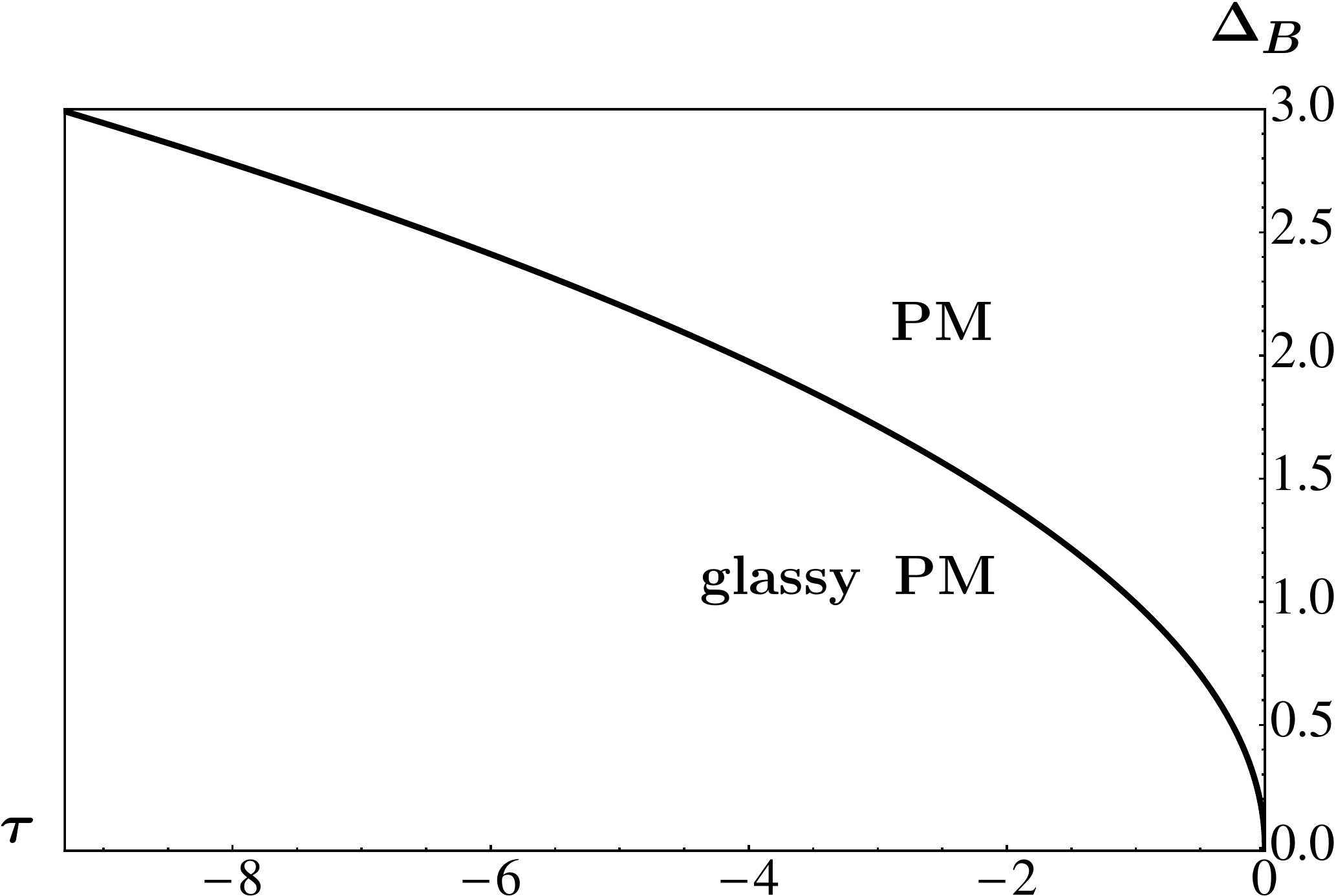}
\caption{(Color on line) Zero-temperature phase diagram of the RAO($N$)M model when $N\to\infty$ obtained from the exact solution in zero applied source  (magnetic field) for $2<d<4$, illustrated here for $d=3$. Same as Fig.~\ref{fig_naive_diagram_b}, except that the unstable SG phase is replaced by a marginal ``glassy'' PM phase.}
\label{fig_exactdiagram_b}
\end{figure}


The critical behavior associated with the transition line between ``cuspy'' or ``glassy'' FM and ``glassy'' PM phases is different from that predicted from the $d\to (d-2)$ dimensional-reduction prediction. From the solution of the 1-PI FRG equations, Eqs. (\ref{eq_solution_Delta_cusp}, \ref{eq_rho_mk}), together with Eq. (\ref{eq_running_Ip_litim0}), one finds for $d<6$ that $\rho_{mk}\sim k^{3(d-4)/2}$ and $\Delta_{mk}\equiv \Delta_{k}(\rho_{mk},1) \sim k^{(d-4)/2}$. From the definition of the anomalous dimensions for a zero-temperature fixed point, {\it i.e.}, $\rho_{mk}\sim k^{d-4+\bar\eta}$ and $\Delta_{mk}\sim k^{-2\eta+\bar\eta}$, one gets that $\eta=0$ and
\begin{equation}
\bar\eta =(d-4)/2. 
\end{equation}
This is in agreement with the FRG results obtained at one loop near $d=4$\cite{tissier06,tissier_2loop} in the limit $N\to \infty$ [and with the direct solution of Eq. (\ref{eq_RG_hatDelta_epsilon})]. The exponent $\beta$ is extracted from Eq.~(\ref{eq_rho_mk}) for $k=0$, which gives that $\rho_m\sim (\lambda-I_{2,0}[0])^{3/2}\sim (1-\Delta_B)^{3/2}$ and therefore that $\beta=3/4$, an unusual value. Similarly, from Eq.~(\ref{eq_solution_I_2_cusp}) for $k=0$ and $\rho=0$, {\it i.e.}, in the ``glassy'' PM phase where $\Delta_B>1$, one finds $I_{2,0}[y(0)]=I_{2,0}[0]/\Delta_B$. Together with the small-$y$ behavior, $I_{2,0}[y]=I_{2,0}[0](1-Ay^{(d-4)/2})$, this leads to $y(0)\sim (\Delta_B-1)^{2/(d-4)}$, {\it i.e.}, $(2-\eta)\nu=2/(d-4)$, and, with the above result $\eta=0$, to $\nu=1/(d-4)$. The critical exponents (for $4<d<6$),
\begin{equation}
\begin{aligned}
\label{eq_exponents}
\eta=0, \nu=1/(d-4), \, \gamma=2/(d-4), \,\beta=3/4\,,
\end{aligned}
\end{equation}
satisfy all the usual relations between exponents, including the hyperscaling one, provided one replaces $d$ by $d-\theta$ with $\theta=2+\eta-\bar\eta=2-(d-4)/2$: {\it e.g.}, $2\beta+\gamma=2-\alpha=\nu(d-\theta)$. This shift of dimension is the signature of the dangerous irrelevance of the temperature at the underlying fixed point.\cite{fisher_zeroT,villain_zeroT} However, dimensional reduction, in the sense that the exponents would be the same as that of the pure model in lower dimension, is {\it not} valid, even with an effective dimensional shift $\theta$.
Note finally that the specific point ($\Delta_B=1$, $\tau=0$) is a multicritical point which we do not find worth analyzing in detail.

For $2<d<4$, we find in the absence of applied source (magnetic field) a ``glassy'' PM phase that replaces the unstable SG phase and transforms into a normal PM phase at the Almeida-Thouless instability line already determined. The phase diagram is illustrated in Fig.~\ref{fig_exactdiagram_b}  for $d=3$. There is no quasi-long range order in the $N\to \infty$ limit, as anticipated from Refs. [\onlinecite{feldman,tissier_2loop}].

\subsection{Phase diagram with nonzero applied sources}

We now discuss what happens in the presence of a nonzero source (or magnetic field), so that $\rho$ is always strictly positive. We start again with $d>4$ because the results for $2<d<4$ can be easily deduced from this case. We find that there are additional transition lines in the presence of nonzero applied sources. For convenience we keep $\rho$ as the additional control variable instead of the applied source $J$, but the passage from one to the other, {\it i.e.}, $y(\rho)^2 \rho=J^2/(4N)$, is fully regular in the region of interest where $y(\rho)>0$. The relevant quantity is then $\rho_*=\rho_*(k=0)$, where $\rho_*(k)$ has been introduced in section \ref{section_solution_runningSD} (see for instance Figs.~\ref{fig_rhostar(k)_d=5} and \ref{fig_rhostar(k)_d=3}). For $\rho=\rho_*(\tau,\Delta_B)$, the replicon eigenvalue $\Lambda_{rep,k}(\rho_*,1)$ vanishes and this defines a whole surface in the ($\rho$, $\tau$, $\Delta_B$) diagram characterized by the solution of the two coupled equations
\begin{equation}
\begin{aligned}
\label{eq_transition_rhostar1}
I_{2,0}[y(\rho_*)]=\lambda-\frac 32(2\mu)^{1/3}\rho_*^{2/3}
\end{aligned}
\end{equation}
and 
\begin{equation}
\begin{aligned}
\label{eq_transition_rhostar2}
 y(\rho_*)=   \tau & +\frac {I_{1,0}[0]}{\lambda} +  \frac w6\Big [\lambda \left (\frac{\rho_*}{2\mu}\right )^{1/3}\hspace{-0.3cm}-\frac{\rho_*}{2}\Big ]-\frac {I_{1,0}[y(\rho_*)]}{I_{2,0}[y(\rho_*)]} \,,
\end{aligned}
\end{equation}
where $\lambda=I_{2,0}[0]/\Delta_B$ and where we have used $\Delta(\rho_*,1)=[\rho_*/(2\mu)]^{1/3}$.

\begin{figure}[h]
\hspace{-0.7cm} 
\centering
\subfigure[\hspace{7cm} ]{
\includegraphics[width=9cm]{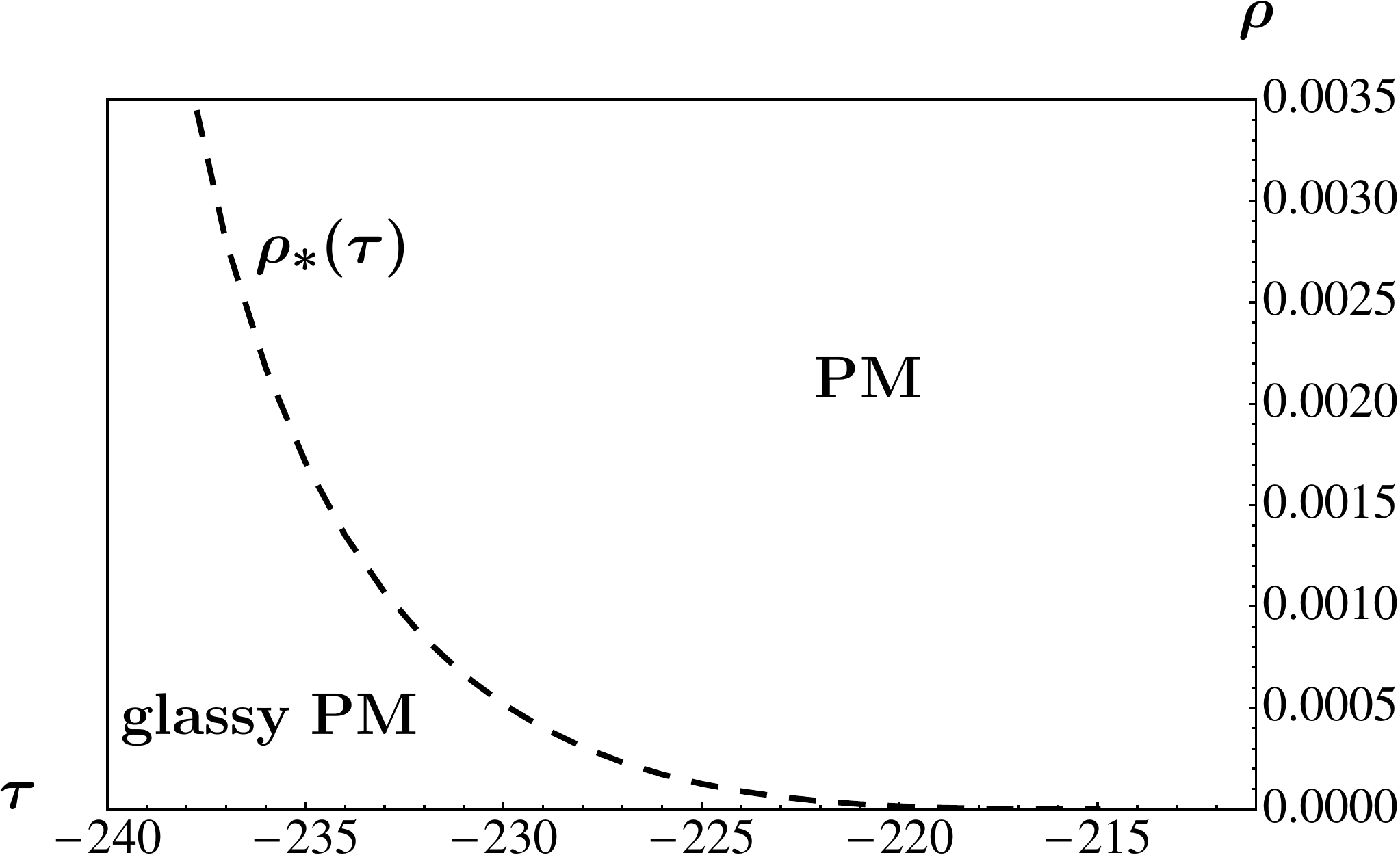}}
\hspace{-0.7cm} 
\subfigure[\hspace{7cm} ]{
\hspace{-0.7cm} \includegraphics[width=\linewidth]{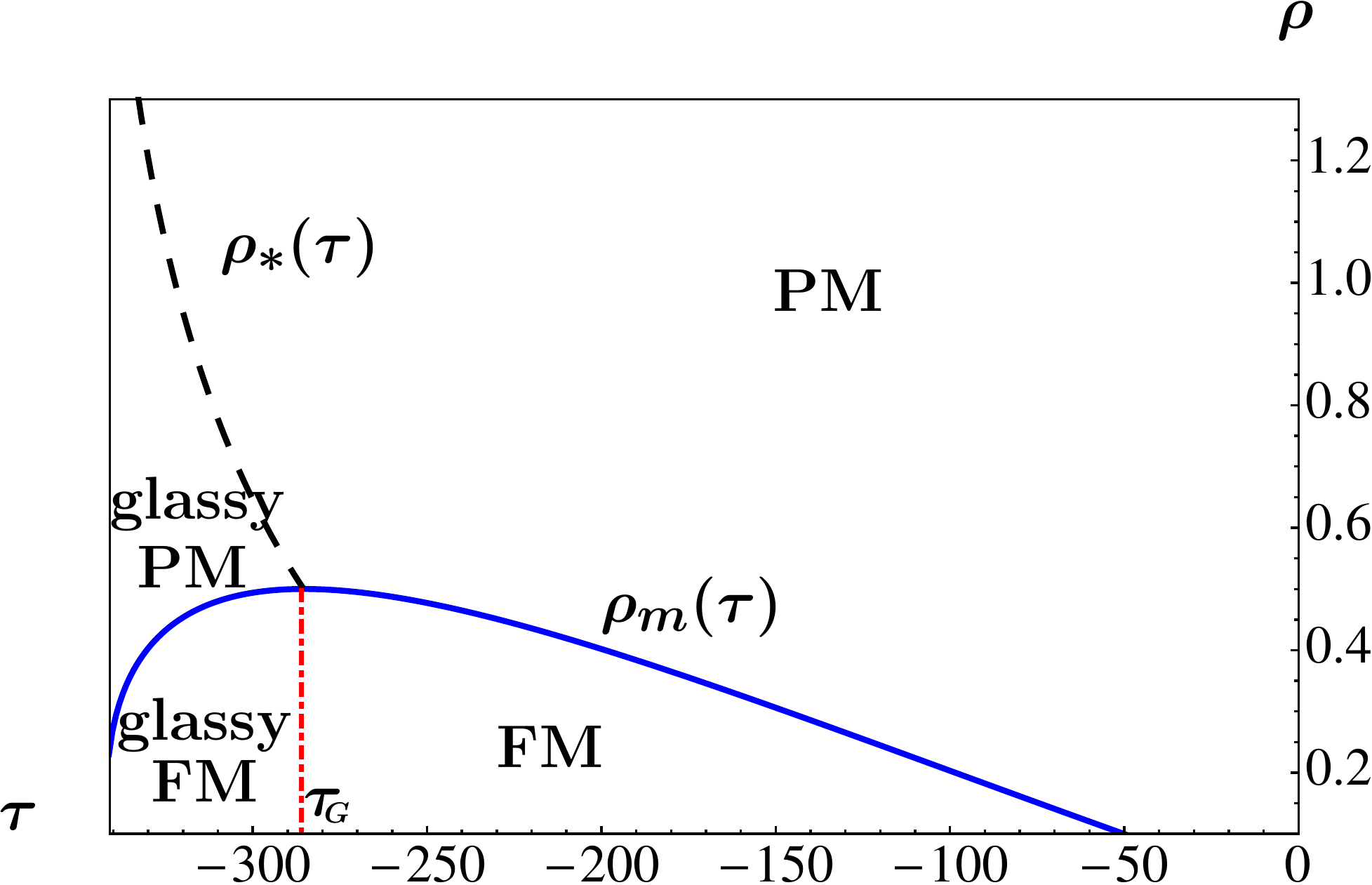}}
\caption{(Color on line) Transition line $\rho_*(\tau)$ between normal and ``glassy'' PM phases at fixed disorder strength $\Delta_B$ in nonzero applied source for $d=5$. (a) $\Delta_B>1$: Only PM phases are present. The approach to the transition point $\tau_{{\rm G}}(\Delta_B)$  in zero applied source goes as $\rho_*(\tau) \sim (\tau_{{\rm G}}-\tau)^3$. Here, $\Delta_B=2$. (b) $\Delta_B<1$: We also plot the FM order parameter $\rho_m(\tau)$, both in the normal FM phase for $\tau\geq \tau_{{\rm G}}$ and in the cuspy or glassy FM phase for $\tau\leq \tau_{{\rm G}}$. The glassy PM phase exists between $\rho_*(\tau)$ and $\rho_m(\tau)$ or $0$. Here, $\Delta_B=0.971$. }
\label{fig_rhostar(tau)_d=5}
\end{figure}

We plot $\rho_*(\tau)$ for fixed $\Delta_B$ and for $d=5$ in Fig.~\ref{fig_rhostar(tau)_d=5}: (a) and (b) respectively illustrate $\Delta_B>1$ where only PM phases are present and $\Delta_B<1$ where an FM phase is present [so that we also show the curve $\rho_m(\tau)$]. In both cases, the line terminates in the points $\tau_{{\rm G}}(\Delta_B)$ that define the Almeida-Thouless-like transition lines in zero applied source (see also section~\ref{section_naive}). For $\Delta_B<1$ the approach to $\tau_{{\rm G}}$ is regular with a finite nonzero slope but for $\Delta_B>1$ one finds that $\rho_*(\tau) \sim (\tau_{{\rm G}}-\tau)^3$. The case where $2<d<4$ is similar to that in Fig.~\ref{fig_rhostar(tau)_d=5} (a) for all values of the disorder strength. (Note that one could distinguish $\tau_{{\rm GPM}}$ for the transition to the PM phase and $\tau_{{\rm GFM}}$ for the transition to the FM phase, see below.)

\begin{figure}[h]
\includegraphics[width=\linewidth]{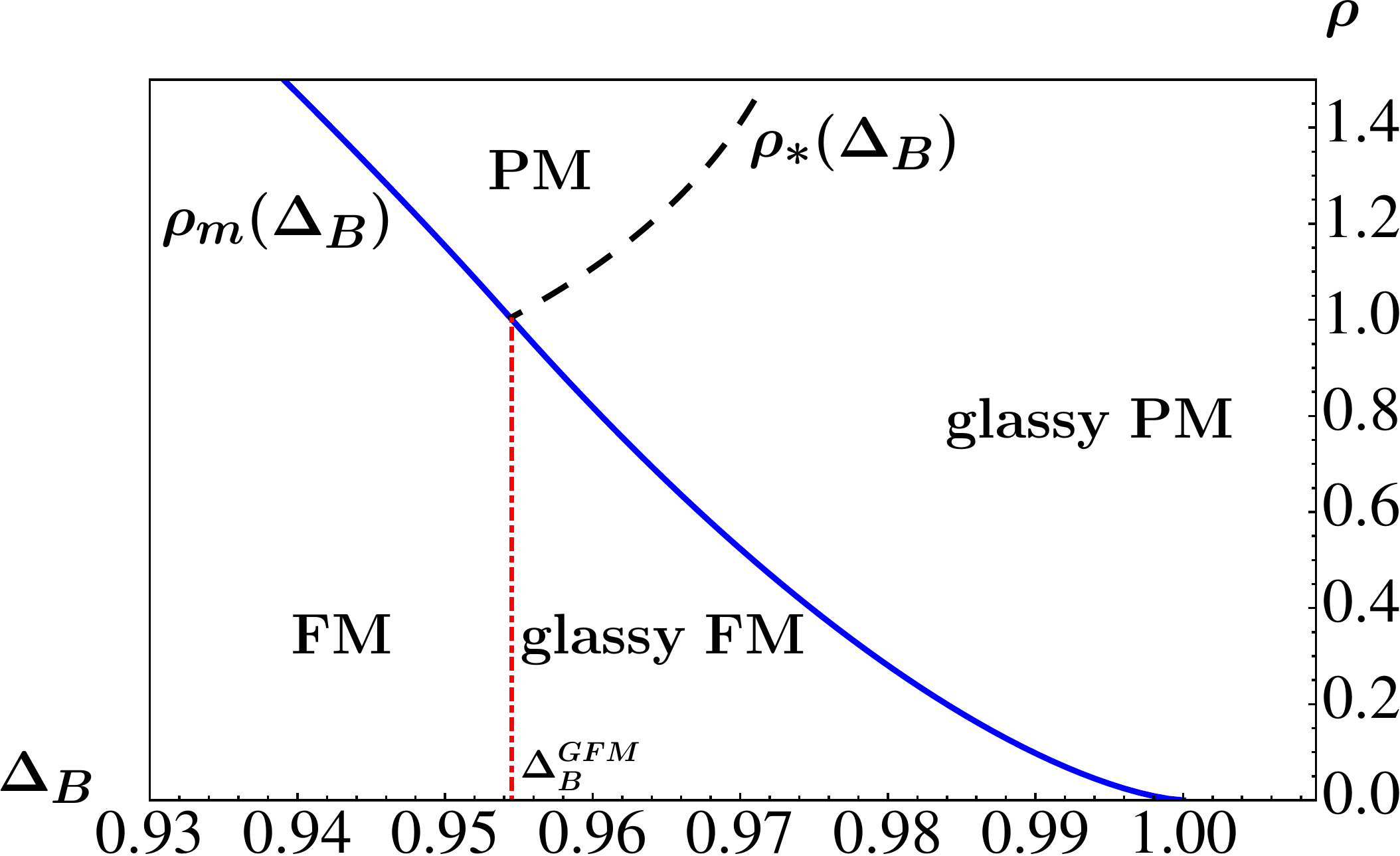}
\caption{(Color on line) Transition line $\rho_*(\Delta_B)$ between normal and ``glassy'' PM phases at fixed  $\tau=-336.24$ in nonzero applied source for $d=5$. We also plot the FM order parameter $\rho_m(\Delta_B)$, both in the normal FM phase for $\Delta_B\leq \Delta_B^{{\rm GFM}}$ and in the cuspy or glassy FM phase for $1\geq \Delta_B\geq \Delta_B^{{\rm GFM}}$. Note that $\rho_m(\Delta_B)$ approaches zero when $\Delta_B\to 1^-$ as $(1-\Delta_B)^{3/2}$. The glassy PM phase exists up to $\Delta_B= \Delta_B^{{\rm GPM}}\simeq 2.25$ near which $\rho_*(\Delta_B)$ approaches zero as $(\Delta_B^{{\rm GPM}}-\Delta_B)^{3}$. However there is an intermediate range of values of $\Delta_B$ for which no physical solution is found and we therefore restrict the plot to the region $\Delta_B \lesssim1$.}
\label{fig_rhostar(delta)_d=5}
\end{figure}

We also display $\rho_*(\Delta_B)$ for fixed $\tau<0$ and for $d=5$, together with $\rho_m(\Delta_B)$, in Fig.~\ref{fig_rhostar(delta)_d=5}. There are two critical values, $\Delta_B^{\rm GFM}$ and $\Delta_B^{\rm GPM}$, which correspond to the Almeida-Thouless-like transition lines between FM phases and between PM phases, respectively, in zero applied source. The approach to $\Delta_B^{\rm GFM}$ is regular, with $\rho_*(\Delta_B)-\rho_*(\Delta_B^{\rm GFM})\sim (\Delta_B^{\rm GFM}-\Delta_B)$, whereas that to $\Delta_B^{\rm GPM}$ is described by $\rho_*(\Delta_B) \sim (\Delta_B^{\rm GPM}-\Delta_B)^3$.

The phase which is delimited by the surfaces $\rho_*(\tau,\Delta_B)$ and $\rho_m(\tau,\Delta_B)$ (with $\rho_m$ replaced by $0$ in the PM region) is 
``glassy'', {\it i.e.}, the replicon is everywhere zero so that the SG-like susceptibility defined as the inverse of the replicon is always infinite. (When approaching the transition line from the normal PM phase, this susceptibility diverges with a critical exponent $\gamma=1$ whereas, as seen in section~\ref{section_naive}, $\gamma=1/2$ in zero applied source.) However, this phase has no spontaneous ordering, as the magnetization and the Edwards-Anderson order parameter are both non zero due to the applied source $J$. As already mentioned, this phase is also characterized by a nonanalytic functional dependence of the renormalized disorder function $\Delta(\rho,z)$ in the form of a cusp in $\sqrt{1-z}$ when $z\to 1$. This cusp in turn is related to the presence of avalanches or shocks in the evolution of the ground state of the system when changing the applied source. In a sense one can consider the amplitude of the cusp as an order parameter for the cuspy behavior. This amplitude is equal to    $-[\rho/(2\mu)]\sqrt{2/3}$  and is therefore strictly different from zero in the whole cuspy, glassy, phase. It is zero in the normal PM or FM phases and its behavior at the transition is governed by a boundary layer, {\it e.g.}, at fixed $\Delta_B$ and $\rho$, in $(1-z)/[\tau-\tau_*(\rho,\Delta_B)]^2$:  
\begin{equation}
\begin{aligned}
\label{eq_Delta_boundary_layer}
\Delta_k(\rho,z)=& \left (\frac{\rho}{2\mu}\right )^{1/3}\bigg[1-\sqrt{D^2\big[\tau-\tau_*(\rho,\Delta_B)\big]^2+\frac 23(1-z)}\\& 
+{\rm O}\Big(1-z,[\tau-\tau_*(\rho,\Delta_B)]\Big)\bigg]\,.
\end{aligned}
\end{equation}
The cusp is therefore rounded in a boundary layer when $\tau \to \tau_*(\rho,\Delta_B)^+$ and $z\to 1$ and it only appears strictly in $\tau =\tau_*(\rho,\Delta_B)$. This completes the description of the zero-temperature phase diagram of the random-anisotropy model in the limit $N\to \infty$.


\section{discussion}

We have provided a unified description of the thermodynamic behavior of the random-anisotropy O($N$) model in the large-$N$ limit. We have focused on the situation at zero temperature, for which one expects the mean-field-like artifacts of the $N\to \infty$ limit to be less severe. There is a whole region of the parameter space where the self-consistent equations for the pair correlations, {\it i.e.}, Schwinger-Dyson-like equations or equations of motion for the 2-PI formalism, no longer have stable solutions, as the so-called replicon eigenvalue of the associated stability operator vanishes and becomes negative. We can get around this apparent impasse by considering the theory in the presence of an IR regulator and looking at its flow as one lowers the IR cutoff, which leads to 1-PI functional RG equations. 

We have found that the flow of some 1-PI quantities exactly stops at a scale, {\it i.e.}, an IR cutoff, where the replicon first vanishes. The full solution down to zero IR cutoff (the exact theory) can then be obtained by using the property that the flow is frozen together with continuity of the 1-PI functions. The corresponding region of the phase diagram is formed by ``glassy'' phases, both in zero and in nonzero applied magnetic field, with a glassy ferromagnetic phase only possible above $d=4$. However, a {\it bona fide} spin-glass phase, associated with a spontaneous symmetry breaking described by a spin-glass order parameter, is nowhere observed. These glassy phases are ``marginal'' because the spin-glass-like susceptibility obtained as the inverse of the replicon eigenvalue is everywhere infinite. They are also characterized by a nonanalytic, ``cuspy'', functional dependence of the renormalized disorder correlations, which can be attributed to the presence of  ``avalanches'' or ``shocks'' in the ground state. Interestingly, the solution of the model and the existence of ``glassy'' phases does not require consideration of a spontaneous breaking of replica symmetry. (As found previously for the random-manifold model,\cite{doussal_largeN,MDW_largeN} it is nonetheless consistent with such a breaking in the present $N\to \infty$ limit , although with some peculiar features due to the $T=0$ limit.)

What are the implications of these findings for the finite-$N$ random-anisotropy case? We expect the zero-temperature phenomenology to carry over to finite $N$, at least at a qualitative level. Cusps and avalanches are rather generic properties of disordered systems at zero temperature, and, as confirmed by the 1- and 2-loop perturbative FRG results of the random-anisotropy model near $d=4$,\cite{tissier06,tissier_2loop} the conclusions reached above in the $N\to \infty$ limit should be robust when considering finite values of $N$. More interesting however is the behavior at finite temperature $T>0$ as it corresponds to the physical realizations of the model (in $d=3$, of course). We have already stressed in several places that the finite-temperature behavior is peculiar in the $N\to \infty$ limit: here, as in the random-manifold model, temperature does {\it not} generate a rounding of the cusp, when the latter is present at $T=0$ in $\Delta(\rho,z)$ (we consider the so-called ``thermodynamic'' regime\cite{MDW_largeN}). Such a thermal rounding is however found in the perturbative FRG near $d=4$ for any finite $N$ and is furthermore related to the physical picture of finite-$T$ droplet excitations.\cite{FRGledoussal-giamarchi,balents-doussal,tissier06,balog-tarjus} 

We therefore take as the most plausible hypothesis that nonanalyticities in the disorder cumulants are rounded by a finite temperature at finite $N$,  {\it unless} the long-distance behavior is controlled by a zero-temperature fixed point. In the latter case, when starting the 1-PI FRG flow from the region of parameter space where a cusp is encountered in, say, the fully transverse renormalized disorder function $\Delta_k(\rho,z)$, at $T=0$, one runs for a small enough bare temperature $T$ into a thermal boundary layer,\cite{FRGledoussal-giamarchi,balents-doussal,tissier06,balog-tarjus} 
\begin{equation}
\begin{aligned}
\Delta_k(\rho,z)\simeq \Delta_k(\rho,1)\bigg[1-T_k \,h\Big(\rho, \frac{1-z}{T_k^2}\Big)\bigg]\,,
\label{eq_thermal_BL}
\end{aligned}
\end{equation}
with $h(\rho,0)=0$ and $h(\rho,y\to\infty)\sim \sqrt y$, when $z\to 1$ and $k\to 0$. The renormalized temperature $T_k$ is defined as $T_k \propto k^2 T/\Delta_{k}(\rho_{mk},1)$. 

The existence of a zero-temperature fixed point means that the (renormalized) temperature is irrelevant at this fixed point and is characterized by an exponent $\theta >0$, {\it i.e.}, $T_k\sim k^\theta$. As a result, when $k=0$, one finds that $\lim_{k\to 0} [\Delta_k(\rho,z)/\Delta_k(\rho,1)] - 1 \sim \sqrt{1-z}$. The renormalized disorder function therefore has a cusp. This is true both in a ferromagnetic (FM) phase with $\rho_m>0$, where $\theta=2$ and $\Delta_0(\rho_m,1)>0$, and along the critical line, where $\theta<2$ is nontrivial and the cusp is only in the dimensionless form, $\Delta_k(\rho,z)/\Delta_k(\rho_{mk},1)$ (since $\Delta_k(\rho_{mk},1)\to 0$ as $k^{\bar\eta-2\eta}$). On the other hand in the normal FM phase for which no cusp appears at zero temperature, no thermal boundary layer is generated and the phase has no ``cuspy'' or  ``glassy'' character at finite temperature either. The transition lines  may of course shift with temperature.

This can be explicitly checked in a toy model in which we consider the $N\to \infty$ limit in the vicinity of $d=4$ in the FM region (see Eq. (\ref{eq_RG_hatDelta_epsilon}) in section \ref{section_1PI}) and add temperature as it appears in the 1-loop FRG for finite $N$:\cite{tissier06,footnote_limits}
\begin{equation}
\begin{aligned}
\partial_t   \widehat \Delta_k(z) =& \, \epsilon \widehat \Delta_k(z)
-C_4 \Big (\big[\widehat \Delta_k(z) -z\widehat \Delta_k(1) \big] \widehat \Delta_k'(z) \\& 
+ \widehat \Delta_k(z) \widehat \Delta_k(1) \Big ) -T_k z \widehat \Delta'_k(z)  \,,
\label{eq_RG_hatDelta_epsilon_T}
\end{aligned}
\end{equation}
with $T_k=k^2 T/\Delta_{mk}(1)$ and $\epsilon=d-4$; $\theta=2-\bar\eta=2-\epsilon/2$ along the critical line and $\theta=2$ in the FM phase. It is easily derived that the boundary-layer function in this case is simply given by $h(y)=-C+\sqrt{C^2+ (2/3)y}$, with $C>0$ a constant.

The above conclusion concerning nonanalyticity and marginality for phases and critical behavior controlled by zero-temperature fixed points at finite $N$ applies to the FM region of the phase diagram, and therefore only to $d>4$. It should also apply to the region of FM QLRO for $d<4$. On the other hand, the situation is more uncertain in the PM region of the phase diagram, which includes the phase(s) that could emerge from the zero-temperature  ``glassy'' PM phase(s) in $d=3$.\cite{footnote_krzakala} There is no obvious reason in favor of the presence of zero-temperature fixed points in this region, presence which seems necessary for the existence of a nontrivial, marginal or spin-glass-like, phase in $d=3$. However, at this stage, we cannot exclude this possibility. Even if one is willing to stay with dimensions larger than $4$, this part of the phase diagram is inaccessible to the perturbative FRG based on the nonlinear sigma model near $d=4$. It seems that an answer can only be brought by a nonperturbative FRG approach, of the type already successfully applied to random-field systems.\cite{tarjus04,tissier06,tissier11} This is however out of the scope of the present study.

\appendix

\section{2-PI formalism for the large-$N$ limit of the RA$O($N$)$M from the Gaussian variational approach}
\label{appendixA}

\subsection{2-PI effective action}

We start from the bare action in replica space given by Eq.~(\ref{eq_RAO($N$)M_replica}). To implement the Gaussian variational method we first redefine the microscopic replica fields as $\boldsymbol \chi_a(x)=\boldsymbol \phi_a +\boldsymbol\varphi_a(x)$ where $\boldsymbol \phi_a\equiv \overline{\langle \boldsymbol \chi_a(x)\rangle }$ (for simplicity we consider average replica fields $\boldsymbol \phi_a$ that are uniform in space, which correspond to uniform sources, but this is not necessary) and we reexpress Eq. (\ref{eq_RAO($N$)M_replica}) in terms of the new fields $\boldsymbol\varphi_a(x)$. The Gaussian variational method relies on introducing a trial Gaussian action,
\begin{equation}
\begin{aligned}
&S_G[\{\boldsymbol\varphi_{a}\},\{\boldsymbol G_{ab}\}] =  \\& {V\over 2} \int_q\sum_{a,b=1}^n  \sum_{\mu,\nu=1}^N[G^{-1}]_{ab}^{\mu\nu}(q)\, {\varphi_{a}^{\mu}(q)}{\varphi_{b}^{\nu}(-q)}\,,
\label{Hg}
\end{aligned}
\end{equation}
where $V$ is the volume of the system and the $[G^{-1}]_{ab}^{\mu\nu}$'s are considered as variational parameters. (As for the replica fields $\boldsymbol \phi_a$, we consider for simplicity translationally invariant functions, but this can be easily generalized.)

The partition function $Z_{rep}$ of the system is then rewritten as
\begin{equation}
\begin{aligned}
\mathcal Z_{rep}=&  \int \prod_{a}  {\cal D} \boldsymbol\varphi_{a}\  {\hbox{exp}} \big(-S_G[\{\boldsymbol\varphi_{a}\},\{\boldsymbol\phi_{a}\}] \big)\times \\&
{\hbox{exp}} \big[-\big(S_{rep}[\{\boldsymbol\varphi_{a}\},\{\boldsymbol G_{ab}\}] -S_G[\{\boldsymbol\varphi_{a}\},\{\boldsymbol G_{ab}\}] \big)\big]
\end{aligned}
\end{equation}
and the associated free energy $F_{rep} =\ln \mathcal Z_{rep}$ is computed  as an expansion in cumulants of  $(S_{rep}-S_G)$, with the average taken with the Gaussian ansatz. Truncating at the first order then leads to
\begin{equation}
F_{rep,1} =  F_G +  \langle S_{rep}-S_G\rangle_{S_G}
\end{equation}
where $F_G=-{V\over 2}\int_q \sum_a  \hbox{Tr}_{N}\, \hbox{ln}\, {\boldsymbol G}_{aa}(q)$ and $\hbox{Tr}_{N}$ denotes a trace over the $N$-vector indices. The method then uses the fact that the above expression is an upper bound of the exact free energy to find the best approximation by minimizing $F_{rep,1}$ with respect to the $\mathbf G_{ab}$'s (and the $\boldsymbol \phi_a$'s).

The first cumulant $\langle S_{rep}-S_G\rangle_{S_G}$ it is easily computed at dominant order in $1/N$.  This implies to determine all terms leading to contributions of order $N$ in the average.  Taking account of the fact that a trace over vector components generates a factor $N$ and that $\phi_a$ is of order $\sqrt{N}$ one finds that the pure (disorder free) part of the free energy is given at dominant order by
\begin{equation}
\begin{aligned}
\frac{F_{1,pure}}{V}& =  \sum_{a} {m^2\over 2T}  {\boldsymbol \phi_a}^2  + {1\over 2}\int_{ q}  \hbox{Tr}\, \hbox{ln}\, {\boldsymbol G}^{-1}( q)    \\&
 + {1\over 2T} \int_{ q} (q^2+m^2)  \hbox{Tr}\  \boldsymbol G(q)  \\&  +{w \over 4! NT}   \sum_{a} \Big(    {\boldsymbol \phi_a}^2  + \int_{q}  \hbox{Tr}_N\  \boldsymbol G_{aa}( q)  \Big)^2
\label{eq_Fpure}
\end{aligned}
\end{equation}
where $\hbox{Tr}$ means a trace over both vector components and replica indices, $\hbox{Tr} G\equiv \sum_a  \hbox{Tr}_{N} {\boldsymbol G}_{aa}$. 

The  disorder contribution to the variational free energy is
\begin{equation}
\begin{aligned}
&{F_{1,dis}\over V}= \\&
-\frac{N}{2T^2}\sum_{a,b}\bigg\langle R \bigg(\frac{ {\boldsymbol \phi_a}\mathbf{.} {\boldsymbol \phi_b}+{\boldsymbol \varphi_a}\mathbf{.} {\boldsymbol \phi_b}+{\boldsymbol \phi_a}\mathbf{.} {\boldsymbol \varphi_b}+{\boldsymbol \varphi_a}\mathbf{.} {\boldsymbol \varphi_b}}{N}\bigg)\bigg\rangle_G
\,.
\end{aligned}
\end{equation}
After expanding the function $R$ in the $\boldsymbol \varphi_a$'s, performing the Gaussian average and using Wick's theorem, and finally resumming all terms, one easily arrives at the expression
\begin{equation}
\begin{aligned}
\frac{F_{1,dis}}{V}=& -\frac N{2T^2}\sum_{a,b} \bigg[R\bigg(\frac{ {\boldsymbol \phi_a}\mathbf{.} {\boldsymbol \phi_b}  + \int_{q}  \hbox{Tr}_N\  \boldsymbol G_{ab}( q)}{N}\bigg)\\&
-  \frac 1N \int_q \hbox{Tr}_N\,\boldsymbol G_{ab}( q) R'\left (\frac{ {\boldsymbol \phi_a}\mathbf{.} {\boldsymbol \phi_b}}{N}\right )\bigg]
\label{eq_Fdis}
\end{aligned}
\end{equation}
where the prime denotes a derivative with respect to the argument of the function.

The variational free energy $F_{rep,1}[\{\boldsymbol \phi_a\},\{\boldsymbol G_{ab}\}]=F_{1,pure}+F_{1,dis}$ coincides with the 2-PI effective action $\Gamma_{2PI}$ in the $N\to \infty$ limit. (This was shown in Ref.~[\onlinecite{mezard-parisi91}] for the random-manifold model but can be easily generalized to the present case.) The minimization equations for the variational free energy then correspond to the equations of motion  associated with the stationarity of the effective-action functional in the 2-PI formalism. By using the expressions in Eqs.~(\ref{eq_Fpure},\ref{eq_Fdis}), one can cast $\Gamma_{2PI}\equiv F_{rep,1}[\{\boldsymbol \phi_a\},\{\boldsymbol G_{ab}\}]$ in the form of Eq.~(\ref{eq_RAO($N$)M_Gamma2}) with $\Gamma_2$ given by
\begin{equation}
\begin{aligned}
\label{eq_Gamma2_GVM}
&{\Gamma_2[\{\boldsymbol{\phi}_a\},\{\mathbf{G}_{ab}\}]\over V}= {1\over 2T} \sum_a \bigg [\int_q (q^2+m^2)\mathrm{Tr}_N \mathbf G_{aa}(q) \\&
+ \frac{2w}{4!N}\bigg( \Big(\boldsymbol{\phi}_a^2 +\int_q \mathrm{Tr}_N\mathbf{G}_{aa}(q)\Big)^2 -\Big(\boldsymbol{\phi}_a^2\Big)^2\bigg)\bigg ]
\\&-\frac N{2T^2}\sum_{a,b} \bigg [R\bigg(\frac{ {\boldsymbol \phi_a}\mathbf{.} {\boldsymbol \phi_b}  + \int_{q}  \hbox{Tr}_N\  \boldsymbol G_{ab}( q)}{N}\bigg)
-  \\&
\frac 1N \int_q \hbox{Tr}_N\,\boldsymbol G_{ab}( q) R'\Big(\frac{ {\boldsymbol \phi_a}\mathbf{.} {\boldsymbol \phi_b}}{N}\Big)-R\Big(\frac{ {\boldsymbol \phi_a}\mathbf{.} {\boldsymbol \phi_b}}{N}\Big)\bigg]\,.
\end{aligned}
\end{equation}

\subsection{Schwinger-Dyson-like equations and expansion in free replica sums}

The equations of motion, $\delta \Gamma_{2PI}/\delta G_{ab}^{\mu\nu}(q)$ [see Eq.~(\ref{eq_RAO($N$)M_stationary2})], which we also refer to as Schwinger-Dyson equations, lead to
\begin{equation}
\begin{aligned}
&[G^{-1}(q)]_{ab}^{\mu\nu}=\frac{\delta^{\mu\nu}}{T}\bigg[ \delta_{ab} (q^2+m^2) + \delta_{ab}  \left ({w\over  6 N}\right )  {\boldsymbol \phi}_a^2 \\&  -\frac 1T R'\Big(\frac{ {\boldsymbol \phi_a}\mathbf{.} {\boldsymbol \phi_b}}{N}\Big) + T\Sigma_{ab}^{\mu\nu}(q)\bigg]\,,
\label{schwinger}
\end{aligned}
\end{equation}
with
\begin{equation}
\begin{aligned}
&T \Sigma_{ab}^{\mu\nu}(q)=\delta^{\mu\nu}\Bigg \{ \delta_{ab}  \Big({w\over  6 N} \Big) \int_{q'}\  \hbox{Tr}_N\  \boldsymbol G_{aa}(q')
 + \\& \frac 1T \bigg [ R'\bigg(\frac{ {\boldsymbol \phi_a}\mathbf{.} {\boldsymbol \phi_b}}{N}\bigg) - R'\bigg(\frac{ {\boldsymbol \phi_a}\mathbf{.} {\boldsymbol \phi_b}  + \int_{q}  \hbox{Tr}_N\  \boldsymbol G_{ab}( q)}{N}\bigg)\bigg]\Bigg \}\,.
\label{schwinger_selfenergy}
\end{aligned}
\end{equation}
As anticipated, the self-energies are purely local functions, or equivalently are independent of $q$, in the $N\to \infty$ limit.

We take advantage of the presence of distinct replica sources which lead to an explicit breaking of the replica symmetry: from the first set of equations of motion in Eq.~(\ref{eq_RAO($N$)M_stationary1}), the replica fields $\boldsymbol \phi_a$ are thus different and can vary independently. We can consider the correlation functions $G_{ab}$ when evaluated at the minimum, {\it i.e.}, when Eqs.~(\ref{schwinger},\ref{schwinger_selfenergy}) are satisfied, as functions of the $\boldsymbol \phi_a$'s.  These functions, and all related ones, can then be expanded in an increasing number of unrestricted or free sums over replicas. The procedure is explained in detail in Refs.~[\onlinecite{tarjus04,tissier11,mouhanna-tarjus}].

Any  matrix  $A_{ab}(\{\boldsymbol\phi_{e}\})$ can be decomposed as
\begin{equation}
A_{ab}(\{\boldsymbol\phi_{e}\})=\widehat{A}_{a}(\{\boldsymbol\phi_{e}\})\, \delta_{a b} + \widetilde{A}_{a b}(\{\boldsymbol\phi_{e}\}),
\label{expansion_matrix}
\end{equation}
where $\widetilde{A}_{ab}$ does not contain any Kronecker symbol and the expansions in free replica sums read
\begin{equation}
\begin{split}
\widehat{A}_{a}(\{\boldsymbol\phi_{e}\}) = &\widehat{A}^{[0]}(\boldsymbol\phi_{a}) \\&+ \sum_{p\geq 1}\frac{1}{p!} \sum_{e_1,...,e_p}\widehat{A}^{[p]}(\boldsymbol\phi_{a}|\boldsymbol\phi_{e_1},...,\boldsymbol\phi_{e_p}),
\label{expansion_hatA}
\end{split}
\end{equation}
\begin{equation}
\begin{split}
\widetilde{A}_{ab}(\{\boldsymbol\phi_{e}\}) =& \widetilde{A}^{[0]}(\boldsymbol\phi_{a},\boldsymbol\phi_{b}) \\&+ \sum_{p\geq 1}\frac{1}{p!} \sum_{e_1,...,e_p}\widetilde{A}^{[p]}(\boldsymbol\phi_{a},\boldsymbol\phi_{b}|\boldsymbol\phi_{e_1},...,\boldsymbol\phi_{e_p}).
\end{split}
\label{expansion_tildeA}
\end{equation}
where the superscripts in square brackets denote the order in the free replica sum expansion. The $\widehat{A}^{[p]}$'s and $\widetilde{A}^{[p]}$'s are independent of the total number of replicas, $n$, and have continuity and symmetry properties ({\it e.g.}, $\widehat{A}^{[p]}$ is invariant under any permutation of the $p$ arguments $\boldsymbol\phi_{e_1},\cdots,\boldsymbol\phi_{e_p}$).

This can be applied to the matrices formed by the pair correlation functions, their inverse, and the self-energies. As a result, the Schwinger-Dyson equations, Eqs.~(\ref{schwinger},\ref{schwinger_selfenergy}), can be solved order by order (this is not an approximation). For the zeroth order, one finds
\begin{equation}
\begin{aligned}
&T \widehat{\Sigma}^{[0]\mu\nu}(q;\boldsymbol\phi_{a}) =\delta^{\mu\nu} \bigg\{ \left ({w\over  6 N}\right )\int_{q'}\  \hbox{Tr}_N \Big[ \widehat{\boldsymbol G}^{[0]}(q';{\boldsymbol \phi}_a)\, + 
\\& \widetilde{\boldsymbol G}^{[0]}(q';{\boldsymbol \phi}_a,{\boldsymbol \phi}_a)\Big ] +  \frac 1T R'\bigg(\frac{ {\boldsymbol \phi_a}^2 + \int_{q'}  \hbox{Tr}_N\  \widetilde{\boldsymbol G}^{[0]}(q';{\boldsymbol \phi}_a,{\boldsymbol \phi}_a) }{N}\bigg) \\& 
-   \frac 1T R'\bigg(\frac{ {\boldsymbol \phi_a}^2 + \int_{q'}  \hbox{Tr}_N \Big[ \widehat{\boldsymbol G}^{[0]}(q';{\boldsymbol \phi}_a) +  \widetilde{\boldsymbol G}^{[0]}(q';{\boldsymbol \phi}_a,{\boldsymbol \phi}_a) \Big] }{N}\bigg)\bigg \}
\label{schwinger2}
\end{aligned}
\end{equation}
and
\begin{equation}
\begin{aligned}
&T^2 \widetilde{\Sigma}^{[0]}(q;\boldsymbol\phi_{a},\boldsymbol\phi_{b})^{\mu\nu} = \delta^{\mu\nu}\bigg \{R'\bigg(\frac{ {\boldsymbol \phi_a}\mathbf{.} {\boldsymbol \phi_b}}{N}\bigg) - \\&  R'\bigg(\frac{  {\boldsymbol \phi_a}\mathbf{.} {\boldsymbol \phi_b} + \int_{q}  \hbox{Tr}_N \widetilde{\boldsymbol G}^{[0]}(q';{\boldsymbol \phi}_a,{\boldsymbol \phi}_b)}{N}\bigg) \bigg\} 
\label{schwinger3}
\end{aligned}
\end{equation}
with
\begin{equation}
\widehat{\boldsymbol G}^{[0]}(q;\boldsymbol\phi_{a})=\widehat{\boldsymbol G^{-1}}^{[0]}(q;\boldsymbol\phi_{a})^{-1}\label{inverse_widehatG0}
\end{equation}
and
\begin{equation}
\begin{aligned}
\widetilde{\boldsymbol G}^{[0]}(q;\boldsymbol\phi_{a},\boldsymbol\phi_{b})=- \widehat{\boldsymbol G}^{[0]}(q;\boldsymbol\phi_{a})\widetilde{\boldsymbol G^{-1}}^{[0]}(q;\boldsymbol\phi_{a},\boldsymbol\phi_{b}) \widehat{\boldsymbol G}^{[0]}(q;\boldsymbol\phi_{b}).
\end{aligned}
\label{inverse_widetildeG0}
\end{equation}

We now consider specific configurations of the fields. We evaluate the 1-replica quantities in a configuration such that $\phi_a^{\mu}=\sqrt{\rho_1 N}\ \delta_{\mu1}$ and the 2-replica ones in a configuration of the two replica fields $\boldsymbol\phi_a$ and $\boldsymbol\phi_b$ that is parametrized by
\begin{equation}
\begin{array}{ll}
&\displaystyle \phi_a^{\mu}=\sqrt{\rho_1 N}\ \delta_{\mu1}\ , \\
\\
&\displaystyle \phi_b^{\mu}=\sqrt{\rho_2  N}\left(\cos\theta\ \delta_{\mu1}+\sin\theta\ \delta_{\mu2}\right).
\label{2replic}
\end{array}
\end{equation}
As in [\onlinecite{mouhanna-tarjus}] we introduce the longitudinal  $L$ ($\mu=\nu=1$) and transverse $T$ ($\mu=\nu\ne1$) components of the 1-replica correlation function $\widehat{\boldsymbol G}^{[0]}(q;\boldsymbol\phi_{a})$.  For the 2-replica correlation functions, there are more components but the only one which is needed in the large-$N$ limit is the fully transverse $TT$ one with $\mu=\nu \neq 1,2$. At leading order in $N$ we have $\hbox{Tr}_N\,  \widehat{\boldsymbol G}^{[0]}\sim N \widehat{G}_T^{[0]}$ and  $\hbox{Tr}_N\, \widetilde{\boldsymbol G}^{[0]}\sim N \widetilde{G}_{TT}^{[0]}$. This immediately leads to
\begin{equation}
\begin{aligned}
&T \widehat{\Sigma}_T^{[0]}(\rho_1) ={w\over  6}\int_{q} \Big[ \widehat{G}_T^{[0]}(q';\rho_1)+ \widetilde{G}_{TT}^{[0]}(q';\rho_1,\rho_1,z=1)\Big]   \\&
+ \frac 1T R'\Big(\rho_1 + \int_{q}  \widetilde{G}_{TT}^{[0]}(q';\rho_1,\rho_1,z=1)\Big) \\&
-   \frac 1T R'\Big(\rho_1 + \int_{q} \Big [\widehat G_T^{[0]}(\rho_1)+\widetilde{G}_{TT}^{[0]}(q';\rho_1,\rho_1,z=1)\Big]\Big)
\label{schwinger_finalhat}
\end{aligned}
\end{equation}
and
\begin{equation}
\begin{aligned}
&T^2\widetilde{\Sigma}^{[0]}(q;\rho_1,\rho_2,z)= R'\Big (\sqrt{\rho_1\rho_2}z\Big) - \\&  R'\Big(\sqrt{\rho_1\rho_2}z+ \int_{q} \widetilde{G}_{TT}^{[0]}(q';\rho_1,\rho_2,z) ] \Big) 
\label{schwinger_finaltilde}
\end{aligned}
\end{equation}
where we have introduced $z=\cos \theta$ and the functions $\widehat G_T^{[0]}(\rho_1)$ and $\widetilde{G}_{TT}^{[0]}(q';\rho_1,\rho_2,z)$ are given in Eqs.~(\ref{eq_RAO(Ninfty)M_SDhat},\ref{eq_RAO(Ninfty)M_SDtilde}) of the main text.

The equation of state can be obtained from Eq.~(\ref{eq_RAO($N$)M_stationary1}) along the same lines and it simply reads $\sqrt \rho\, y(\rho)=\vert J \vert/(2\sqrt N)$ with $y(\rho)$ defined in and above Eq.~(\ref{eq_RAO(Ninfty)M_self-energy_hat}).

\subsection{Replicon eigenvalue}

Finally, we give some indications on how to obtain the replicon eigenvalue. We first consider the 2-PI effective action as a functional of the self-energies in place of the correlation functions. The stability of the equations of motion (or Schwinger-Dyson equations) is then governed by the second functional derivative of $\Gamma_{2PI}[\{\boldsymbol{\phi}_a\},\{\Sigma_{ab}\}]$, evaluated for the replica symmetric solution (when $n\to 0$) or alternatively at the zeroth order of the expansion in free replica sums. One finds
\begin{equation}
\begin{aligned}
\label{eq_RAO(N)M_secondderivative}
&\frac{\delta^2(2\, \Gamma_2/V)}{\delta \Sigma_{ab}^{\mu\mu}(q=0)\delta \Sigma_{cd}^{\nu\nu}(q=0)}= \\&\delta_{ac}\delta_{cd}\int_q\widehat G_T^{[0]}(q)^2 \bigg [\delta_{\mu\nu} -
\frac{R''\Big(\rho+\int_q\widetilde G_{TT}^{[0]}(q)\Big)}{NT^2}\int_q\widehat G_T^{[0]}(q)^2 \bigg] + \\&
(\delta_{ac}+\delta_{cd})\int_q\widehat G_T^{[0]}(q)\widetilde G_{TT}^{[0]}(q) \bigg [\delta_{\mu\nu} -2\frac{R''\Big(\rho+\int_q\widetilde G_{TT}^{[0]}(q)\Big)}{NT^2} \\& \times \int_q\widehat G_T^{[0]}(q)^2 \bigg ] +
 \int_q\widetilde G_{TT}^{[0]}(q)^2 \bigg [\delta_{\mu\nu} -2\frac{R''\Big(\rho+\int_q\widetilde G_{TT}^{[0]}(q)\Big)}{NT^2} \\& \times
 \int_q\widehat G_T^{[0]}(q)^2 \bigg ] 
\end{aligned}
\end{equation}
for $a<b$ and $c<d$. The replicon eigenvalue is obtained from a linear combination of such terms: symbolically, ``12,12'' - 2 $\times$ ``12,13'' + ``12,34''.\cite{SGbook,almeida78} Only the first term of the above expression then survives, which leads to Eq.~(\ref{eq_replicon_z=1}) of the main text.

\section{Graphical representation of the $2$-PI formalism for the large $N$ limit of the RA$O($N$)$M}
\label{appendixB}

In this appendix we  derive the explicit expression of $\Gamma_{2PI}$ at leading order in $1/N$ in the case where the bare disorder function $R$ is restricted to:
\begin{equation}
R(u)=\Delta_{2} {u^{2}\over 2} + \Delta_{4} {u^{4}\over 4}
\end{equation}
Higher-order terms can be treated along the same lines.

 We recall that  $\Gamma_{2PI}$  generalizes  the 1PI  effective action   in the sense that it is  a  functional  of  both  the  local order parameter fields 
and the connected two-points  correlation  functions.  To build this quantity one introduces  two  sources that  couple linearly and quadratically  to the microscopic field $\{\boldsymbol\chi_{a}\}$:
\begin{equation}
\begin{array}{ll}
\displaystyle Z_{rep}[ \{\boldsymbol J_{a}\}, \{\boldsymbol  K_{ab}\}]&= \displaystyle \int \prod_{a=1}^n  {\cal D}  \boldsymbol\chi_{a}\  {\hbox{exp}} \Big( - S_{rep}[\{\boldsymbol \chi_{a}\}]   \\
\\
& \hspace{-3cm}\displaystyle + \sum_{a=1}^n \int_{ x}  \boldsymbol J_{a}(x)  \boldsymbol \chi_{a}( x)  + \frac{1}{2} \sum_{a,b}^n \int_{ x} \int_{ y} \boldsymbol \chi_{a}( x) \boldsymbol K_{ab}( x, y) \boldsymbol \chi_{b}( y)\Big)\  .
\label{ZjKrep}
\end{array}
\end{equation}From this expression one defines as usual  the free energy $W_{rep}[ \{\boldsymbol J_{a}\}, \{\boldsymbol K_{ab}\}]=\ln  Z_{rep}[ \{\boldsymbol J_{a}\}, \{\boldsymbol K_{ab}\}]$, from which follows the expectation value 
\begin{equation}
{\delta W_{rep}[ \{\boldsymbol J_{e}\}, \{\boldsymbol K_{e f}\}]\over\delta J_{a}^{\mu}( x)}= \langle \chi_{a}^{\mu}( x)\rangle =\displaystyle  \boldsymbol\phi_{a}^{\mu}( x) 
\label{phi2}
\end{equation}
  and the correlation function
\begin{equation}
\begin{array}{ll}
\displaystyle{\delta W_{rep}[ \{\boldsymbol J_{e}\}, \{\boldsymbol K_{e f}\}]\over\delta  K_{ab}^{\mu\nu}(x,y)}&=\displaystyle{1\over 2} \langle\chi_{a}^{\mu}(x)\chi_{b}^{\nu}(y)\rangle \\
\\
& =\displaystyle {1\over 2} \big [ G_{ab}^{\mu\nu}(x,y)+\phi_{a}^{\mu}(x)\phi_{b}^{\nu}(y)\big ] \ .
\label{G}
\end{array}
\end{equation}
One then performs  a  double Legendre  transform of  the free energy  with respect  to the  sources  $J_{a}$ and $K_{ab}$.  This defines   the 2-PI effective action $\Gamma_{2PI}[\{\boldsymbol\phi_{a}\},\{\boldsymbol G_{ab}\}]$:
\begin{equation}
\begin{array}{ll}
&\displaystyle  \Gamma_{2PI}[\{\boldsymbol\phi_{a}\},\{\boldsymbol G_{ab}\}] =\\
\\& \displaystyle -W_{rep}[\{\boldsymbol J_{a}\},  \{\boldsymbol K_{ab}\}] + \sum_{a=1}^n  \int_x  \boldsymbol J_{a} (x) \boldsymbol \phi_{a}(x) \\
\\& \displaystyle  + \frac{1}{2} \sum_{a,b=1}^n   \int_x \int_y \boldsymbol K_{ab}(x,y) \big [\boldsymbol G_{ab}(x,y) + \boldsymbol\phi_{a}(x) \boldsymbol\phi_{b}(x) \big ],
\label{gamma2PI}
\end{array}
\end{equation}
where we have used Eqs.~(\ref{phi2}) and (\ref{G}).

The 2-PI effective action is generally parametrized as in Eq.~(\ref{eq_RAO($N$)M_Gamma2}). The classical (bare) action $S_{rep}[\{\boldsymbol\phi_{a}\}]$ is here given by
\begin{equation}
\begin{aligned}
& S_{rep}[\{\boldsymbol\phi_{a}\}] = {1\over T} \int_x  \Bigg\{  \sum_{a=1}^n \bigg[{1\over 2} (\partial \boldsymbol \phi_{a})^2 + {m^2\over 2}  {\boldsymbol\phi_{a}^2} + {w\over 4!  N}   (\boldsymbol \phi_{a}^2)^2 \\&
-  {N\over 2T} \sum_{a,b=1}^n  \left[{\Delta_2\over 2}\Big({\boldsymbol\phi_a.\boldsymbol\phi_b\over N}\Big)^2+{\Delta_4\over 4}\Big({\boldsymbol\phi_a.\boldsymbol\phi_b\over N}\Big)^4\right]\ \Bigg\}\,.
\label{replic2}
\end{aligned}
\end{equation}
The factors of $N$ are such that if $\phi_a\propto \sqrt N$, $S_{rep}$ is of order $N$. The classical inverse propagator $\boldsymbol G_0^{-1}({ q};\{\boldsymbol\phi_{a}\})$ is defined by 
\begin{equation}
\displaystyle \left[ G_{0}^{-1}\right]_{ab}^{\mu\nu}(x, x';\{\boldsymbol \phi_{e}\})={\delta^2S_{rep}[\{\boldsymbol \phi_{e}\}]\over \delta \phi_a^{\mu}( x)\delta \phi_b^{\nu}( x')} \,,
\nonumber
\end{equation}
which, in momentum space and  for uniform replica field configurations, reads
\begin{equation}
\begin{array}{ll}
\displaystyle [G_0^{-1}]_{ab}^{\mu\nu}( q, q';\{\boldsymbol \phi_{e}\})&=\displaystyle  {1\over T}\bigg\{\bigg({q}^2+m^2+ {w\over 6NT}   \boldsymbol\phi_a^2\bigg)\,  \delta^{\mu\nu}\  \delta_{ab} \\
\\
&\hspace{-3.5cm} + \displaystyle  {w\over 3NT}\  \phi_a^{\mu} \phi_b^{\nu}\  \delta_{ab}  - {{\Delta_2}\over N T^2}\bigg(\boldsymbol\phi_a .\boldsymbol\phi_b\  \delta^{\mu\nu}+ \phi^{\mu}_b\  \phi^{\nu}_a  + \\
\\
&\hspace{-3.5cm} \displaystyle \delta_{ab} \sum_{c}  \phi^{\mu}_c \phi^{\nu}_c \bigg)- {\Delta_4\over  N^3 T^2}\bigg( (\boldsymbol\phi_a.\boldsymbol\phi_b)^3\ \delta^{\mu\nu}+ 3(\boldsymbol\phi_a .\boldsymbol\phi_b)^2\  \phi^{\mu}_b\  \phi^{\nu}_a  \\
\\
&\hspace{-3.5cm}  \displaystyle + 3\  \delta_{ab} \sum_{c}  \phi^{\mu}_c \phi^{\nu}_c\   (\boldsymbol \phi_a.\boldsymbol\phi_c)^2\bigg)\bigg\}\delta( q+ q') \ .
\label{bare}
\end{array}
\end{equation}

 In practice the sum of all  2-PI  contributions to   the effective action, $\Gamma_2[\{\boldsymbol\phi_{a}\},\{\boldsymbol G_{ab}\}]$, is computed by considering the microscopic  action $S_{rep}[\{\boldsymbol\chi_{a}\}]$ 
 and decomposing each  replica field as $\boldsymbol\chi_a=\boldsymbol\phi_a+\boldsymbol\varphi_a$  where $\boldsymbol\phi_a$ is the  expectation value.  The cubic and higher-order terms in $\boldsymbol\varphi_a$    
 appearing  in the expression of $S_{rep}[\{\boldsymbol\phi_a+\boldsymbol\varphi_a\}]$ define the vertices of the theory where the $\boldsymbol\phi_a$'s  are  considered as  external fields  and  the propagator lines identify with $\boldsymbol {G}_{ab}$. $\Gamma_2[\{\boldsymbol\phi_{a}\},\{\boldsymbol G_{ab}\}]$ is then obtained by considering all 2-PI diagrams built from the vertices  with dressed propagators $\boldsymbol {G}_{ab}$  and insertions of composite operators $\boldsymbol\phi_{a}$  at the considered order. Also, since the 2-PI effective action is a singlet under the rotation group O($N$) it must be constructed from  O($N$) invariants. If one defines the 2-replica matrix $\mathbf \Phi_{ab}$ by $\Phi_{ab}^{\mu\nu}=\phi_a^{\mu}\phi_b^{\nu}$, these invariants are given by  Tr$\boldsymbol \Phi_{ab}= \boldsymbol \phi_a.\boldsymbol{\phi}_b$,  Tr$(\boldsymbol {G}_{ab}^p)$ -- this term being generated by contraction of $\boldsymbol\varphi_a$'s --  and  Tr($\boldsymbol \Phi_{ab} \boldsymbol {G}_{ab}^p)$.  Note that this implies in particular that  only terms even in $\boldsymbol\phi_a$ contribute. 
 
Let us now   count the powers of $N$ involved in a given diagram.  Each trace involving the fluctuating field  $\boldsymbol\varphi_a$ -- more precisely the propagator $\boldsymbol G_{ab}$ --  or the expectation value $\boldsymbol\phi_a$   generates a factor $N$  while  the vertices provide  factors  of $1/N$  -- for vertices proportional to $w$ or $\Delta_2$  -- or $1/N^3$ -- for vertices proportional to $\Delta_4$: see  Eq.~(\ref{replic2}).  

We consider successively the $\boldsymbol\phi$-independent and the $\boldsymbol\phi$-dependent part  of $\Gamma_2[\{\boldsymbol\phi_{a}\},\{\boldsymbol G_{ab}\}]$.   The  fluctuating field  $\boldsymbol\varphi_a$   is represented by  a single line and its expectation value $\boldsymbol\phi_a$ by a double line. 

\begin{figure}[h]
\hskip0cm
\includegraphics[scale=0.5]{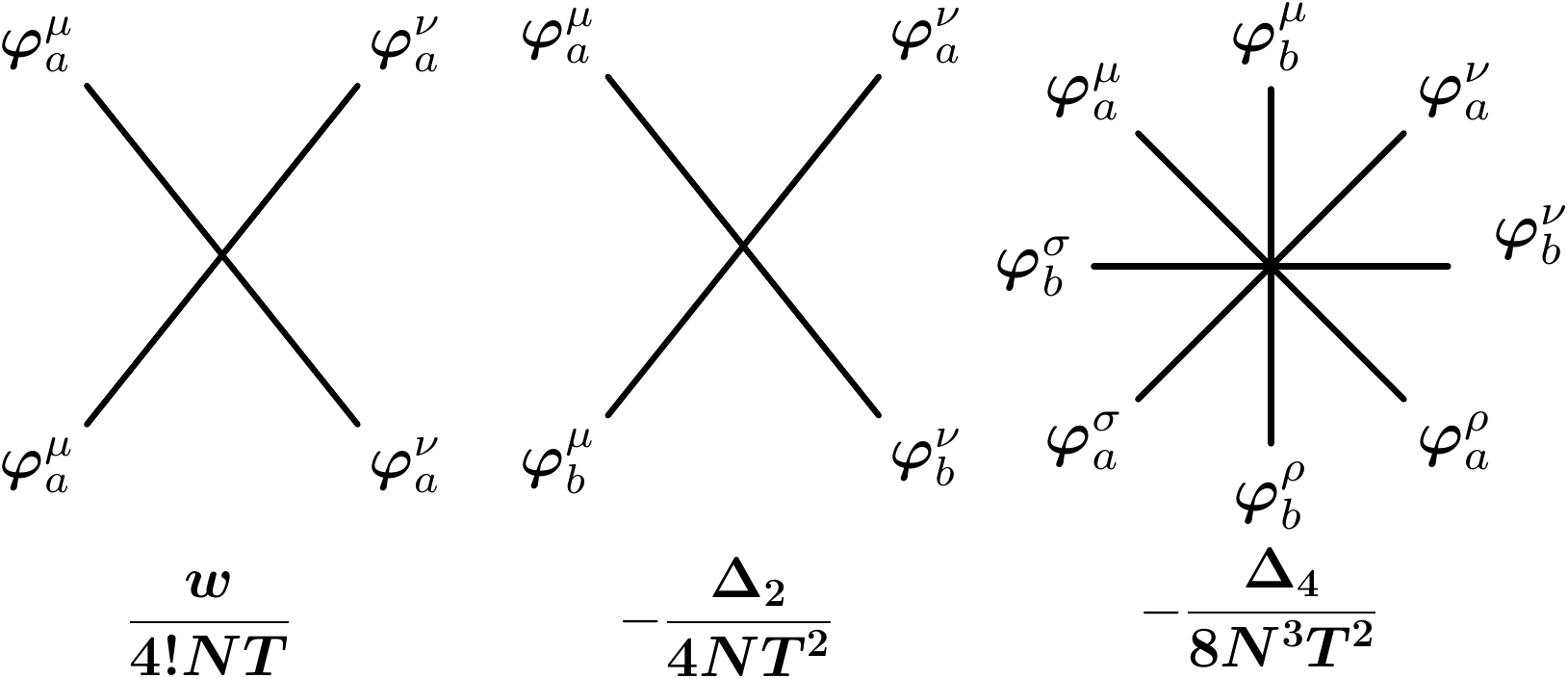}
\caption{Vertices  associated  to the $\boldsymbol\phi$-independent part of $\Gamma_2[\{\boldsymbol\phi_{a}\},\{\boldsymbol G_{ab}\}]$  contributing to the leading order in $1/N$.}
\label{fig0}
\end{figure}

The diagrams contributing to the $\boldsymbol\phi$-independent part of $\Gamma_2$ are  built with the  vertices obtained from the expansion of $S_{rep}[\{\boldsymbol\phi_a+\boldsymbol\varphi_a\}]$ that are independent of the $\boldsymbol\phi_{a}$'s. They are given  in Fig.~\ref{fig0}.


\begin{widetext}
\begin{figure}[h]
\hskip0cm
\includegraphics[scale=0.5]{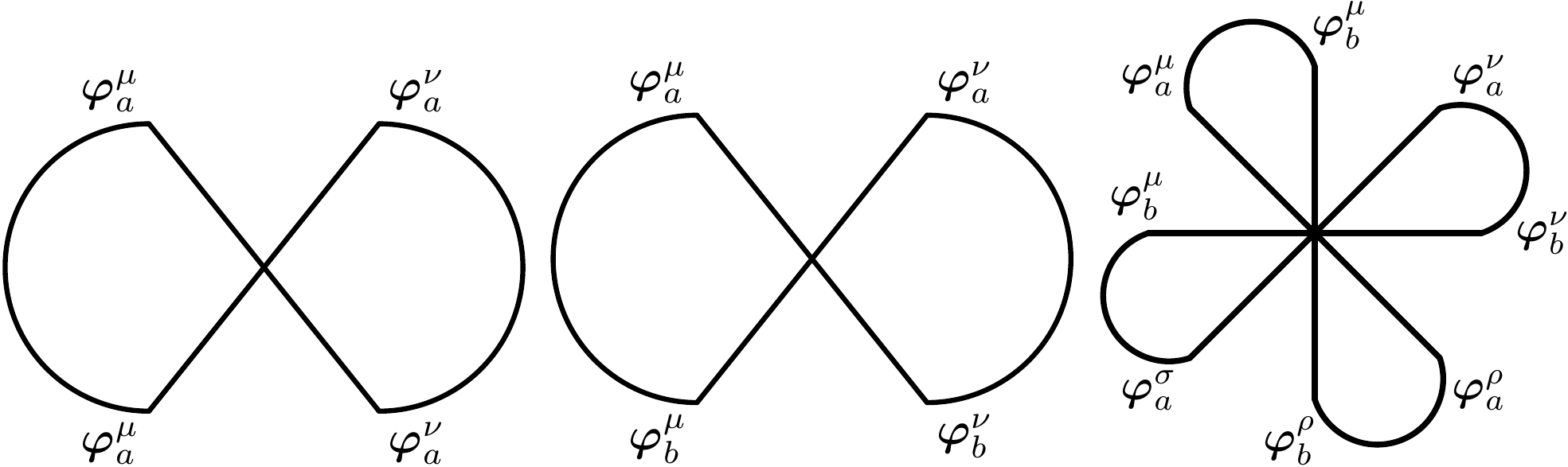}
\caption{ Diagrams contributing to the $\boldsymbol\phi$-independent part of $\Gamma_2[\{\boldsymbol\phi_{a}\},\{\boldsymbol G_{ab}\}]$ at leading order in $1/N$.}
\label{fig1}
\end{figure}
\end{widetext}

As for the diagrams contributing to the leading order they should form closed loops, each providing a factor of $N$, which implies contractions of the {\it same} vectorial indices. They are displayed in Fig.~\ref{fig1}.

\vspace{2cm}

This leads  to the $\boldsymbol\phi$-independent contribution to $\Gamma_2$:  
\begin{equation}
\begin{aligned}
&\Gamma_{2,ind}^{LO}[\{\boldsymbol\phi_{a}=0\},\{\boldsymbol G_{ab}\}] = {w\over 4! N T}\sum_a \int_x \big [\hbox{Tr}_N\,  \boldsymbol G_{aa}(x,x)\big]^2 \\&-\frac1{2NT^2}\sum_{a,b} \int_x\ \bigg({\Delta_2 \over 2} \big[\hbox{Tr}_N\,\boldsymbol G_{ab}(x,x)\big]^2  \\& +  {\Delta_4 \over 4 N^2} \big [\hbox{Tr}_N\, \boldsymbol G_{ab}(x,x) \big]^4\bigg) \,.
\end{aligned}
\end{equation}

\begin{figure}[h!]
\hskip0cm
\includegraphics[scale=0.5]{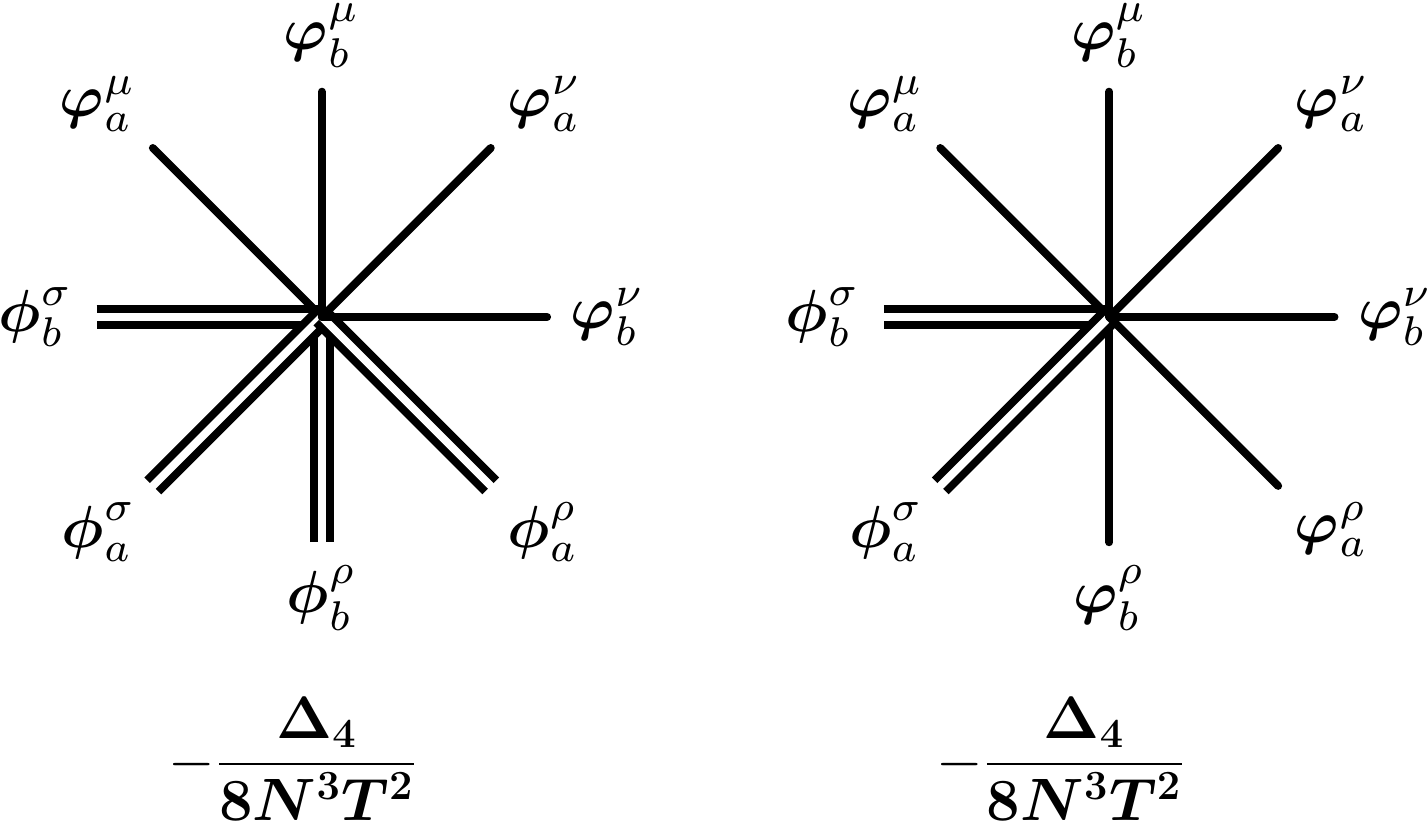}
\caption{Vertices  associated  to the $\boldsymbol\phi$-dependent part of $\Gamma_2[\{\boldsymbol\phi_{a}\},\{\boldsymbol G_{ab}\}]$  contributing to the leading order in $1/N$.}
\label{fig2}
\end{figure}

The  $\boldsymbol\phi$-dependent part of   $\Gamma_2$ is generated by the diagrams built from vertices that mix the fields $\boldsymbol{\varphi}_a$ and $\boldsymbol{\phi}_a$.  Vertices proportional to $\Delta_2$  or $w$  contribute to the next-to-leading order and thus, at leading  order, only vertices proportional to $\Delta_4$ contribute. They are displayed in Fig.~\ref{fig2}.


\begin{figure}[!h]
  \hspace{-0.75cm}
 \includegraphics[scale=0.5]{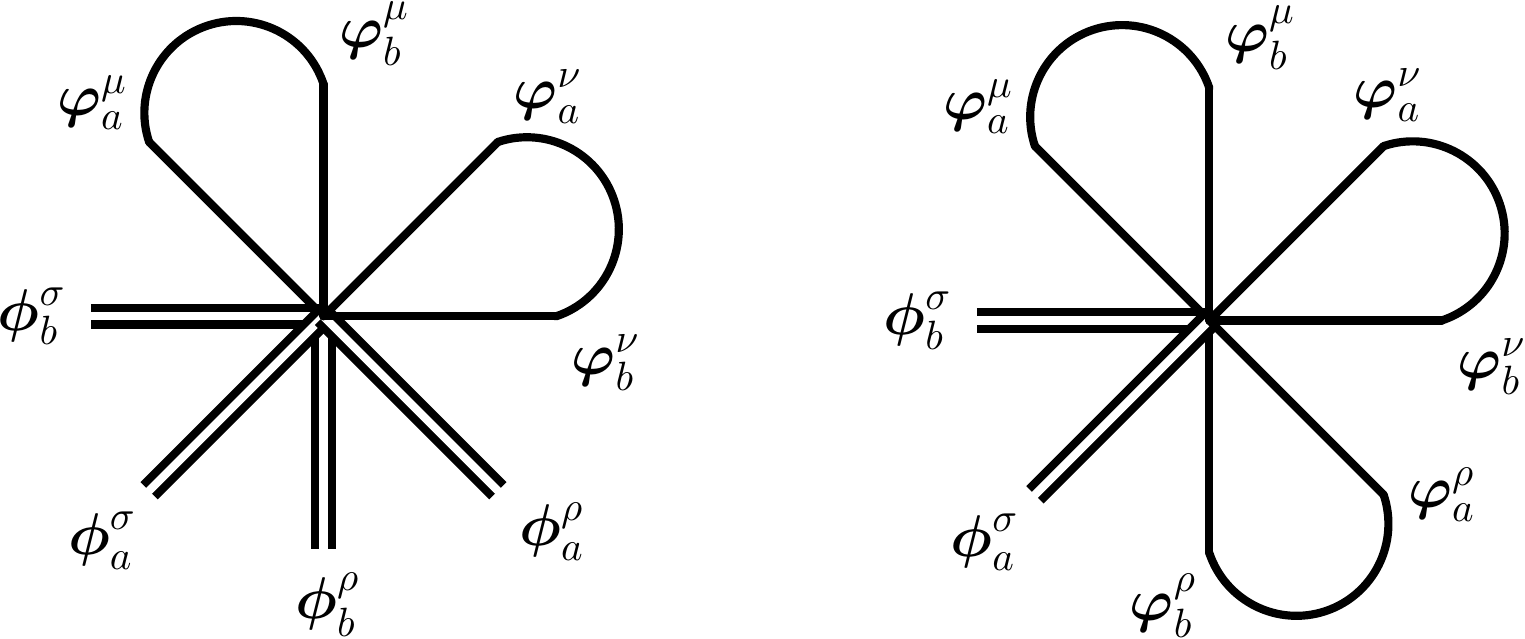}
 \caption{Diagrams contributing  to the $\boldsymbol\phi$-dependent $\Gamma_2[\{\boldsymbol\phi_{a}\},\{\boldsymbol G_{ab}\}]$ at   leading order in $1/N$. Contraction of the indices of the composite operators are implicit. }
\label{fig3}
\end{figure}

The diagrams contributing  to the $\boldsymbol\phi$-dependent part of $\Gamma_2$ at  leading order should  form closed loops  providing  each a factor $N$, which again implies contractions of the {\it same}  vectorial indices. They are displayed in Fig.~\ref{fig3}.  

 
Combined with numerical coefficient coming from the expansion of $S_{rep}[\{\boldsymbol\phi_a+\boldsymbol\varphi_a\}]$ these diagrams sum up to
\begin{equation}
\begin{array}{lll}
&\Gamma_{2,dep}^{LO}[\{\boldsymbol\phi_{a}\},\{\boldsymbol G_{ab}\}]=
\\
\\
&-\displaystyle{3\Delta_4 \over 4 N^3 T^2}  \sum_{a,b} \int_x\  \big [\hbox{Tr}_N\  \boldsymbol G_{ab}(x,x) \big ]^2 \times 
\displaystyle \big[\boldsymbol\phi_a(x)\, . \, \boldsymbol\phi_b(x)\big]^2\\
\\
&\displaystyle - {\Delta_4 \over 2 N^3T^2}  \  \sum_{a,b} \int_x\  \big [\hbox{Tr}_N\  \boldsymbol G_{ab}(x,x) \big ]^3  \boldsymbol\phi_a(x)\, .\, \boldsymbol\phi_b(x) \ . 
\end{array}
\end{equation}

After gathering all terms and taking uniform expectations values $\boldsymbol\phi_{a}$, one finds 
\begin{equation}
\begin{array}{lll}
&\displaystyle \Gamma_2[\{\boldsymbol\phi_{a}\},\{\boldsymbol G_{ab}\}]/V=\\
\\
&\displaystyle{w \over 4! N T}  \sum_{a} \Big[ \int_{ q}  \hbox{Tr}_N\  \boldsymbol G_{aa}(  q) \Big]^2\\
\\
&\displaystyle - {\Delta_2 \over  4 N T^2} \sum_{a,b} \Big[ \int_{  q}  \hbox{Tr}_N\  \boldsymbol G_{ab}(  q) \Big]^2 \\
\\
&\displaystyle -{\Delta_4 \over  8 N^3 T^2} \sum_{a,b} \Big[\int_{  q}  \hbox{Tr}_N\  \boldsymbol G_{ab}(  q ) \Big]^4 \\
\\
&\displaystyle -{3\Delta_4 \over  4 N^3 T^2} \sum_{a,b} \Big[\int_{  q}    \hbox{Tr}_N\  \boldsymbol G_{ab}(  q)  \Big]^2 (\boldsymbol\phi_a . \boldsymbol \phi_b)^2 \\
\\
&\displaystyle -{\Delta_4 \over  2 N^3 T^2} \sum_{a,b} \Big[\int_{  q}   \hbox{Tr}_N\  \boldsymbol G_{ab}(  q)  \Big]^3\ \boldsymbol\phi_a . \boldsymbol \phi_b   \ . 
\label{gamma2total}
\end{array}
\end{equation}

Finally we give the contribution of the term $ (1/2)\int_{ q} \hbox{Tr}\,\boldsymbol G({ q})\,\boldsymbol G_0^{-1}({ q};\{\boldsymbol\phi_{a}\})$ entering in Eq.~(\ref{eq_RAO($N$)M_Gamma2}). From the expression in Eq.~(\ref{bare}), and keeping only dominant terms, we  get:
\begin{equation}
\begin{array}{lll}
&\displaystyle  {1\over 2}\int_{ q} \hbox{Tr}\,  \boldsymbol G({ q})\,\boldsymbol G_0^{-1}({ q};\{\boldsymbol\phi_{a}\}) =\\
\\
&\displaystyle    {1\over 2T}\int_{ q}  ({ q}+m^2)  \hbox{Tr}\  \boldsymbol G ({  q} )  \\
\\
&\displaystyle     +{w \over  12 NT}   \sum_a {\boldsymbol \phi}_a^2 \   \int_{q}  \hbox{Tr}_N\  \boldsymbol G_{aa}(  q)   \\
\\
&\displaystyle - {\Delta_2 \over  2 N T^2}   \sum_{a,b}  {\boldsymbol \phi}_a.{\boldsymbol \phi}_b    \int_{q}  \hbox{Tr}_N\  \boldsymbol G_{ab}(q)  \\
\\
&\displaystyle -{\Delta_4 \over  2 N^3 T^2} \sum_{a,b}  ({\boldsymbol \phi}_a.{\boldsymbol \phi}_b)^3  \int_q \hbox{Tr}_N\  \boldsymbol G_{ab}(q) \ . 
\label{traceG0G}
\end{array}
\end{equation}

By using the expression Eqs.~(\ref{eq_RAO($N$)M_Gamma2}), (\ref{gamma2total}) and (\ref{traceG0G})  we obtain the following expression for $\Gamma_{2PI}$:
\begin{equation}
\begin{aligned}
&\frac{\Gamma_{2PI}[\{\boldsymbol\phi_{a}\},\{\boldsymbol G_{ab}\}]}{V}=\sum_{a=1}^n {m^2\over 2T}  {\boldsymbol \phi_a}^2 + {1\over 2}\int_{  q}  \hbox{Tr}\, \hbox{ln}\, {\boldsymbol G}^{-1}(  q) \\&
+ {1\over 2T} \int_{  q} ({ q}^2+m^2)  \hbox{Tr}\  \boldsymbol G( q) 
\\& + {w \over 4! N T}   \sum_{a=1}^n \Big(\boldsymbol\phi_{a}^2+ \int_{{ q}}  \hbox{Tr}_N\  \boldsymbol G_{aa}(  q)  \Big)^2
\\& -{N\over {2T^2}}\sum_{a,b=1}^n \Bigg [ {\Delta_2 \over {2}}  \Bigg({\boldsymbol\phi_{a}.\boldsymbol\phi_{b}+\int_{  q}  \hbox{Tr}_N\  \boldsymbol G_{ab}(q)\over N} \Bigg)^2
\\&  +  {\Delta_4 \over 4}  \Bigg({\boldsymbol\phi_{a}.\boldsymbol\phi_{b}+\int_{  q}  \hbox{Tr}_N\  \boldsymbol G_{ab}(q)\over N} \Bigg)^4\Bigg ]\,,
\label{eq_gamma2PIfinal}
\end{aligned}
\end{equation}
where the sum of the last two terms can be rewritten as  $-[N/(2T^2)]\sum_{a,b} R\big([\boldsymbol\phi_{a}.\boldsymbol\phi_{b}+\int_{  q}  \hbox{Tr}_N\  \boldsymbol G_{ab}(q)]/ N \big)$. It is easy to check that the above expression in Eq.~(\ref{eq_gamma2PIfinal}), when rewritten more generally in terms of the bare disorder function, coincides with the result of the Gaussian variational method given in the previous appendix.

\section{Derivation of the exact  $1$-PI FRG equations}
\label{appendixC}

Consider first the second Schwinger-Dyson equation, Eq.~(\ref{eq_RAO(Ninfty)M_self-energy_tilde}) and take its first derivative with respect to $k$, to $z$ and to $\rho$. The ratio of $\partial_k   \widetilde\Delta_k(\rho,z)$ over $\partial_z   \widetilde\Delta_k(\rho,z)$ immediately gives
\begin{equation}
\frac{\partial_k  \widetilde\Delta_k (\rho,z)}{ \partial_z \widetilde\Delta_k(\rho,z)}= {1\over \rho} \widetilde\Delta_k(\rho,z)\,  \partial_k I_{2k}[y_k(\rho)]\,,
\label{app_derivdelta_z}
 \end{equation}
which is Eq.~(\ref{eqderivdelta_z}) of the main text. Similarly, one has 
\begin{equation}
\frac{\partial_\rho  \widetilde\Delta_k(\rho,z)}{ \partial_z \widetilde\Delta_k(\rho,z)}= {1\over \rho} \Big(z+\widetilde\Delta_k(\rho,z)\,  \partial_\rho I_{2k}[y_k(\rho)]\Big )\,,
\label{app_derivdelta_1}
 \end{equation}
which leads to Eq.~(\ref{eq_I2partial}). When considering Eq.~(\ref{eq_RAO(Ninfty)M_self-energy_tilde}) in $z=1$ and taking first derivatives with respect to $k$ and $\rho$, one also obtains 
\begin{equation}
\frac{\partial_\rho  \widetilde\Delta_k(\rho,1)}{ \partial_\rho \widetilde\Delta_k(\rho,1)}= \frac{\widetilde\Delta_k(\rho,1)\,  \partial_k I_{2k}[y_k(\rho)]}{1+\widetilde\Delta_k(\rho,1)\,  \partial_\rho I_{2k}[y_k(\rho)]}\,,
\label{app_I2}
\end{equation}
which corresponds to Eq.~(\ref{eqderivdelta_1}).

We can now use the first Schwinger-Dyson equation, Eq.~(\ref{eq_RAO(Ninfty)M_self-energy_hat}), and take its first derivative with respect to $k$ and to $\rho$. Comparing again the two derivatives one arrives at
\begin{equation}
\begin{array}{ll}
\displaystyle \left(\partial_{\rho}y_k(\rho)-{1\over T}\partial_{\rho} \widetilde\Delta_k(\rho,1)\right) \bigg(\partial_k \widetilde\Delta_k(\rho,1)\,  I_{2k}[y_k(\rho)]+\\
\\
\widetilde\Delta_k(\rho,1)\, \partial_k I_{2k}[y_k(\rho)]+ T\,  \partial_k I_{1k}[y_k(\rho)]\bigg)\\
 =
 \\
 \displaystyle \left(\partial_k  y_k(\rho)-{1\over T}\partial_k \widetilde\Delta_k(\rho,1)\right) \bigg(1+\partial_{\rho}\widetilde\Delta_k(\rho,1)\,  I_{2k}[y_k(\rho)]+\\
\\
\widetilde\Delta_k(\rho,1)\, \partial_{\rho} I_{2k}[y_k(\rho)]+T\,  \partial_{\rho} I_{1k}[y_k(\rho)]\bigg)\,.
\label{eqyrhok}
\end{array}
\end{equation}
Note that in the above equations, we have managed to get rid of all bare quantities and make use of 1-PI quantities only. This is a prerequisite for deriving 1-PI FRG equations.

To go further, we first introduce the decomposition
\begin{equation}
\partial_k I_{p,k}[y_k(\rho)]= \widehat \partial_k I_{p,k}[y_k(\rho)]-p\,  I_{p+1 k}[y_k(\rho)]\, \partial_k  y_k(\rho) \,,
\label{derivII}
\end{equation}
where the notation $\widehat \partial_k$ indicates that the derivative acts only on the IR regulator $R_k(q)$ and not on $y_k(\rho)$, and we also use
\begin{equation} 
\partial_{\rho} I_{p,k}[y_k(\rho)]= -p\,  I_{p+1 k}[y_k(\rho)]\, \partial_{\rho}  y_k(\rho)\,.
\label{derivI}
\end{equation}

Next, after inserting Eqs.~(\ref{app_derivdelta_1}), (\ref{derivII}), (\ref{derivI}) in Eq.~(\ref{eqyrhok}), we derive a flow equation for $y_k(\rho)$:
\begin{equation}
\begin{aligned}
\partial_k y_k(\rho)=& \partial_{\rho} y_k(\rho)\Big (T\,  \widehat \partial_k I_{1k}[y_k(\rho)]+ \widetilde\Delta_k(\rho,1)  \widehat \partial_k I_{2k}[y_k(\rho)]\Big)\\&
 -\   \widehat \partial_k I_{1k}[y_k(\rho)]\, \partial_{\rho}\widetilde\Delta_k(\rho,1)
\label{eqy}
\end{aligned}
\end{equation}
Setting $T=0$ in the above equation leads to Eq.~(\ref{eq_RG_y_T0}).

Finally, the flow equation for $\Delta_k(\rho,z)$ follows from Eqs.~(\ref{app_derivdelta_z}), (\ref{app_I2}), (\ref{derivII}), (\ref{derivI}),  and (\ref{eqy}). It reads
\begin{equation}
\begin{aligned}
&\partial_k   \widetilde\Delta_k(\rho,z) = {1\over \rho}\,  \widehat\partial_k I_{2k}[y_k(\rho)]
\bigg [\Big(\widetilde\Delta_k(\rho,z)-z\,  \widetilde\Delta_k(\rho,1)\Big) \\&
\times \partial_z  \widetilde\Delta_k(\rho,z)  +   \widetilde\Delta_k(\rho,1)\, \rho\,  \partial_{\rho} \widetilde\Delta_k(\rho,z)\bigg]  + {T\over \rho}  \widehat\partial_k I_{1k}[y_k(\rho)]\\&
\times \left(\rho\, \partial_{\rho}\,  \widetilde\Delta_k(\rho,z) -z\, \partial_z  \widetilde\Delta_k(\rho,z)\right)- {1\over \rho}\,   \widehat\partial_k I_{1k}[y_k(\rho)]\\&
\times  {\partial_{\rho}  \widetilde\Delta_k(\rho,1)\over \partial_{\rho}y_k(\rho)} \left(\rho\, \partial_{\rho}\,  \widetilde\Delta_k(\rho,z) -z\, \partial_z  \widetilde\Delta_k(\rho,z)\right)\,,
\label{eqrgdelta}
\end{aligned}
\end{equation}
and when setting $T=0$ we recover Eq.~(\ref{eq_RG_Delta_T0}).

\section{RSB solution when $T \to 0$ for the ferromagnetic and the spin-glass regions}
\label{appendixD}

We study the region of parameter space where the replicon eigenvalue, when calculated from the replica-symmetric solution in the case where all sources are equal or from the analytic solution in $z$ when the sources explicitly break replica symmetry, is zero or negative. This corresponds to taking $\rho_1=\rho_2=\rho$ and $z=1$ in the $2$-replica part, and the replicon eigenvalue is given by Eq.~(\ref{eq_replicon_z=1}) of the main text.
It is negative both in what naively appears as a spin-glass (SG) phase, where the minimum of the potential is equal to $\rho_m=0$ and the associated  mass $y(\rho=0)>0$, and in part of the ferromagnetic phase, where the minimum $\rho_m>0$ and the associated mass $y(\rho_m)=0$: see Figs.~\ref{fig_naive_diagram_a} and \ref{fig_naive_diagram_b}.

In what follows, to be more explicit, we consider the model with $R'^{-1}(Y)=\lambda Y-\mu Y^3$ at the bare level, but the conclusion applies more generally. To make contact with the solution of the IR-regularized SD equations and the 1-PI FRG ones, we study the theory in the presence of an IR cutoff $k$ (but this does not introduce any additional difficulty). In this case, the running replicon eigenvalue, calculated from the replica-symmetric solution, is given by Eq.~(\ref{eq_replicon_k_z}) which we reproduce here:
\begin{equation}
\begin{aligned}
\label{eq_replicon_k_RSB}
\Lambda_{rep,k}(\rho,1)= \lambda- I_{2,k}\left [y_k(\rho)\right ]-3\mu \Delta_k(\rho,z=1)^2\,.
\end{aligned}
\end{equation}

We allow for a spontaneous replica-symmetry breaking (RSB) in the continuous Parisi-like form for the solution of the cutoff-dependent SD equations obtained from Eqs.~(\ref{eq_RAO($N$)M_stationary2}) and (\ref{schwinger},\ref{schwinger_selfenergy}), {\it i.e.}, for the purely transverse correlation matrix in replica space, 
\begin{equation}
\begin{aligned}
\label{eq_parisi}
& G_{ab,k}(q;\rho) \rightarrow \widetilde G_k(q;\rho;u)\\&
G_{aa,k}(q;\rho) \rightarrow \widehat G_{T,k}(q;\rho) + \overline{\widetilde G}_k(q;\rho)\,,
\end{aligned}
\end{equation}
where $u \in [0,1]$ is the index that labels the distance between replicas in the ultrametric structure associated with the continuous RSB,\cite{SGbook} and
\begin{equation}
\begin{aligned}
\label{eq_overlinetilde_RSB}
\overline{\widetilde G}_k(q;\rho)= \int_0^1 du \,  \widetilde G_k(q;\rho;u))\,.
\end{aligned}
\end{equation}
Similar expressions hold for the self-energies.

The SD equations then read (see also the Appendix~\ref{appendixA})
\begin{equation}
\begin{aligned}
\label{eq_hat_RSB}
y_k(\rho)= & m^2+\frac{w}{6}\Big (\rho + T I_{1,k}\left [y_k(\rho)\right ]+\int_q \overline{\widetilde G}_k(q;\rho)\Big ) \\&
+ \frac{1}{T} \int_0^1 du \Big [ R'\Big(\rho+ \int_q \widetilde G_k(q;\rho;u)\Big)- \\&
R'\Big(\rho+T I_{1,k}[y_k(\rho)]+ \int_q \overline{\widetilde G}_k(q;\rho)\Big)\Big ]\, ,\\&
\end{aligned}
\end{equation}
where $I_{p,k}$ is defined in Eq.~(\ref{eq_running_Ip}), and, after using $R'^{-1}(Y)=\lambda Y-\mu Y^3$,
\begin{equation}
\label{eq_tilde_RSB}
\rho + \int_q \widetilde{G}_k(q;\rho;u)=\lambda \Delta_k \left (\rho, u \right )-\mu \Delta_k \left (\rho, u \right )^3\,.
\end{equation}
Moreover, from the algebra of ultrametric matrices,\cite{mezard-parisi91} one also has
\begin{equation}
\begin{aligned}
\label{eq_OZ_RSB_0}
\widetilde G_{k}(q;\rho;u=0)=\frac1{T^2} \widehat G_k(q;\rho)^2 \Delta_k (\rho;u=0)\,,
\end{aligned}
\end{equation}
and
\begin{equation}
\begin{aligned}
\label{eq_OZ_RSB}
\widehat G_k(q;\rho)& -[\widetilde G_k(q;\rho)](u)=\\& \frac{T}{q^2+\widehat R_k(q^2)+y_k(\rho) +[\Delta_k (\rho)](u)}\,,
\end{aligned}
\end{equation}
where by definition $[A](u)=uA(u)-\int_0^udv A(v)$ for any function $A(u)$.

Consider first Eqs.~(\ref{eq_tilde_RSB}) and (\ref{eq_OZ_RSB}). Deriving them with respect to $u$  gives
\begin{equation}
\label{eq_tilde_RSBderiv}
\int_q \partial_u \widetilde{G}_k(q;\rho;u)=\Big (\lambda -3\mu \Delta_k \left (\rho, u \right )^2\Big )\partial_u \Delta_k \left (\rho, u \right )
\end{equation}
and, for $u\neq 0$,
\begin{equation}
\begin{aligned}
\label{eq_OZ_RSB_T0deriv}
\partial_u \widetilde G_k(q;\rho;u)= \frac{\partial_u \Delta_k (\rho;u)}{\left [q^2+\widehat R_k(q^2)+y_k(\rho) +[\Delta_k (\rho)](u)\right ]^2} \,,
\end{aligned}
\end{equation}
which can be combined in
\begin{equation}
\begin{aligned}
\label{eq_RSB_T0deriv}
&\partial_u \Delta_k(\rho;u)\Big [\lambda -I_{2,k}\left [y_k(\rho)+[\Delta_k (\rho)](u)\right ]-3\mu \Delta_k \left (\rho, u \right )^2\Big ]\\&=0
\end{aligned}
\end{equation}
with, we recall,
\begin{equation}
\begin{aligned}
\label{eq_I_2k(tau)}
I_{2,k} [y_k(\rho)+&[\Delta_k (\rho)](u) ]=\\& 
\int_q \, \frac{1}{\left [q^2+\widehat R_k(q^2)+y_k(\rho)+[\Delta_k (\rho)](u)\right ]^2}\,.
\end{aligned}
\end{equation}
The solution to Eq.~(\ref{eq_RSB_T0deriv}) is either $\partial_u \Delta_k(\rho;u)=0$ or
\begin{equation}
\begin{aligned}
\label{eq_RSB_replicon}
\lambda -I_{2,k}\left [y_k(\rho)+[\Delta_k (\rho)](u)\right ]-3\mu \Delta_k \left (\rho, u \right )^2=0\,,
\end{aligned}
\end{equation}
which is nothing but the usual marginality condition of the full RSB, \textit{i.e.}, that the replicon eigenvalue calculated with the RSB solution [compare with Eq.~(\ref{eq_replicon_k_RSB})] is zero. 
On the basis of previous studies,\cite{SGbook,mezard-parisi91} one thus expects a solution for $\Delta_k (\rho;u)$ that is constant for $0\leq u\leq u_{0k}(\rho)$ and for $u_{1k}(\rho)\leq u\leq 1$ and satisfies the marginality condition, Eq.~(\ref{eq_RSB_replicon}), for an interval $u_{0k}(\rho)\leq u\leq u_{1k}(\rho)$ with $u_{0k}(\rho)$ and $u_{1k}(\rho)$ yet to be determined. Therefore, $\Delta_k (\rho;u)=\Delta_{0k} (\rho)$  for $0\leq u\leq u_{0k}(\rho)$ and  $\Delta_k (\rho;u)=\Delta_{1k} (\rho)$  for $u_{1k}(\rho)\leq u\leq 1$ with  $\Delta_k (\rho;u)$ (but not necessarily its derivative) continuous in $u_{0k}(\rho)$ and $u_{1k}(\rho)$.

From Eq.~(\ref{eq_RSB_replicon}) considered in the interval $0\leq u\leq u_{0k}(\rho)$, one finds
\begin{equation}
\begin{aligned}
\label{eq_RSB_replicon_0}
\lambda -I_{2,k}\left [y_k(\rho)\right ]-3\mu \Delta_{0k} \left (\rho\right )^2=0\,.
\end{aligned}
\end{equation}
Similarly, one has
\begin{equation}
\begin{aligned}
\label{eq_RSB_replicon_1}
\lambda -I_{2,k}\left [y_k(\rho)+[\Delta_k (\rho)](u_{1k}(\rho))\right ]-3\mu \Delta_{1k} \left (\rho\right )^2=0\,.
\end{aligned}
\end{equation}
In addition, after deriving Eq.~(\ref{eq_RSB_replicon}) for $u_{0k}(\rho)<u<  u_{1k}(\rho)$ one obtains
\begin{equation}
\label{eq_RSB_replicon_deriv}
u \,I_{3,k}\left [y_k(\rho)+[\Delta_k (\rho)](u)\right ]=3\mu \Delta_k \left (\rho, u \right ) ,
\end{equation}
which is an implicit equation for $\Delta_k \left (\rho, u \right )$ that depends on $y_k(\rho)$.  For $u=u_{0k}(\rho)^+$ and $u=u_{1k}(\rho)^{-}$ this also implies that
\begin{equation}
\begin{aligned}
\label{eq_RSB_replicon_deriv01}
&u_{0k}(\rho)=3\mu \frac{\Delta_{0k} \left (\rho\right )}{I_{3,k}\left [y_k(\rho)\right ]}\,,\\&
u_{1k}(\rho)=3\mu \frac{\Delta_{1k} \left (\rho\right )}{I_{3,k} [y_k(\rho)+[\Delta_k (\rho)](u_{1k}(\rho)) ]}\,.
\end{aligned}
\end{equation}
Furthermore, from Eqs.~(\ref{eq_tilde_RSB}), (\ref{eq_OZ_RSB_0}) and (\ref{eq_RSB_replicon_0}), one infers the solution for $\widetilde G_{0k}(q;\rho)=\widetilde G_{k}(q;\rho;u)$ for $0\leq u\leq u_{0k}(\rho)$:
\begin{equation}
\begin{aligned}
\label{eq_OZ_RSB_u0}
\widetilde G_{0k}(q;\rho)= \frac 1{T^2} \widehat G_k(q;\rho)^2 \Delta_{0k} (\rho)
\end{aligned}
\end{equation}
with
\begin{equation}
\begin{aligned}
\label{eq_solution_DeltaI2_0_RSB}
&\Delta_{0k}(\rho)=\left (\frac{\rho}{2\mu}\right )^{1/3}\,,\\&
I_{2,k} [y_k(\rho)]=\lambda-\frac 32(2\mu)^{1/3}\rho^{2/3}\,.
\end{aligned}
\end{equation}
The renormalized mass can be determined by solving the above implicit equation. It is then easy to check that the solution to all the above equations is a seemingly replica-symmetric solution which satisfies the marginality condition, with $\Delta_{1k}(\rho)=\Delta_{0k}(\rho)=$ and $u_{1k}=u_{0k}$, yet with $u_{0k}\neq 1$. 

To better understand this odd behavior we need to also consider the additional SD equation for $y_k(\rho)$, Eq.~(\ref{eq_hat_RSB}).  This equation {\it a priori} involves an explicit dependence on the temperature. As we will see, the seemingly anomalous behavior found above results from a boundary-layer mechanism involving the temperature. To study under which conditions the limit of Eq.~(\ref{eq_hat_RSB}) exists when $T \to 0$, let us write $\widetilde G_k(q;\rho;u)=\overline{\widetilde G}_k(q;\rho) +\delta \widetilde G_k(q;\rho;u)$ and treat $\delta \widetilde G_k$ as a perturbation. The term in the square bracket with the $1/T$ factor in front can be expanded as
\begin{equation}
\begin{aligned}
\label{eq_hat_RSB_T0}
& -TI_{1,k}\left [y_k(\rho)\right ]R''\Big(\rho+ \int_q \overline{\widetilde G}_k(q;\rho)\Big)+\\&
\frac 12 R'''\Big(\rho+ \int_q \overline{\widetilde G}_k(q;\rho)\Big)\int_0^1 du \Big(\int_q \delta \widetilde G_k(q;\rho;u) \Big)^2 \\& + {\rm O}\Big(\Big(\int \delta \widetilde G_k\Big)^3,T^2\Big).
\end{aligned}
\end{equation}
A nontrivial but well-defined limit is then obtained if $\delta \widetilde G_k(q;\rho;u)={\rm O}(\sqrt T)$, which from Eq.~(\ref{eq_OZ_RSB}) also implies that $\Delta_k (\rho;u)-\overline \Delta_k (\rho)={\rm O}(\sqrt T)$.

This immediately tells us that $\Delta_{1k} \left (\rho\right )-\Delta_{0k} \left (\rho\right )={\rm O}(\sqrt T)$ and, from Eqs.~(\ref{eq_RSB_replicon_deriv01}), that $u_{1k}(\rho)-u_{0k}(\rho)={\rm O}(\sqrt T)$ [while $u_{0k}(\rho)={\rm O}(1)$]. This is consistent with the result found above at $T=0$. As a consequence, one finds a solution of Eq.~(\ref{eq_RSB_replicon_deriv}) in the form $\Delta_k (\rho;u)-\Delta_{0k} (\rho) = \sqrt T \delta_k(\rho;[u-u_{0k}(\rho)]/\sqrt T)$ where $\delta_k(\rho;\alpha)$ is given for $0\leq \alpha\leq \alpha_{1k}(\rho)$ [{\it i.e.}, for $u_{0k}(\rho)\leq u\leq u_{1k}(\rho)$] by
\begin{equation}
\begin{aligned}
\label{eq_RSB_solution}
\delta_k(\rho;\alpha) = \frac{I_{3,k}\left [y_k(\rho)\right ]}{3\left (\mu+ u_{0k}(\rho)^2 I_{4,k}\left [y_k(\rho)\right ]\right )}\, \alpha\,,
\end{aligned}
\end{equation}
where we have used that $[\Delta_k (\rho)](u)=\sqrt T u_{0k}(\rho)\delta_k(\rho;\alpha)+{\rm O}(T)$, and $\Delta_{0k} (\rho)$, $u_{0k}(\rho)$ are solutions of Eqs.~(\ref{eq_RSB_replicon_0},\ref{eq_RSB_replicon_deriv01}). Similarly,  from Eq.~(\ref{eq_OZ_RSB}), the solution for $\widetilde G_k(q;\rho;u)$ is found of the form $\widetilde G_k(q;\rho;u)-\widetilde G_{0k}(q;\rho)=\sqrt T \widetilde  g_k(q;\rho;[u-u_{0k}(\rho)]/\sqrt T)$ where $\widetilde  g_k(q;\rho;\alpha)$ is given for $0\leq \alpha\leq \alpha_{1k}(\rho)$ by
\begin{equation}
\begin{aligned}
\label{eq_RSB_solution_g}
\widetilde  g_k(q;\rho;\alpha)= \frac 1{T^2} \widehat G_k(q;\rho)^2 \delta_k(\rho;\alpha)  \,.
\end{aligned}
\end{equation}

The renormalized mass $y_k(\rho)$ is then given in the $T \to 0$ limit by
\begin{equation}
\begin{aligned}
\label{eq_tau_RSB_T0}
&y_k(\rho)= m^2+\frac{w}{6}\left (\rho + \Delta_{0k} (\rho)I_{2,k}\left [y_k(\rho)\right ]\right ) - \frac{I_{1,k}\left [y_k(\rho)\right ]}{\lambda-3\mu \Delta_{0k} (\rho)^2} \\&
+ 3\mu u_{0k}(\rho)[1-u_{0k}(\rho)] \delta_k(\rho;\alpha_{1k}(\rho))^2\frac{ \Delta_{0k} (\rho) I_{2,k}\left [y_k(\rho)\right ]^2}{\left [\lambda-3\mu \Delta_{0k} (\rho)^2\right ]^3}
\end{aligned}
\end{equation}
where we have used that $\int_q \overline{\widetilde G}_k(q;\rho)=\int_q \widetilde G_{0k}(q;\rho)+{\rm O}(\sqrt T)=\Delta_{0k} (\rho)I_{2,k}\left [y_k(\rho)\right ] +{\rm O}(\sqrt T)$ as well as the explicit expressions of $R''$ and $R'''$ [see Eq.~(\ref{eq_hat_RSB_T0})]. Eq.~(\ref{eq_tau_RSB_T0}) can be further simplified by using the marginality condition in Eq.~(\ref{eq_RSB_replicon_0}) and made more explicit by introducing the expressions in Eq.~(\ref{eq_solution_DeltaI2_0_RSB}). At this point,  $\alpha_{1k}$ is still unknown. It can be determined by requiring consistency of Eq.~(\ref{eq_tau_RSB_T0}) with $I_{2,k} [y_k(\rho)]=\lambda-(3/2)(2\mu)^{1/3}\rho^{2/3}$ [see Eq.~(\ref{eq_solution_DeltaI2_0_RSB})].

Note that due to the boundary-layer mechanism in $[u-u_{0k}(\rho)]/\sqrt T$ the SD equation for $y_k(\rho)$ is modified from the replica-symmetric form given by Eq.~(\ref{eq_y_k}) even in the limit $T\to 0$, despite the fact that the correction to $\Delta_{0k} (\rho)$ is of order ${\rm O}(\sqrt T)$ and therefore vanishes.
\\


\begin{thebibliography}{15}
\expandafter\ifx\csname natexlab\endcsname\relax\def\natexlab#1{#1}\fi
\expandafter\ifx\csname bibnamefont\endcsname\relax
  \def\bibnamefont#1{#1}\fi
\expandafter\ifx\csname bibfnamefont\endcsname\relax
  \def\bibfnamefont#1{#1}\fi
\expandafter\ifx\csname citenamefont\endcsname\relax
  \def\citenamefont#1{#1}\fi
\expandafter\ifx\csname url\endcsname\relax
  \def\url#1{\texttt{#1}}\fi
\expandafter\ifx\csname urlprefix\endcsname\relax\def\urlprefix{URL }\fi
\providecommand{\bibinfo}[2]{#2}
\providecommand{\eprint}[2][]{\url{#2}}

\bibitem{harris73}
R. Harris, M. Plischke, and M. J. Zuckermann, Phys. Rev. Lett. {\bf 31}, 160 (1973).

\bibitem{harris_review}
R. W. Cochrane, R. Harris, and M. J. Zuckermann, Phys. reports {\bf 48}, 1 (1978).

\bibitem{pelcovits78}
R. A. Pelcovits, E. Pytte, and J. Rudnick, Phys. Rev. Lett. {\bf 40}, 476 (1978); Phys. Rev. Lett. {\bf 48}, 1297 (1982).

\bibitem{aharony80}
A. Aharony and E. Pytte, Phys. Rev. Lett. {\bf 45}, 1583 (1980).

\bibitem{goldschmidt83}
Y. Y. Goldschmidt, Nucl. Phys. B {\bf 225}, 123 (1983); Phys. Rev. B {\bf 30}, 1632 (1984).

\bibitem{dudka_review}
M. Dudka, R. Folk, and Yu. Holovatch, J. Magn. Magn. Mater {\bf 294}, 305 (2005).

\bibitem{feldman}
D. E. Feldman, Phys. Rev. B {\bf 61}, 382 (2000);  Int. J. Mod. Phys. B {\bf 15}, 2945 (2001).

\bibitem{itakura03}
M. Itakura, Physical Review B {\bf 68}, 100405 (2003).

\bibitem{tissier_2loop}
M. Tissier and G. Tarjus, Phys. Rev. B {\bf 74}, 214419 (2006).

\bibitem{boyanovsky83}
D. Boyanovsky, Nucl. Phys. B {\bf 225}, 523 (1983).

\bibitem{goldschmidt84}
Y. Y. Goldschmidt, Phys. Rev. B {\bf 30}, 1632 (1984).

\bibitem{khurana84}
A. Khurana, A. Jagannathan, and J. M. Kosterlitz, Nucl. Phys. B {\bf 240}, 1 (1984); 
A. Jagannathan, M. Schaub, and J. M. Kosterlitz, Nucl. Phys. B {\bf 265}, 324 (1986).

\bibitem{fisher85}
D. S. Fisher, Phys. Rev. B {\bf 31}, 7233 (1985).

\bibitem{fisher_pathologies}
D. S. Fisher, Physica A {\bf 177}, 84 (1991).

\bibitem{toldin06}
F. P. Toldin, A. Pelissetto and E. Vicari, J. Stat. Mech. P06002 (2006).

\bibitem{doussal_largeN}
P. Le Doussal and K. Wiese, Phys. Rev. Lett. {\bf 89}, 125702 (2002); Phys. Rev. B {\bf 68}, 17402 (2003).

\bibitem{MDW_largeN}
P. Le Doussal, M. M\"uller, and K. Wiese, Phys. Rev. B {\bf 77}, 064203 (2008).

\bibitem{fisher86b}
D. S. Fisher, Phys. Rev. Lett. {\bf 56}, 1964 (1986).

\bibitem{nattermann92}
 T. Nattermann, S. Stepanow, L.-H. Tang, and H. Leschhorn, J. Phys. II (Paris) 2, 1483 (1992). 

\bibitem{narayan92}
O. Narayan and D. S. Fisher, Phys. Rev. B {\bf 46}, 11520 (1992); Phys. Rev. B {\bf 48}, 7030 (1993).

\bibitem{FRGledoussal-giamarchi}
P. Chauve, T.Giamarchi and P. Le Doussal, Phys. Rev. B 62, 6241 (2000).

\bibitem{FRGledoussal}
P. Le Doussal, K. J. Wiese, and P. Chauve, Phys. Rev. B \textbf{66}, 174201 (2002); Phys. Rev. E \textbf{69}, 026112 (2004). 
P. Le Doussal and K. J. Wiese, Phys. Rev. E \textbf{79}, 051106 (2009). 

\bibitem{ledoussal-wiese_review}
K. J. Wiese and P. Le Doussal, Markov Processes Relat. Fields {\bf 13}, 777 (2007).
 
\bibitem{tarjus04}
G. Tarjus and M. Tissier, Phys. Rev. Lett. {\bf 93}, 267008 (2004); Phys. Rev. B {\bf 78}, 024203 (2008).

\bibitem{tissier06}
M. Tissier and G. Tarjus, Phys. Rev. Lett. {\bf 96}, 087202 (2006); Phys. Rev. B {\bf 78}, 024204 (2008).

\bibitem{tissier11}
M. Tissier and G. Tarjus, Phys. Rev. Lett. {\bf 107}, 041601 (2011); Phys. Rev. B {\bf 85}, 104202 (2012); \textit{ibid}, 104203 (2012).

\bibitem{balog-tarjus}
I. Balog and G. Tarjus, Phys. Rev. B {\bf 91}, 214201 (2015).

\bibitem{fisher_review}
D. S. Fisher, J. Appl. Phys. 61, 3672 (1987).

\bibitem{balents-doussal}
L. Balents and P. Le Doussal, Phys. Rev. E {\bf 69}, 061107 (2004); P. Le Doussal, Ann. Phys. {\bf 325}, 49 (2010).

\bibitem{mezard-parisi91}
M. M\'ezard and G. Parisi, J. Phys. I (France) \textbf{1}, 809 (1991). 

\bibitem{luttinger-ward}
J. M. Luttinger and J. C. Ward, Phys. Rev. {\bf 118}, 1417 (1960); G. Baym, Phys. Rev. {\bf 127}, 1391 (1962).

\bibitem{cornwall74}
J. M. Cornwall, R. Jackiw and E. Tomboulis, Phys. Rev. D {\bf 10}, 2428 (1974).

\bibitem{mouhanna-tarjus}
D. Mouhanna and G. Tarjus, Phys. Rev. E \textbf{81}, 051101 (2010).

\bibitem{SGbook}
M. M\'ezard, G. Parisi, and M. A. Virasoro, {\it Spin glass theory and beyond} (World Scientific, Singapore, Singapore,1987).

\bibitem{almeida78}
J. R. L. Almeida and D. J. Thouless, J. physique A: Math. Gen.  \textbf{11}, 983 (1978).

\bibitem{footnote_firstorder}
For this reason too, we have not considered the possibility of a first-order SG-FM transition.

\bibitem{BBM96}
L. Balents L, J.-P. Bouchaud and M. Mezard, J. physique I  \textbf{6}, 1007 (1996).

\bibitem{mouhanna-tarjus_new}
D. Mouhanna and G. Tarjus, in preparation (2016).

\bibitem{berges02}
J. Berges, N. Tetradis, and C. Wetterich, Phys. Rep. \textbf{363}, 223 (2002).

\bibitem{litim}
D. Litim, Phys. Lett. B {\bf 486}, 92 (2000).

\bibitem{fisher_zeroT}
D. S. Fisher, Phys. Rev. Lett. {\bf 56}, 416 (1986).

\bibitem{villain_zeroT}
J. Villain, Phys. Rev. Lett. {\bf 52}, 1543 (1984).

\bibitem{footnote_limits}
This corresponds to first considering the 1-loop FRG equation at finite $N$ and finite $T$ and then taking the large-$N$ limit.

\bibitem{footnote_krzakala}
Note that the rigorous argument, put forward in [\onlinecite{krzakala}], that forbids the divergence of the spin-glass susceptibility when the ferromagnetic susceptibility does not itself diverge in ferromagnetic systems with local quenched disorder is valid for Ising spins and does not carry over to spins with continuous symmetry. A spin-glass phase in the RAO($N$)M is thus not {\it a priori} forbidden.

\bibitem{krzakala}
F. Krzakala, F. Ricci-Tersenghi, D. Sherrington, and L. Zdeborov\'a, J. Phys. A: Math. Theor. {\bf 44}, 042003 (2011).







\end{thebibliography}
\end{document}